\documentclass{article}
\usepackage{amsmath,amsfonts,amsthm,amssymb,amscd,latexsym, bm,bbm}
\usepackage{hyperref}
\usepackage[usenames,dvipsnames]{color}
\hypersetup{colorlinks,
citecolor=Blue,
linkcolor=Blue,
urlcolor=Blue}

\usepackage[T2A]{fontenc}
\usepackage{graphicx}
\usepackage{mathrsfs}

{\theoremstyle {definition} \newtheorem {defi} {Definition} [section] }
{\theoremstyle {plain}  \newtheorem {theo} [defi] {Theorem}}
{\theoremstyle {plain}  \newtheorem {cor} [defi]{Corollary}}
{\theoremstyle {plain}  \newtheorem {lem} [defi]{Lemma}}
{\theoremstyle {plain}  \newtheorem {prop} [defi]{Proposition}}
{\theoremstyle {remark} \newtheorem {rem}[defi] {Remark}}
{\theoremstyle {remark} \newtheorem {example}[defi] {Example}}

\numberwithin{equation}{section}
\newtheorem*{defi*}{Definition}
\newtheorem*{problem*}{Problem}

\newtheorem*{rem*}{Remark}
\newtheorem*{note*}{Note}

\makeatletter \@addtoreset{equation}{section} \makeatother

\newcommand{\mC}{\mathbb{C}}
\newcommand{\mR}{\mathbb{R}}
\newcommand{\mT}{\mathbb{T}}
\newcommand{\mZ}{\mathbb{Z}}
\newcommand{\mN}{\mathbb{N}}

\newcommand{\mI}{\mathbb{I}}

\newcommand{\AAA}{{\cal A}}
\newcommand{\BB}{{\cal B}}
\newcommand{\CC}{{\cal C}}
\newcommand{\DD}{{\cal D}}
\newcommand{\EE}{{\cal E}}
\newcommand{\FF}{{\cal F}}
\newcommand{\GG}{{\cal G}}
\newcommand{\HH}{{\cal H}}
\newcommand{\LL}{{\cal L}}

\newcommand{\SSS}{{\cal S}}
\newcommand{\VV}{{\cal V}}

\newcommand{\UU}{{\cal U}}
\newcommand{\YY}{{\cal Y}}
\newcommand{\RR}{{\cal R}}

\newcommand{\JJ}{{\cal J}}
\newcommand{\TT}{{\cal T}}

\newcommand{\KK}{{\cal K}}

\newcommand{\eps}{\varepsilon}
\newcommand{\ph}{\varphi}

\newcommand{\Imm}{\operatorname{Im}}
\newcommand{\Ree}{\operatorname{Re}}

\newcommand{\Id}{\operatorname{Id}}
\newcommand{\diag}{\operatorname{diag}}

\newcommand{\Om}{\Omega}
\newcommand{\tht}{\theta}
\newcommand{\al}{\alpha}
\newcommand{\ga}{\gamma}

\newcommand{\La}{\Lambda}
\newcommand{\del}{\delta}

\newcommand{\ov}{\overline}
\newcommand{\wid}{\widetilde}

\newcommand{\ssk}{\smallskip}
\newcommand{\msk}{\medskip}

\newcommand{\chp}{\partial}
\newcommand{\MO}{\mbox{\bf E}\,}
\newcommand{\MOOM}{\mbox{\bf E}_{\Omega_R}\,}
\newcommand{\MOF}{\mbox{\bf E}_{\FF_l\cup\Omega_R}\,}
\newcommand{\PR}{\mbox{\bf P}\,}
\newcommand{\ds}{\displaystyle{}}
\newcommand{\ra}{\rightarrow}
\newcommand{\ran}{\rangle}
\newcommand{\lan}{\langle}
\newcommand{\volna}{\thicksim}
\newcommand{\sm}{\setminus}
\newcommand{\raw}{\rightharpoonup}

\newcommand{\bee}{\begin{equation}}
\newcommand{\eee}{\end{equation}}
\newcommand{\btt}{\begin{theo}}
\newcommand{\ett}{\end{theo}}
\newcommand{\bl}{\begin{lem}}
\newcommand{\el}{\end{lem}}
\newcommand{\bpp}{\begin{prop}}
\newcommand{\epp}{\end{prop}}
\newcommand{\bcc}{\begin{cor}}
\newcommand{\ecc}{\end{cor}}
\newcommand{\bdd}{\begin{def}}
\newcommand{\edd}{\end{def}}
\newcommand{\brr}{\begin{rem}}
\newcommand{\err}{\end{rem}}

\begin{document}

\large
\title
{Statistical mechanics of nonequilibrium systems of rotators with alternated spins}
\author{A. Dymov}
\date{Universit\'e de Cergy-Pontoise, CNRS, Mathematics Department, \\
F-95000 Cergy-Pontoise \\ e-mail: adymov88@gmail.com}
\maketitle

\begin{abstract}
We consider a finite region of a d-dimensional lattice of nonlinear Hamiltonian rotators, where neighbouring rotators  have opposite (alternated) spins and are coupled by a small potential of size $\eps^a,\, a\geq1/2$. We weakly stochastically perturb the system in such a way that each rotator interacts with its own stochastic thermostat with a force of order $\eps$. Then we introduce action-angle variables for the system of uncoupled rotators ($\eps=0$) and note that the sum of actions over all nodes is conserved by the purely Hamiltonian dynamics of the system with $\eps>0$.  We investigate the limiting (as $\eps \ra 0$) dynamics of actions for solutions of the $\eps$-perturbed system on time intervals of order $\eps^{-1}$. It turns out that the limiting dynamics is governed by a certain autonomous (stochastic) equation for the vector of actions. This equation has a completely non-Hamiltonian nature. This is a consequence of the fact that the system of rotators with alternated spins do not have resonances of the first order. 

The $\eps$-perturbed system has a unique stationary measure $\wid \mu^\eps$ and is mixing. Any limiting point of the family $\{\wid \mu^\eps\}$ of stationary measures as $\eps\ra 0$ is an invariant measure of the system of uncoupled integrable rotators. There are plenty of such measures. However, it turns out that only one of them describes the limiting dynamics of the $\eps$-perturbed system: we prove that a limiting point of $\{\wid\mu^\eps\}$ is unique, its projection to the space of actions is the unique stationary measure  of the autonomous equation above, which turns out to be  mixing, and its projection  to the space of angles is the normalized Lebesque measure on the torus $\mT^N$.  

The results and convergences, which concern the behaviour of actions on long time intervals, are uniform in the number $N$ of rotators. Those, concerning the stationary measures, are uniform in $N$ in some natural cases. 
\end{abstract}

\tableofcontents


\bigskip

\section{Introduction}
\label{nlsec:intro}

Investigation of the energy transport in crystals is one of the main problems  in the  non-equilibrium statistical mechanics (see \cite{Leb}).  It is closely related to the
derivation of autonomous equations which describe a flow of quantities, conserved by the Hamiltonian  (for example, the flow of energy and the corresponding heat equation). In the classical setting one looks for the energy transport in a Hamiltonian system, coupled with thermal baths which have different temperatures. This coupling is weak in geometrical sense: the thermal baths interact with the Hamiltonian system only through its boundary.  
Unfortunately, for the moment of writing this problem turns out to be too difficult due to the weakness of the coupling. In this case even the existence of a stationary state in the system is not clear (see \cite{Eck, RBT}, and \cite{Tr, Dym12} for a similar problem in a deterministic setting).  That is why usually one modifies the system in order to get some additional ergodic properties. Two usual ways to achieve that are {\it i)} to consider a weak perturbation of the hyperbolic system of independent particles (\cite{DL, R}); {\it ii)} to perturb each particle of the Hamiltonian system by stochastic dynamics of order one (\cite{BLL,BO,BBO,BLLO,BeKLeLu,LO}).

In particular, 
in \cite{DL} the authors consider a finite region of a lattice of weakly interacting geodesic flows on manifolds of negative curvature. In \cite{LO} the authors investigate that of weakly interacting anharmonic oscillators perturbed by energy preserving stochastic exchange of momentum  between neighbouring nodes. Then in the both papers the authors rescale the time appropriately and, tending the strength of interaction in the Hamiltonian system to zero, show that the limiting dynamics of local energy is governed by a certain autonomous (stochastic) equation, which turns out to be the same in the both papers.
 
In all works listed in {\it i)} and {\it ii)} above,  a source of the additional ergodic properties (the hyperbolicity of unperturbed system and the coupling of Hamiltonian system with stochastic dynamics) stays of order one.  
It is natural to investigate what happens when its intensity goes to zero. Such situation was studied in \cite{BOS},\cite{BeHu} and \cite{BeHuLeLO}. In \cite{BOS} the authors consider the FPU-chain with the nonlinearity replaced by energy preserving stochastic exchange of momentum between neighbouring nodes. They investigate the energy transport under the limit when the rate of this exchange tends to zero.
In \cite{BeHu} the authors study a pinned disordered harmonic chain, where each oscillator is weakly perturbed by energy preserving noise and an anharmonic potential.  They investigate behaviour of an upper bound for the Green-Kubo conductivity  under the limit when the perturbation vanishes.
In  \cite{BeHuLeLO}  the authors consider an infinite chain of weakly coupled cells, where each cell is weakly perturbed by energy preserving noise. They formally find a main term of the Green-Kubo conductivity and  investigate its limiting behaviour when strength of the noise tends to zero.

In the present paper we weakly couple each particle of a Hamiltonain system with its own Langevin-type stochastic thermal bath and study the energy transport when this coupling goes to zero (note that such stochastic perturbation does not preserve the energy of the system). 
So, as in the classical setting given above, we study the situation when the coupling of the Hamiltonian system with the thermal baths is weak, but the weakness is understood  in a different, non-geometrical sense. This setting seems to be natural: one can think about a crystal put in some medium and weakly interacting with it. 


However, as in a number of works above, we have to assume the coupling of particles in the Hamiltonian system also to be  sufficiently weak. Namely, we  rescale the time and let the strength of interaction in the Hamiltonian system go to zero in the appropriate scaling with the coupling between the Hamiltonian system and the thermal baths.  We prove that under this limit the local energy of the system satisfies a certain autonomous (stochastic) equation, which turns out to be mixing, 
\footnote{I.e. the this equation has a unique stationary measure, and its solutions converge weakly in distribution to this  measure.}
 and show that the limiting behaviour of steady states of the system is governed by a unique stationary measure of this equation. 
 
Since the systems of statistical physics are of very high dimension, then it is crusial to control the dependence of the systems on their size. Our work satisfies this physical requirement: most of results we obtain are uniform in the size of the system.
 

More specifically, we consider a $d$-dimensional lattice of $N$  nonlinear Hamiltonian rotators. The neighbouring rotators have opposite spins and interact weakly via a potential (linear or nonlinear) of size  $\eps^a$, $a\geq 1/2$. We couple each rotator with its own stochastic Langevin-type thermostat of arbitrary positive temperature by a coupling of size $\eps$. 
We introduce action-angle variables for the uncoupled Hamiltonian, corresponding to $\eps$=0, and note that a sum of actions is conserved by the Hamiltonian dynamics with $\eps> 0$. That is why the actions play for us the role of the local energy.
In order to feel the interaction between rotators and the influence of thermal baths, we consider time interval of order $t\volna \eps^{-1}.$ 
We let $\eps$ go to zero and obtain that the limiting dynamics of actions is given by equation which describes their autonomous (stochastic) evolution. It has completely non-Hamiltonian nature (i.e. it does not feel the Hamiltonian interaction of rotators) and describes a non-Hamiltonian flow of actions. 
Since we consider a time interval of order $t\volna \eps^{-1}$, 
in the case $a=1/2$ we have $t\volna(\mbox{Hamiltonian interaction})^{-2}$.
In \cite{DL} and \cite{LO} scalings of time and of the Hamiltonian interaction satisfy the same relation. Since the autonomous equations for energy obtained there feel the Hamiltonian interaction, in our setting one could expect to obtain an autonomous equation for actions which also feels it. However, it is not the case.  

For readers, interested in the limiting dynamics of energy, we note that it can be easily expressed in terms of the limiting dynamics of actions.

The system in question (i.e. the Hamiltonian system, coupled with the thermal baths) is mixing.
 We show that its stationary measure $\wid\mu^\eps$, written in action-angle variables, converges, as $\eps\ra 0$, to the product of the unique stationary measure $\pi$ of the obtained autonomous equation for actions and the normalized Lebesgue measure on the torus $\mT^N$. 

We prove that the convergence as $\eps\ra 0$ of the vector of actions to a solution of the autonomous equation  is uniform in the number of rotators $N$. 
The convergence of the stationary measures is also uniform in $N$, in some natural cases.

We use Khasminski-Freidlin-Wentzell-type averaging technics in the form developed in \cite{KuPi, Kuk10, Kuk12}. For a general Hamiltonian these methods are applied when the interaction potential is of the same order as the coupling with the thermal baths, i.e. $a=1$. However, we find a large natural class of Hamiltonians such that the results stay the same even if $1/2 \leq a <1$, i.e. when the interaction potential is stronger. This class consists of Hamiltonians which describe lattices of rotators with alternated spins, when neighbouring rotators rotate in opposite directions. 
It has to do with the fact that such systems of rotators do not have resonances of the first order. To apply the methods above in the case $1/2 \leq a <1$ we kill the leading term of the interaction potential by a global canonical transformation which is $\eps^a$-close to the identity. 
The resulting autonomous equation for actions has the non-Hamiltonian nature since the averaging eliminates Hamiltonian terms. 

Note that a similar (but different) problem was considered in \cite{FW06} (see also Chapter \nolinebreak 9.3 of \cite{FW}). There the authors study a system of oscillators, weakly interacting via couplings of size $\eps$. Each oscillator is weakly perturbed by its own stochastic Langevin-type thermostat, also of the size $\eps$. The authors consider time interval of order $\eps^{-1}$ and using the averaging method show that under the limit $\eps\ra 0$ the local energy satisfies an autonomous (stochastic) equation. 
Compare to our work, in this study the authors do not investigate the limiting (as $\eps\ra 0$) behaviour of stationary measures as well as the dependence of the results on the number of particles in the system.   

The present paper is a full version of the article \cite{Dym}. Compare to the latter, here we present a complete proof of Theorem \ref{nllem:prest} (while in \cite{Dym} we only sketch it, since it is too long and technical), and give some generalizations of our results (see Section \ref{nlsec:generalizations}).

\section{Set up and main results}
\label{nlsec:introduction}
\subsection{Set up}
\label{nlsetup}

We consider a lattice $\CC \subset \mZ^d$, $d\in\mN$, which consists of 
$N$  nodes $j\in\CC,\, j=(j_1,\ldots,j_d).$ In each node we put an integrable nonlinear Hamiltonian rotator which is coupled through a small potential with rotators in neighbouring positions. The rotators are described by complex variables $u=(u_j)_{j \in\CC} \in \mC^{N}$. Introduce the  symplectic structure by the $2$-form $\frac{i}{2}\sum\limits_{j\in\CC}\,d u_j \wedge d\ov u_j=\sum\limits_{j\in\CC} d x_j\wedge d y_j$, if $u_j=x_j+iy_j$. Then the system of rotators is given by the Hamiltonian equation
\bee
\label{nlhe}
\dot u_j = i\nabla_{j} H^\eps (u), \quad j\in\CC,
\eee
where the dot means a derivative in time $t$ and $\nabla_j H^\eps = 2 \chp_{\ov u_j} H^\eps $  is the gradient of the Hamiltonian $H^\eps$ with respect to the Euclidean scalar product $\cdot$ in $\mC \simeq \mR^{2}:$ 
\bee
\label{nlscalarproduct}
\mbox{for} \quad z_1,z_2 \in \mC \quad z_1\cdot z_2 :=  \Ree z_1 \Ree z_2+\Imm z_1\Imm z_2= \Ree z_1 \ov z_2.
\eee
 The Hamiltonian has the form
\begin{equation}
\label{nlham0}
H^\eps=\frac12 \sum\limits_{j \in\CC} F_j (|u_j|^2) + \frac{\eps^a}{4} \sum\limits_{j,k \in\CC: |j-k|=1} G(|u_j-u_{k}|^2), 
\end{equation}
where $|j|:=|j_1|+\ldots+|j_d|$, $a\geq 1/2$ and $F_j, G: [0,\infty) \rightarrow \mR$ are sufficiently smooth functions with polynomial bounds on the growth at infinity (precise assumptions are given below). 

We weakly couple each rotator with its own stochastic thermostat of arbitrary temperature $\TT_j$, satisfying 
$$0<\TT_{j}\leq C<\infty, $$
where the constant $C$ does not depend on $j,N,\eps$. More precisely, we consider the system
\begin{equation}
\label{nlini}
\dot u_j = i\nabla_{j} H^\eps (u) + \eps g_j(u) + \sqrt{\eps  \TT_j} \dot\beta_j, \quad u_j(0) = u_{0j},\quad j\in\CC,
\end{equation} 
where $\beta = (\beta_j)_{j \in\CC}\in \mC^{N}$  are standard complex independent Brownian motions. That is, their real and imaginary parts are standard real independent Wiener processes. Initial conditions $u_0=(u_{0j})_{j\in\CC}$ are random variables, independent from $\beta$.  They are the same for all $\eps$. Functions $g_j$, which we call "dissipations", have some dissipative properties, for example,  $g_j(u)=-u_j$ (see Remark \ref{nlrem:g_j} below). They couple only neighbouring rotators, i.e. $g_j(u)=g_j\big((u_k)_{k\in\CC:|k-j|\leq 1}\big)$.

The scaling of the thermostatic term in equation (\ref{nlini}) is natural since,  in view of the dissipative properties of $g_j$, the only possibility for solution of equation
$\dot u_j=\eps g_j(u) + \eps^b \sqrt{ \TT_j} \dot\beta_j, \; j\in\CC$, to stay of the order 1 for all $t\geq 0$ as $\eps\ra 0$ is $b=1/2$.

The case $a=1/2$ is the most difficult, so further on we consider only it, the other cases are similar.
Writing the corresponding equation (\ref{nlini}) in more details, we obtain
\begin{align}
\label{nlini_e}
\dot u_j &= if_j(|u_j|^2)u_j + i\sqrt\eps \sum\limits_{k \in\CC: |j-k|=1} G^{\prime}(|u_j-u_k|^2)(u_j-u_k) + \eps  g_j(u) + \sqrt {\eps \TT_j}\dot \beta_j, \\
\label{nlini_c}
u_j(0)&=u_{0j}, \; j \in \CC, 
\end{align}
where $f_j(x):=F^{\prime}_j(x)$ and the prime denotes a derivative in $x$.
\begin{rem}
\label{nlrem:g_j}
Our principal example is the case of {\it diagonal} dissipation, when   $g_j(u)=-|u_j|^{p-2}u_j$ for all $j\in\CC$ and some $p\in\mN,\,p\geq 2$. In particular, the linear diagonal dissipation when $p=2$ and $g_j(u)=-u_j$.  The diagonal dissipation does not provide any interaction between rotators. In this case each rotator is  just coupled with a Langevin-type thermostat. The results become more interesting if we admit functions $g_j$ of a more involved structure which not only introduces dissipation, but also provides some non-Hamiltonian interaction between the rotators. 
If for the reader the presence of the non-Hamiltonian interaction seems unnatural, he can simply assume that the dissipation is diagonal. 
\end{rem}

We impose on the system assumptions {\it HF, HG, Hg} and {\it HI}. Their exact statements are given at the end of the section. Now we briefly summarize them. We fix some $p\in\mN,\,p\geq 2$, and
assume that $f_j(|u_j|^2)=(-1)^{|j|}f(|u_j|^2)$, where $f(|u_j|^2)$ is separated from zero and has 
at least a polynomial growth of a power $p$  ({\it HF}). It means that the leading term of the Hamiltonian $H^\eps$ is a nonlinearity which rotates the neighbouring rotators in opposite directions sufficiently fast. We call it the "alternated spins condition". The function $G'(|u_j|^2)$ is assumed to have at most the polynomial growth of the power $p-2$, i.e. the interaction term in (\ref{nlini_e}) has  the growth at most of the  power   $p-1$ ({\it HG}). The functions $g_j(u)$ have some dissipative properties and have the polynomial growth of the power $p-1$  ({\it Hg}). The functions $f, G$ and $g_j$ are assumed to be sufficiently smooth. In {\it HI} we assume that the initial conditions are "not very bad", this assumption is not restrictive. For an example of functions $f,G$ and $g_j$ satisfying assumptions  {\it HF, HG} and {\it Hg}, see  Example \nolinebreak \ref{nlexam:conditions}.
 In the case $a\geq 1$ the assumptions get weaker, see Remark \ref{nlrem:a>1}. In particular, the rotators are permitted to rotate in any direction.         

\subsection{Main results}
\label{nlmainresults}

For a vector $u=(u_k)_{k\in\CC}\in\mC^N$ we define the corresponding vectors of {\it actions} and {\it angles}  
$$I=I(u)=(I_k(u_k))_{k\in\CC},\; I_k=\frac12 |u_k|^2 \quad \mbox{and}\quad  \ph=\ph(u)=(\ph_k(u_k))_{k\in\CC}, \; \ph_k=\arg u_k,$$ 
where we put $\ph_k(0)=0$. Thus, $(I,\ph)\in\mR^N_{+0}\times\mT^N$, where $\mR^N_{+0}=\{I=(I_k)_{k\in\CC}\in\mR^N:\, I_k\geq 0\; \forall k\in\CC\}$, and $u_k=\sqrt{2I_k}e^{i\ph_k}$. \nolinebreak
\footnote{Usually, for a vector from $\mC^N$, denoted by the letter $u$, we write its actions and angles as above, and for a vector, denoted by $v$, we write them as $(J,\psi)$, $J=J(v),\;\psi=\psi(v)$.} 
The variables $(I,\ph)$ form the action-angle coordinates for the uncoupled Hamiltonian (\ref{nlham0})$|_{\eps=0}$. 

The direct computation shows that the 
sum of actions $\sum\limits_{k\in\CC} I_k$ is a first integral of the Hamiltonian $H^\eps$ for every $\eps>0$. 
That is why for our study the actions will play the role of the local energy, and we will examine their limiting behaviour as $\eps\ra 0$ instead of the limiting behaviour of energy. Moreover, the reader, interested in the limiting dynamics of energy, will easily express it in terms of the limiting dynamics of actions, since in view of (\ref{nlham0}), the energy of a $j$-th rotator tends to $\frac12 F_j(2I_j)$ as $\eps\ra 0$, see Corollary \ref{nllocen} for details. 
\ssk  

Let us write a function $h(u)$ in the action-angle coordinates, $h(u)=h(I,\ph)$. Denote its averaging in angles as
$$\lan h \ran(I):=\int\limits_{\mT^{N}} h(I,\ph) \, d\ph.$$
Here and further on $d\ph$ denotes the normalized Lebesgue measure on the torus $\mT^{N}$.
Let
\bee\label{nlusre}
\RR_j(I):=\lan g_j(u)       \cdot u_j \ran,
\eee
where we recall that the scalar product $\cdot$ is given by (\ref{nlscalarproduct}).
It is well known that under our assumptions a solution $u^\eps(t)$ of system (\ref{nlini_e})-(\ref{nlini_c}) exists, is unique and is defined for all $t\geq 0$ (\cite{Khb}).
Let $I^\eps(t)$ and $\ph^\eps(t)$ be the corresponding vectors of actions and angles, i.e. $I^\eps(t)=I(u^\eps(t))$, $\ph^\eps(t)=\ph(u^\eps(t))$. We fix arbitrary $T\geq 0$ and examine the dynamics of actions $I^\eps$ on the long-time interval $[0,T/\eps]$ under the limit $\eps\ra 0$.   It is useful to pass to the slow time $\tau=\eps t$, then the interval  $t\in[0,T/\eps]$ corresponds to $\tau\in[0,T]$. We prove
\btt
\label{nltheo:avIintro}
In the slow time the family of distributions of the actions $\DD(I^{\eps}(\cdot))$  with $\eps\rightarrow 0$ converges weakly  on $C([0,T], \mR^N)$ to a distribution $\DD(I^0(\cdot))$ of a unique weak solution $I^0(\tau)$ of the system
\begin{align}
\label{nlav_I}
&d I_{j} = (\RR_j(I) +\TT_{j} )\,d\tau + \sqrt{2I_{j} \TT_{j}}\,d\wid\beta_{j}, \quad j\in\CC,\\
\label{nlav_Ic}
&\DD(I(0))=\DD(I(u_0)), 
\end{align}
where $\wid\beta_{j}$ are standard real independent Brownian motions.  
The convergence is uniform in $N$.  
\ett
The limiting measure satisfies some estimates, for details see Theorem \ref{nltheo:fin_dyn}. 
In order to speak about the uniformity in $N$ of convergence, we assume that the set $\CC$ depends on the number of rotators $N$ in such a way that $\CC(N_1)\subset\CC(N_2)$ if $N_1<N_2$. The functions $G, F_j$ and the temperatures $\TT_j$ are assumed to be independent from $N$, while the functions $g_j$ are assumed to be independent from $N$ for $N$ sufficiently large (depending on $j$). 
\footnote{We can not assume that $g_j$ is independent from $N$ for all $N\in\mN$ since for small $N$ the $j$-th rotator may have fewer neighbours then for large $N$.}
The initial conditions $u_0$ are assumed to agree in $N$, see assumption {\it HI(ii)}.
The uniformity of convergence of measures through all the text  is understood in the sense of finite-dimensional projections. For example, for Theorem \ref{nltheo:avIintro}  it means that for any $\La\subset\mZ^d$ which does not depend on $N$ and satisfies $\La\subset\CC(N)$ for all $N\geq N_\La,\,N_\La\in\mN$, we have
\footnote{We recall that the weak convergence of measures is metrisable (see \cite{Dud}, Theorem 11.3.3), so it makes sense to talk about its uniformity.}
$$\DD\big((I^\eps_j(\cdot))_{j\in\La}\big)\raw\DD\big((I^0_j(\cdot))_{j\in\La}\big)\quad\mbox{as}\quad \eps \ra 0 \quad\mbox{uniformly in } N\geq N_\La.$$

Note that in the case of diagonal dissipation $g_j(u)=-u_j|u_j|^{p-2}$  equation (\ref{nlav_I}) turns out to be diagonal
\bee\label{nlavdiag}
d I_{j} = (-(2I_j)^{p/2} +\TT_{j} )\,d\tau + \sqrt{2I_{j} \TT_{j}}\,d\wid\beta_{j}, \quad j\in\CC. 
\eee
For more examples see Section \ref{nlsec:example}.

Relation (\ref{nlav_I}) is an autonomous equation for actions which describes their transport under the limit $\eps\ra 0$. Since it is obtained by the avergaing method we call it the {\it averaged equation}. Note that the averaged equation does not depend on a precise form of the potential $G$.  It means that the limiting dynamics does not feel the Hamiltonian interaction between rotators and provides a flow of actions between nodes only if the dissipation is not diagonal.  
\ssk

In Section \ref{nlsec:measure} we investigate the limiting behaviour, as $\eps\ra 0$, of  averaged in time joint distribution of actions and angles $I^\eps, \ph^\eps$. See Theorem \ref{nltheo:measure}.
\ssk

 Recall that a stochastic differential equation  is {\it mixing} if it has a unique stationary measure and all solutions of this equation weakly converge to this stationary measure in distribution.
It is well known that equation (\ref{nlini_e}) is mixing (see \cite{Khb,Ver,VerPol}). Denote its stationary measure by $\wid\mu^{\eps}$.
Denote the projections to spaces of actions and angles by $\Pi_{ac}:\,\mC^N\ra\mR^N$ and $\Pi_{ang}:\,\mC^N\ra\mT^N$  correspondingly. 
Let $$\CC^\infty:=\cup_{N\in\mN}\CC(N).$$
We will call  equation (\ref{nlav_I}) for the case $N=\infty$, i.e. with $\CC$ replaced by $\CC^\infty$, the "averaged equation for the infinite system of rotators". Let $\mR^\infty$ ($\mC^\infty$) be the space of real (complex) sequences provided with the Tikhonov topology.
 \begin{theo}
\label{nltheo:stmintro}
{\it (i)} The averaged equation (\ref{nlav_I}) is mixing.

{\it (ii)} For the unique stationary measure $\wid\mu^\eps$ of (\ref{nlini_e}), written in the action-angle coordinates, we have
\bee\label{nlintroconv}
(\Pi_{ac}\times\Pi_{ang})_*\wid\mu^\eps \rightharpoonup \pi\times\, d\ph \quad \mbox{as} \quad \eps\ra 0,
\eee
where $\pi$ is a unique stationary measure of the averaged equation   (\ref{nlav_I}). 
If the averaged equation for the infinite system of rotators has a unique stationary measure $\pi^\infty$ in the class of measures defined on the Borel $\sigma$-algebra $\BB(\mR^\infty)$ and satisfying $\sup\limits_{j\in\CC^\infty}\langle\pi^\infty , I_j \rangle <\infty$, then convergence (\ref{nlintroconv}) is uniform in $N$.

{\it (iii)} The vector of actions $I^\eps(\tau)$, written in the slow time, satisfies
\bee
\label{nlieeeeee}
\lim\limits_{\tau\ra\infty}\lim\limits_{\eps\ra 0} \DD(I^\eps(\tau))=\lim\limits_{\eps\ra 0}\lim\limits_{\tau\ra \infty} \DD(I^\eps(\tau))=\pi.
\eee
\end{theo}
We prove this theorem in Section \ref{nlsec:st_measures}. 
Each limiting point (as $\eps\ra 0$) of the family of measures  $\{\wid\mu^\eps, 0<\eps\leq 1\}$  is an invariant measure of the system of uncoupled integrable rotators, corresponding to (\ref{nlhe})$|_{\eps=0}$. It has plenty of invariant measures. Theorem \ref{nltheo:stmintro} ensures that only one of them is a limiting point, and distinguishes it.

Arguing as when proving Theorem \ref{nltheo:stmintro}, we can show that the averaged equation for the infinite system of rotators has a stationary measure belonging to the class of measures above, but we do not know if it is unique. However, it can be proven that it is unique if this equation is diagonal.
In this case the convergence (\ref{nlintroconv}) holds uniformly in $N$. In particular, this happens when  the dissipation is diagonal. For more examples see Section \nolinebreak \ref{nlsec:example}.

\subsection{Strategy}
\label{nlsec:strategy}
In this section we describe the main steps of proofs of Theorems \ref{nltheo:avIintro} and \ref{nltheo:stmintro}.  
\ssk

First we need to obtain uniform in $\eps$, $N$ and time $t$ estimates for solutions of (\ref{nlini_e}). For a general system of particles there is no reason why all the energy could not concentrate at a single position, forming a kind of delta-function as $N\rightarrow \infty$. It is remarkable that in our system this does not happen, at least on time intervals of order $1/\sqrt{\eps}$, even without alternated spins condition and in absence of dissipation. One can prove it working with the family of norms $\|\cdot\|_{j,q}$ (see {\it Agreements.6}). But for a dissipative system with alternated spins the concentration of energy also does not happen as $t\ra \infty$. To see this, we make  first one step of the perturbation theory. The alternated spins condition provides that the system does not have resonances of the first order. Then in Theorem \ref{nllem:prest} we find a global  canonical change of variables in $\mC^N$, transforming  $u\rightarrow v,\,(I,\ph)\rightarrow (J,\psi),$ which is $\sqrt\eps$-close to identity uniformly in $N$ and kills in the Hamiltonian the term of order $\sqrt\eps$.   
We rewrite equation (\ref{nlini_e}) in the new variables $v$ and call the result "$v$-equation" (see (\ref{nlv})). Using the fact that in the new coordinates the interaction potential has the same size as the dissipation and working with the family of norms  $\|\cdot\|_{j,q}$, we obtain desired estimates for solutions of the $v$-equation. 

Then we pass to the limit $\eps \ra 0$.
 In the action-angle coordinates $(J,\psi)$ the  $v$-equation takes the form 
\begin{align}
\label{nlintro_ac}
d J&=X(J, \psi,\eps)\,d\tau + \sigma (J,\psi,\eps) d \beta+\ov \sigma (J,\psi,\eps) d\ov \beta, \\
\label{nlintro_an}
d\psi&=\eps^{-1}Y(J,\eps)\,d\tau + \ldots,
\end{align}
where the term $\ldots$ and $X,Y, \sigma$ are of order $1$. For details see (\ref{nlac'})-(\ref{nlang'}). So the angles rotate fast, while the actions change slowly. The averaging principle for systems of the type (\ref{nlintro_ac})-(\ref{nlintro_an}) was established in \cite{Kh},\cite{FW03},\cite{FW06},\cite{FW} and, more recently,  in \cite{KuPi},\cite{Kuk12}. Our situation is similar to that in \cite{KuPi},\cite{Kuk12}, and we follow the scheme suggested there. Let $v^\eps(\tau)$ be a solution of the $v$-equation, written in the slow time, and   $J^\eps(\tau)=J(v^\eps(\tau))$ be the corresponding vector of actions.
We prove Theorem \ref{nltheo:av}, stating that the family of measures $\DD(J^{\eps}(\cdot))$ converges weakly as $\eps\ra 0$
to a distribution of a unique weak solution of the averaged in angles equation (\ref{nlintro_ac})$|_{\eps=0}$, which has the form (\ref{nlav_I}). To prove that this convergence is uniform in $N$, we use the uniformity of estimates obtained above and the fact that the averaged equation   for the infinite system of rotators has a unique weak solution.   Since the change of variables is $\sqrt\eps$-close to identity, the behaviours of actions $J^\eps$ and $I^\eps$ as $\eps\ra 0$ coincide, and we get Theorem \ref{nltheo:avIintro}. 
The averaged equation   (\ref{nlav_I}) does not feel the Hamiltonian interaction of rotators since the averaging eliminates the   Hamiltonian terms.

The  averaged equation   (\ref{nlav_I}) is irregular: its dispersion matrix is not Lipschitz continuous.  To study it we use the method of {\it effective equation}, suggested  in \cite{Kuk10},\cite{Kuk12} (in our case its application simplifies).   
The effective equation (see (\ref{nlef})) is defined in the complex coordinates $v=(v_k)_{k\in\CC}\in\mC^N$. If $v(\tau)$ is its  solution then the actions $J(v(\tau))$ form a weak solution of equation (\ref{nlav_I}) and vice versa (see Proposition \ref{nltheo:lifting}). The effective equation is well posed and  mixing. This implies item {\it (i)} of Theorem {\ref{nltheo:stmintro}}. 
The proof of  item {\it (ii)} is based on the averaging technics developed in Theorem \ref{nltheo:avIintro}.

Note that the convergence (\ref{nlintroconv}) is equivalent to 
\bee
\label{nlstmef1}
\wid\mu^\eps\raw m \quad \mbox{as}\quad\eps\ra 0,
\eee
where $m$ is the unique stationary measure of the effective equation,  see Remark \ref{nlrem:stmef1}.
Item \nolinebreak {\it (iii)} of Theorem \ref{nltheo:stmintro} follows from the first two items and Theorem \ref{nltheo:avIintro}.
\ssk

{\bf Acknowledgments}

I am very grateful to  my PhD supervisors S. Kuksin and A. Shirikyan for formulation of the problem, guidance, encouragement and all-round help. Also I would like to thank J. Bricmont, P. Collet, V. Jaksic, S. Olla and  C.-A. Pillet for useful discussions  concerning physical meaning of the problem. This research was carried out within the MME-DII Center of 
Excellence (ANR-11-LABX-0023-01) and supported by the ANR grant STOSYMAP 
(ANR 2011 BS01 015 01) and the RFFI grant \#13-01-12462.

\subsection{Agreements and assumptions} \label{nlsec:aa}
{\bf Agreements}
\ssk  

$1)$ We refer to item 1 of Theorem \ref{nllem:prest} as Theorem \ref{nllem:prest}.1, etc.   

$2)$ By $C,C_1,C_2,\ldots$ we denote various positive constants and by $C(b),C_1(b),\ldots$ we denote positive constants which depend on the parameter $b$. We do not indicate their dependence on the dimension $d$, power $p$ and time $T$ which are fixed through all the text and {\it always} indicate if they depend on the number of rotators $N$, times $t,s,\tau,\ldots$, positions $j,k,l,m,\ldots\in\CC$ and small parameter $\eps$. 
Constants $C,C(b),\ldots$ can change from formula to formula.

$3)$ Unless otherwise stated, assertions of the type "$b$ is sufficiently close to $c$" and "$b$ is sufficiently small/big" always suppose  estimates independent from $N$,  positions $j,k,l,m,\ldots\in\CC$ and times $t,s,\tau,\ldots$.  
  
$4)$  We use notations $b \wedge c:=\min(b,c),\; b \vee c=\max(b,c)$.

$5)$ For vectors $b=(b_k),\, c=(c_k)$, $b_k,c_k\in\mC$, we denote
$$\quad a\cdot b:=\sum\limits a_k\cdot b_k=\sum\limits \Ree a_k \ov b_k.$$

$6)$ For $1/2<\gamma<1$, $j\in\CC$ and $q>0$  we introduce a family of scalar products and a family of norms on $\mC^N$ as  
\footnote{For details see Section \ref{nlsec:norms}. We will fix $\ga$, so we do not indicate the dependence on it.}
$$( u \cdot u^1)_j: = \sum\limits_{k\in\CC} \ga^{|k-j|}u_k\cdot u^1_k, \;
\|u\|_{j,q}^q:=\sum\limits_{k\in\CC}\ga^{|k-j|}|u_k|^q,\mbox{ where }u=(u_k)_{k\in\CC},\,u^1=(u^1_k)_{k\in\CC}\in\mC^N. $$  

$7)$ For a metric space $X$ by $\LL_b(X)$ ($\LL_{loc}(X)$) we denote the space of bounded Lipschitz continuous (locally Lipschitz continuous) functions from $X$ to $\mR$.

$8)$ Convergence of measures we always understand  in the weak sense.

$9)$ 
We suppose $\eps$ to be sufficiently small, where it is needed. 
\bigskip

{\bf Assumptions}
\medskip

Here we formulate our assumptions. In Example \ref{nlexam:conditions} we give examples of functions $F_j,G$ and $g_j$ satisfying them. 
 
Fix $p\in\mN,\,p\geq 2$. Assume that there exists $\varsigma>0$ such that the following holds.

{\bf HF}. (Alternated spins condition). 
{\it For every $j\in\CC$ and some function $f$ we have $f_j=(-1)^{|j|} f$. Function $f:(-\varsigma,\infty)\mapsto\mR_+$ 
is $C^3$-smooth and its derivative $f'$ has only isolated zeros. Moreover, for any $x\geq 0$ we have } 
$$f(x) \geq C(1+x^{p/2}) \quad\mbox{and}\quad |f'(x)|x^{1/2}+|f''(x)|x + |f'''(x)|x^{3/2} \leq Cf(x). $$

{\bf HG}. {\it Function $G:(-\varsigma,\infty)\mapsto\mR$ is $C^4$-smooth. Moreover, for any $x\geq 0$ it satisfies  
$$|G'(x)|x^{1/2}+ |G''(x)|x +  |G'''(x)|x^{3/2}\leq C(1+ x^{(p-1)/2}). $$ }

{\bf Hg}.(i) {\it Functions $g_l:\mC^N\mapsto\mC,\,l\in\CC$ are $C^2$-smooth and depend  on $u=(u_k)_{k\in\CC}$ only through $(u_k)_{k:|k-l|\leq 1}$. For any $u\in\mC^N$ and $l,m\in\CC$ they satisfy
$$|g_l(u)|,|\chp_{u_m} g_{l}(u)|,|\chp_{\ov u_m} g_{l}(u)|\leq C\big(1+\sum\limits_{k:|k-l|\leq 1}|u_k|^{p-1}\big),$$ 
while all the second derivatives are assumed to have at most a polynomial growth at infinity, which is uniform in $l\in\CC$.}

(ii) (Dissipative condition)
 {\it There exists a constant $C_g>0$, independent from  $N$, such that for  any $j\in\CC$ and  $1/2<\gamma<1$ sufficiently close to one, for any $(u_k)_{k\in\CC}\in\mC^N$}
$$
( g(u)\cdot u )_j \leq -C_g \|u\|^p_{j,p} + C(\gamma), \quad \mbox{{\it where}}\quad g:=(g_{l})_{l\in\CC}, 
$$
{\it and the scalar product $(\cdot)_j$ and the norm $\|\cdot\|_{j,p}$ are defined in {\it Agreements.6}. Recall that they depend on $\ga$.}

{\bf HI}.(i) { \it For some constant $\al_0>0$, independent from $N$, and every $j \in\CC$ we have} 
$$\MO e^{\al_0 |u_{0j}|^2}\leq C.$$
(ii) {\it The  initial conditions $u_0=u_0^N$ agree in $N$ in the sense that there exists a $\mC^\infty$-valued random variable $u_0^\infty=(u_{0j}^\infty)_{j\in\CC^\infty}$ satisfying  for any $N\in\mN$ the relation} 
$$
\DD ((u_{0j}^N)_{j\in\CC(N)})=\DD ((u^\infty_{0j})_{j\in\CC(N)}). 
$$

In what follows,  we suppose the assumptions above to be held.
\ssk

\begin{example}
\label{nlexam:conditions}
As an example of functions $f$ and $G$ satisfying conditions {\it HF} and {\it HG}, we propose $f(x)=1+x^k$ for any $ \mN\ni k \geq p/2$, and $G(x)=\hat G(\sqrt{x+\varsigma})$ for any constant $\varsigma>0$ and any $C^4$-smooth function $\hat G:\mR_{+}\mapsto\mR$ satisfying
$$
|\hat G'(x)|+|\hat G''(x)|+|\hat G'''(x)| \leq C(1+x^{p-1})\quad\mbox{for all }x\geq \sqrt{\varsigma}.
$$

The simpliest example of functions $g_l$ satisfying assumption {\it Hg} is the diagonal dissipation 
$g_l(u)=-u_l|u_l|^{p-2}$. As an example of functions $g_l$ providing non-Hamiltonian interaction between rotators, we propose $g_l(u)=-u_l|u_l|^{p-2}+\wid g_l(u)$, where $\wid g_l$ satisfies {\it Hg(i)} and
$|\wid g_l(u)| \leq  \wid C\sum\limits_{k:|k-l|\leq 1} |u_k|^{p-1} + C$, where the constant $\wid C$ satisfies
\footnote{This constant is not optimal, one can improve it.} $\wid C< \frac{1}{8d(2d+1)^2}$.
\end{example}
For more examples see Section \ref{nlsec:example}. 
\ssk

\brr\label{nlrem:a>1}
In the case $a\geq 1$ assumptions {\it HF} and {\it HG} simplify.

{\bf HF'-HG'}. {\it Functions $f_j,G:(-\varsigma,\infty)\mapsto\mR$ 
are $C^1$- and $C^4$-smooth correspondingly, $f_j'$ have only isolated zeros and $|G'(x)|x^{1/2}\leq C(1+x^{(p-1)/2})$ for any $x\geq 0$}. 
\err

\section{Preliminaries}
\subsection{Norms}
\label{nlsec:norms}
Since $\sum\limits_{j\in\CC} |u_j|^2$ is conserved by the Hamiltonian flow, it would be natural to work in the $l_2$-norm. However, the $l_2$-norm of solution of (\ref{nlini_e}) diverges as $N\ra\infty$. To overcome this difficulty and obtain  uniform in $N$ estimates for the solution,  
 we introduce  the family of $l_q$-weighted norms with exponential decay:  for each $q>0$ and every $j\in\CC$, for  $v=(v_k)_{k\in\CC}\in\mC^N$ we set
$$
\|v\|_{j,q}=\Big(\sum\limits_{k\in\CC} \gamma^{|k-j|} |v_k|^q\Big)^{1/q}, \quad\mbox{where  the constant $1/2<\gamma<1$ will be chosen later.} 
$$
Similar norms were considered, for example, in \cite{DZ}, Section 3.12.
Define the family of
$l_2$-weighted scalar products on $\mC^{N}$,
$$
( v^1 \cdot v^2 )_j= \sum\limits_{k\in\CC} \gamma^{|k-j|} v_k^1 \cdot v_k^2,
$$
corresponding to the norms 
$
\|v\|_j^2: = \|v\|_{j,2}^2= (v \cdot v )_j.
$
It is easy to see that the Holder inequality holds: for any  $m,n>0$, satisfying $m^{-1}+n^{-1}=1$, we have
\begin{equation}
\label{nlholder}
|( v^1\cdot v^2 )_j|\leq \|v^1\|_{j,m}\|v^2\|_{j,n}.
\end{equation}
Moreover, since for any $m\geq n$ we have $|v_k|^n\leq |v_k|^m +1 $, then we get 
\begin{equation}
\label{nlmn}
\|v\|_{j,n}^{n}\leq \|v\|_{j,m}^{m} + \sum\limits_{k\in\CC}\ga^{|j-k|} \leq \|v\|_{j,m}^{m} + C(\gamma) \quad \mbox{for} \quad m\geq n,
\end{equation}
where the constant $C(\ga)$ does not depend on $N$ since the geometrical series converges.
\subsection{The change of variables}
\label{nlsec:res}
Consider the complex variables $v=(v_j)_{j\in\CC}\in\mC^N$ and the corresponding vectors of actions and angles $(J,\psi)\in\mR^N_{+0}\times\mT^N.$ 
Define a vector $B:=(\beta,\ov \beta)^T\in\mC^{2N}$, where $\beta$ is a complex $N$-dimensional Brownian motion as before and $T$ denotes the transposition.
Recall that by $\lan\cdot\ran$ we denote the averaging in angles, see Appendix \ref{nlapp:aver} for its properties. Let $\nabla:=(\nabla_j)_{j\in\CC}$ and $g:=(g_j)_{j\in\CC}$.
\btt
\label{nllem:prest}
There exists a $C^2$-smooth $\sqrt\eps$-close to identity canonical change of variables of $\mC^N$, transforming $u\ra v,\,(I,\ph)\ra (J,\psi)$   such that the Hamiltonian $H^\eps$ in the new coordinates takes the form
\begin{eqnarray}
\label{nlham1'}
\HH^\eps(J,\psi)
=H^\eps_0(J) + \eps H_2(J,\psi) + \eps\sqrt \eps H^\eps_> (J,\psi),  
\end{eqnarray}  
where 
\bee\label{nlH00}
H^\eps_0(v)=\frac12\sum\limits_{j\in\CC} F_j (|v_j|^2)+ \frac{\sqrt\eps}{4}\sum\limits_{|j-k|=1} \lan G(|v_j-v_k|^2)\ran
\eee
is $C^4$-smooth and the functions $H_2(v)$ and $H_>^\eps(v)$ are $C^2$-smooth.
System  (\ref{nlini_e})-(\ref{nlini_c}) written in $v$-variables has the form
\begin{align}
\label{nlv}
\dot v&=i\nabla H^\eps_0(v) + \eps i \nabla H_2(v) + \eps g(v) + \eps\sqrt\eps r^\eps(v) + \sqrt\eps W^\eps(v)\dot B, \\
\label{nlv_c} 
v(0)&=v(u_{0})=:v_{0}, 
\end{align}
where $r^\eps=(r^\eps_j)_{j\in\CC}:\mC^N\mapsto\mC^N$ is a continuous vector-function  and $W^\eps$ is a new dispersion matrix. The latter has  the size $N\times 2N$ and  consists of two blocks, $W^\eps=(W^{\eps1},W^{\eps2}),$ so that $W^\eps \dot B=W^{\eps1} \dot \beta+W^{\eps2} \dot {\ov\beta}.$ The blocks have the form $W^{\eps1,2}=(W^{\eps1,2}_{kl})_{k,l\in\CC}$, where $W_{kl}^{\eps1}=\sqrt{\TT_l}\chp_{u_l}v_k$, $W_{kl}^{\eps2}=\sqrt{\TT_l}\chp_{\ov u_l}v_k$.   
Moreover, for any $j\in\CC$ and $1/2 <\ga< 1$ we have 
\ssk

{\bf 1.} $|( i\nabla H_
2\cdot v)_j| \leq (1-\gamma)C\|v\|_{j,p}^p + C(\gamma)$.  
\ssk

{\bf 2.}{\bf a.} $\nabla_{j} H_2$ depends only on $v_n$ such that $|n-j|\leq 2$, and $|\nabla_{j} H_2|\leq C\sum\limits_{n:|n-j|\leq 2}|v_n|^{p-1}+C.$

\quad {\bf b.} For any $q\geq 1$ we have 
$\|r^\eps\|_{j,q}^{q}\leq C(\gamma,q)+C(q)\|v\|_{j,q(p-1)}^{q(p-1)}.$

{\bf 3.} 
The functions $d_{kl}^{1,2}$, defined as in (\ref{nlddd}),  satisfy
$|d^{1}_{kl}-\del_{kl}\TT_k|, \,|d^{2}_{kl}|\leq C\sqrt\eps$ for all  $k,l\in\CC$.

{\bf 4.} We have  $|u_j-v_j|\leq C\sqrt\eps$ and $|I_j-J_j|\leq C\sqrt\eps$. 
\ett

Further on we will usually skip the upper index $\eps.$ If $\ga=1$, then the norm  $\|u\|_j=\big(\sum\limits_{j\in\CC} |u_j|^2\big)^{1/2}$ is the first integral of the Hamiltonian $H^\eps$. Consequently, the norm $\|u\|_j$ with $\gamma$ close to one is an approximate integral of the Hamiltonian flow. Item {\it 1} of Theorem \ref{nllem:prest}  means that the change of variables preserves this property in the order $\eps$, modulo constant $C(\gamma)$. 
This is crucial for  deriving of uniform in $N$ estimates for solutions of (\ref{nlv}). 
 
In equation (\ref{nlini_e}) all functions, except the rotating nonlinearity $if_j(|u_j|^2)u_j$,  have at most a polynomial  growth of a power $p-1$. Item {\it 2} affirms, in particular,  that this property is conserved by the transformation.
\ssk

The proof of the theorem is technically rather complicated and is given in Section  \ref{nlsec:change}. Since the potential $G$ is not a differentiable function of actions,  we have to work in the $v$-coordinates despite that the transformation is  constructed in the action-angle variables. This rises some difficulties since the derivative of $\psi_j$ with respect to $v_j$ have a singularity when $v_j=0$. Moreover, we have to work in rather inconvenient norms $\|\cdot\|_{j,q}$ and estimate not only Poisson brackets, but also non-Hamiltonian terms of the $v$-equation. 

Let us briefly explain why the alternated spins condition {\it HF} provides that system (\ref{nlini_e}) does not have resonances of the first order.
Writing equation (\ref{nlini_e}) in the action-angle coordinates, we find that the angles satisfy $\dot\ph_j\volna f_j(|u_j|^2)$, $j\in\CC$. It is not difficult to see that  the interaction potential 
$G(|u_j-u_k|^2)$ depends on the angles only through their difference $\ph_j-\ph_k$, see (\ref{nlG}). Due to assumption {\it HF}, the corresponding combination of rotation frequences is separated from zero. Indeed, 
$f_j-f_k=2(-1)^{|j|}f$, where we recall that the function $f$ is assumed to be strictly positive.
\ssk

\subsection{Estimates for solution}
\label{nlsec:est}

System (\ref{nlv})-(\ref{nlv_c}) has a unique solution since system (\ref{nlini_e})-(\ref{nlini_c}) does. Let us denote it by $v(t)=(v_k(t))_{k\in\CC}$. 
\begin{lem}
\label{nllem:est}
For any $1/2<\ga<1$ sufficiently close to one there exists  $\al=\al(\ga)>0$ such that for all  $j\in\CC$, $t\geq 0$ and $\eps$ sufficiently small  we have
\bee
\label{nlestimates}
\MO\sup\limits_{s\in [t,t+1/\eps]} e^{\al \|v(s)\|_j^2} < C(\gamma).
\eee
\end{lem}
Let us emphasize that estimate (\ref{nlestimates}) holds uniformly in $N,j,t$ and $\eps$ sufficiently small.
\begin{cor}
\label{nlcor:est} 
There exists $\al>0$ such that
for any $m>0$, $t\geq 0$, $j\in\CC$ and  $\eps$ sufficiently small we have
$$\MO \sup\limits_{s\in [t,t+1/\eps]} e^{\al |v_j(s)|^2} < C,\quad\quad \MO \sup\limits_{s\in [t,t+1/\eps]} |r_j(v(s))|^m < C(m), $$
where $r=(r_j)_{j\in\CC}$ is the reminder in (\ref{nlv}).
\end{cor}
{\it Proof of Corollary \ref{nlcor:est}.} Fix any $\gamma$ and $\al$ such that (\ref{nlestimates}) holds true. 
By the definition of $\|\cdot\|^2_j$ we have 
$|v_j|^2 \leq \|v\|^2_j$,
so Lemma \ref{nllem:est} implies the first inequality.
Let us prove the second one. Without loss of generality we assume that $m\geq 2$. Theorem \ref{nllem:prest}.2b implies
\bee
\label{nlrrrrr}
|r_j|^m\leq \|r\|_{j,m}^m\leq C(\ga,m) + C(m)\|v\|_{j,m(p-1)}^{m(p-1)} \leq C(\ga,m) + C(m,\kappa) e^{\kappa\|v\|_{j,m(p-1)}^2}
\eee
for any $\kappa>0$. Using that $2/m(p-1)\leq 1$ and the Jensen inequality, we get
\bee\label{nlrrrrr1}
 e^{\kappa\|v\|_{j,m(p-1)}^2}\leq  e^{\kappa\sum\limits_{k\in\CC}\ga^{\frac{2|j-k|}{m(p-1)}}|v_k|^2}\leq \sum\limits_{k\in\CC}\ga^{\frac{2|j-k|}{m(p-1)}}(C(\ga))^{-1} e^{\kappa C(\ga) |v_k|^2},
\eee
where $C(\ga)=\sum\limits_{k\in\CC}\ga^{\frac{2|j-k|}{m(p-1)}}$. Choosing $\kappa$ in such a way that $\kappa C(\ga)\leq\al$ and combining (\ref{nlrrrrr}), (\ref{nlrrrrr1}) and the first estimate of the corollary, we get the desired inequality.
\qed
\ssk

{\it Proof of Lemma \ref{nllem:est}.}
{\bf Step 1.} Take some $1/2<\ga<1$ and $0<\al_1<1$.
Further on we present only formal computation which could be justified by standard stopping-time arguments (see, e.g., \cite{KarShr}).
Applying  the Ito formula in complex coordinates (see Appendix \nolinebreak\ref{nlcomplexIto}) to $e^{\al_1 \|v\|^2_j}$ and noting that $i\nabla_{j} H_0\cdot v_j =0$ since $H_0$ depends on $v$ only through $J(v)$, we get
\begin{eqnarray}
\nonumber
\frac{d}{ds}e^{\al_1 \|v(s)\|^2_j} &=&
 2\al_1 \eps e^{\al_1 \|v\|^2_j} \Big( ( i\nabla H_2\cdot v )_j + ( g\cdot v )_j + \sqrt\eps ( r\cdot v )_j  +  \sum\limits_{k\in\CC} \gamma^{|j-k|}d_{kk}^1  \\
\label{nlexp}
&+&
\al_1 \sum\limits_{k,l\in\CC} \gamma^{|j-k|+|j-l|}
\big(v_k \ov v_l d^1_{kl} + \Ree  (\ov v_k \ov v_l d^2_{kl})   \big) \Big) + 2\al_1\sqrt\eps \dot M_s,
\end{eqnarray}
where we recall that $d^{1,2}_{kl}$ are calculated in (\ref{nlddd}), and  the martingal
\bee\label{nlmartingal} 
M_s:=\int\limits_{s_0}^s e^{\al_1 \|v\|^2_j}( v \cdot WdB )_j \quad \mbox{for some} \quad s_0<s. 
\eee
First we estimate $ ( r \cdot v)_j$.
Theorem \ref{nllem:prest}.2b implies
$$
 \|r\|_{j,p/(p-1)} \leq \big( C \|v\|_{j,p}^p+C(\gamma)\big)^{(p-1)/p}\leq C_1 \|v\|_{j,p}^{p-1} + C_1(\gamma).
$$ 
Then, the Holder inequality (\ref{nlholder}) with $m=p/(p-1)$ and $n=p$, jointly with (\ref{nlmn})  implies
 \begin{equation}
\label{nlrv}
|(r\cdot v )_j|\leq \|r\|_{j,p/(p-1)}\|v\|_{j,p}\leq C_1\|v\|_{j,p}^p + C_1(\ga) \|v\|_{j,p} \leq C_2(\ga)(\|v\|_{j,p}^p + 1). 
\end{equation}
Secondly we estimate Ito's term.  By Theorem \ref{nllem:prest}.3 we get
\bee\label{nlitoO}
\left|\sum\limits_{k\in\CC} \ga^{|j-k|}d_{kk}^1\right|\leq C(\ga).
\eee
Note that
$$
\sum\limits_{k,l\in\CC} \gamma^{|j-k|+|j-l|}
|v_k| |v_l| \leq \sum\limits_{k,l\in\CC} \gamma^{|j-k|+|j-l|}
(|v_k|^2 + |v_l|^2) \leq C(\ga)\| v\|^2_j.
$$
Consequently, due to Theorem \ref{nllem:prest}.3, we have
\begin{align}
\nonumber
\left|\sum\limits_{k,l\in\CC} \gamma^{|j-k|+|j-l|}\big(v_l \ov v_k d^1_{kl} + \Ree  (\ov v_k \ov v_l d^3_{kl})   \big)\right|
&\leq
\sum\limits_{k\in\CC} \gamma^{2|j-k|}\TT_k|v_k|^2 + \sqrt\eps C(\ga) \|v\|_j^2\\
\label{nlxvost}
\leq \big(C+ \sqrt\eps C_1(\ga)\big)\|v\|^2_j&\leq\big(C+ \sqrt\eps C_1(\ga)\big)\|v\|^p_{j,p}+C_2(\ga),
\end{align}
where we have used  (\ref{nlmn}). 
Now Theorem \ref{nllem:prest}.1, assumption {\it Hg(ii)}, (\ref{nlrv}), (\ref{nlitoO}) and (\ref{nlxvost}), applied to (\ref{nlexp}), imply that for $\ga$ sufficiently close to one  we have
\begin{equation}
\label{nlexp1}
\frac{d}{ds}e^{\al_1 \|v\|^2_j} \leq 2\al_1\eps e^{\al_1 \|v\|^2_j} \Big( -\big(C_{g}-(1-\gamma)C-\al_1 C-\sqrt\eps C(\gamma) \big)\|v\|^p_{j,p} + C_1(\gamma) \Big) + 2\al_1\sqrt\eps \dot M_s.
\end{equation}
We take $1/2<\gamma<1$ sufficiently close to one,  then choose  $\al_1(\ga)>0$ and $\eps_0(\ga)>0$,  sufficiently small, in such a way that 
\bee
\label{nlddelta}
\Delta:=C_{g}-(1-\gamma)C- \al_1 C-\sqrt\eps_0 C(\gamma) > 0.
\eee 
For any constant $C$ there exists a constant $C_1$ such that for all $x\geq 0$ we have
$$
2\al_1 e^{\al_1 x}(-\Delta x + C) \leq -  e^{\al_1 x} + C_1. 
$$ 
 Consequently, (\ref{nlexp1}) jointly  with (\ref{nlmn}) implies that for $\eps<\eps_0$ we have
\begin{equation}
\label{nlexp2}
\frac{d}{ds}e^{\al_1 \|v(s)\|^2_j} \leq -\eps e^{\al_1 \|v(s)\|^2_j} + \eps C(\ga)+ 2\al_1\sqrt\eps \dot M_s.
\end{equation}
Fixing $s_0=0$ (which is defined in (\ref{nlmartingal})), taking expectation and applying the Gronwall-Bellman inequality to (\ref{nlexp2}), we have
$$
\MO e^{\al_1 \|v(s)\|^2_j} \leq \MO e^{\al_1 \|v_0\|^2_j} e^{-\eps s} + C(\ga).
$$
Due to assumption {\it HI(i)} and Theorem \ref{nllem:prest}.4, we have $ \MO e^{\al_1 |v_{0j}|^2} \leq C$ for all $j\in\CC$. Then the Jensen inequality implies that $ \MO e^{\al_1 \|v_0\|^2_j} \leq C(\ga),$ if $\al_1$ is  sufficiently small.
Thus  we obtain  
\bee
\label{nl1est}
\MO e^{\al_1 \|v(s)\|_j^2}\leq C(\ga) \quad\mbox{for all }s\geq 0 \mbox{ and }j\in\CC. 
\eee   

{\bf Step 2.} We fix the parameters $\gamma$ and $\al_1$ as above. Accordingly, the constants, which depend only on them, will be denoted just $C,C_1,\ldots$. 

 Now we will prove (\ref{nlestimates}). Take any $0<\al<\al_1/2$ and fix $s_0=t$. Integrating inequality (\ref{nlexp2}) with $\al_1$ replaced by $\al$  over the interval $ t \leq s\leq t+1/\eps$  and  using (\ref{nl1est}), we have
\begin{eqnarray}
\nonumber
\MO \sup\limits_{s\in[t,t+1/\eps]} e^{\al \|v(s)\|^2_j} &\leq& 
\MO e^{\al \|v(t)\|^2_j} + C+ 2\al \sqrt\eps\,\MO \sup\limits_{s\in[t,t+1/\eps]} M_s \\
\label{nle2}
&\leq& C_1+  2\al \sqrt\eps\,\MO \sup\limits_{s\in[t,t+1/\eps]} M_s.
\end{eqnarray}
Now we turn to the martingal part. The definition of $M_s$ implies
$$
 \sup\limits_{s\in[t,t+1/\eps]} M_s \leq \sum\limits_{k\in\CC} \sup\limits_{s\in[t,t+1/\eps]} M_{ks},
$$
where 
$ M_{ks} = \int\limits_{t}^s e^{\al \|v\|^2_j}\gamma^{|j-k|}v_k\cdot  (WdB)_k$.
The Doob-Kolmogorov inequality implies that 
$$\MO \sup\limits_{s\in[t,t+1/\eps]} M_{ks} \leq C \MO \sqrt{[ M_k]_{t+1/\eps}} \leq C \sqrt {\MO [ M_k]_{t+1/\eps} },$$
where $[ M_k]_s$ denotes the quadratic variation of $M_{ks}$.  
Similarly to (\ref{nlvariation}), we obtain 
\begin{equation*}
 [ M_k]_{t+1/\eps} = \int\limits_{t}^{t+1/\eps} e^{2\al \|v\|^2_j} \gamma^{2|j-k|} 
S^J_{kk}\, ds\leq C(\kappa)\gamma^{|j-k|}\int\limits_{t}^{t+1/\eps} e^{2(\al+\kappa) \|v\|^2_j} (|d_{kk}^1|+|d_{kk}^2|)\, ds
\end{equation*}
for any $\kappa>0$, where $S_{kk}^J$ is defined in (\ref{nlSSS}). Take $0<\kappa<\al_1/2-\al$. Then, using Theorem  \ref{nllem:prest}.3 and  (\ref{nl1est}), we get 
\begin{equation}
\nonumber
\MO \sup\limits_{s\in[t,t+1/\eps]} M_s \leq
C\sum\limits_{k\in\CC} \sqrt{\MO  [ M_k]_{t+1/\eps} } \leq 
C(\kappa)\sum\limits_{k\in\CC} \gamma^{|j-k|/2}\left( \int\limits_{t}^{t+1/\eps} \MO e^{2(\al+\kappa) \|v\|^2_j} \, ds \right)^{1/2}
\leq \frac{C_1(\kappa)}{\sqrt\eps}.
\end{equation}
Now (\ref{nlestimates})  follows from (\ref{nle2}).
\qed
\ssk

\section{The limiting dynamics}
\label{nlsec:limitsfull}
In this section we investigate the limiting (as $\eps\ra 0$) behaviour of system (\ref{nlini_e}). We prove Theorems \ref{nltheo:fin_dyn}, \ref{nltheo:measure} and \ref{nltheo:stmintro} which are  our main results.
\subsection{Averaged equation}
\label{nlsec:limits}
Here we prove Theorem \ref{nltheo:fin_dyn}, which describes the limiting dynamics of actions on long time intervals of order $\eps^{-1}$.
In the slow time $\tau=\eps t$ system (\ref{nlv})-(\ref{nlv_c}) has the form
\begin{equation}
\label{nlvs}
d v_j=( \eps^{-1} i\nabla_{j} H_0 + i \nabla_{j} H_2 + g_j + \sqrt\eps r_j )\,d\tau + ( W d B)_j,  \quad v_j(0)=v_{0j},\quad j\in\CC.
\end{equation}
Let us write equation (\ref{nlvs}) in the action-angle variables $J=J(v),\psi=\psi(v)$. Due to (\ref{nlitoforactions}) and the equalities $i\nabla_j H_0\cdot v_j=0$ and $i\nabla_j H_0\cdot\frac{iv_j}{|v_j|^2}=\chp_{J_j}H_0$, we have
\begin{align}
\label{nlac'}
dJ_j &= A_j^J\,d\tau +  v_j\cdot  ( W d B)_j, \\
\label{nlang'}
d \psi_j &= \left( \eps^{-1}\frac{\chp H_0}{\chp J_j} +\frac{A_j^\psi}{|v_j|^2} \right)\,d\tau+ \frac{ iv_j} {|v_j|^2}\cdot  ( W d B)_j, \quad j\in\CC,
\end{align}
where 
\bee
\label{nlone}
A_j^J:=A_j\cdot v_j+ d_{jj}^1, \quad 
A_j^\psi:= A_j\cdot (iv_j) -\Imm(\ov v_j v_j^{-1} d^2_{jj}), \quad A_j:= i \nabla_{j} H_2  +g_j + \sqrt\eps r_j, 
\eee
and $d^{1,2}_{jj}$ are calculated in (\ref{nlddd}).
In view of (\ref{nlH00}), Proposition \ref{nlprop:aver} implies that 
\bee\label{nlH000}
\mbox{for each $j\in\CC$ the function $\chp_{J_j}H_0$ is  $C^1$-smooth with respect to $J=(J_k)_{k\in\CC}$}.
\eee 
Theorem \ref{nllem:prest}.2a,3 jointly with Corollary \ref{nlcor:est} implies that for all $j,k,l\in\CC$ and every $m>0$  we have 
\begin{equation}
\label{nlest'}
\MO\sup\limits_{0\leq \tau \leq T} \big( |A_j| + |A^J_j| + |A^\psi_j| + |S^J_{kl}|\big)^m \leq C(m),
\end{equation}
where $S^J_{kl}$ is the element of the diffusion matrix for equation (\ref{nlac'}) with respect to the real Brownian motion; it is calculated in (\ref{nlSSS}).

Note that the quadratic vatriations of the martingales from the r.h.s. of (\ref{nlac'}) and (\ref{nlang'}) are calculated in (\ref{nlvariation}).

Let $v^{\eps}(\tau)$ be a solution of (\ref{nlvs}). Then $J^{\eps}(\tau):=J(v^\eps(\tau)),\psi^{\eps}(\tau):=\psi(v^\eps(\tau))$ satisfy (\ref{nlac'})-(\ref{nlang'}).   
Due to estimate (\ref{nlest'}) and slow equation (\ref{nlac'}), using Arzela-Ascoli theorem, we get
\begin{prop}
\label{nlprop:tight}
The family of measures $\{\DD(J^{\eps}(\cdot)), \; 0<\eps\leq 1\}$ is tight on $C([0,T],\, \mR^N)$. 
\end{prop}

Let $Q_0$ be a weak limiting point of $\DD(J^{\eps}(\cdot))$:\begin{equation}
\label{nllp}
\DD(J^{\eps_k}(\cdot)) \rightharpoonup Q_0 \quad\mbox{as}\quad k\rightarrow \infty \quad\mbox{on}\quad  C([0,T],\, \mR^N),
\end{equation}
where $\eps_k\ra 0$ as $k\ra\infty$ is a suitable sequence. Now we are going to show that the limiting point $Q_0$ does not depend on the sequence $(\eps_k)$ and is governed by the main order in $\eps$ of the averaging of equation (\ref{nlac'}). Let us  begin with writing down this equation.
Since by Theorem \nolinebreak\ref{nllem:prest}.3 we have $d_{jj}^1=\TT_j+O(\sqrt\eps)$, the main order of the drift of equation (\ref{nlac'}) is $ i\nabla_{j} H_2 \cdot v_{j} + g_j(v)\cdot v_j +\TT_{j}.$  
Since for any real-valued $C^1$-smooth function $h(v)$ we have $i\nabla_j h\cdot v_j=-\chp_{\psi_j} h$, then periodicity of the function $h$ with respect to $\psi_j$ implies $\lan i\nabla_{j} h \cdot v_{j} \ran=0$. So that, in particular, $\lan i\nabla_{j} H_2 \cdot v_{j} \ran=0$. Thus the main order of the averaged drift takes the form 
\begin{equation}
\label{nlavdrift}
\lan i\nabla_{j} H_2 \cdot v_{j}+ g_j(v)\cdot v_j +\TT_{j}\ran =\RR_{j}(J) +\TT_{j},
\end{equation}
where $\RR_j$ is defined in (\ref{nlusre}).
Proposition \ref{nlprop:diffusion} jointly with Theorem \ref{nllem:prest}.3 implies that the main order of the diffusion matrix of (\ref{nlac'}) with respect to the real Brownian motion $(\Ree \beta_k,\Imm \beta_k)_k$  is $\diag(\TT_{k}|v_{k}|^2)_{k\in\CC}=\diag(2\TT_{k}J_{k})_{k\in\CC}.$ It does not depend on angles, so the averaging does not change it. Choose its square root as $\diag(\sqrt{2\TT_{k}J_{k}})_{k\in\CC}$. Then in the main order the averaging of equation (\ref{nlac'}) takes the form
\bee\label{nlaver}
d J_{j} = (\RR_{j}(J)+\TT_{j}) \,d\tau + \sqrt{2J_{j}\TT_{j}}\,d\wid\beta_{j}, \quad {j}\in\CC,
\eee
where $\wid\beta_{j}$ are  independent standard real Brownian motions. The averaged equation (\ref{nlaver}) has a weak singularity: its dispersion matrix is not Lipschitz continuous. However, its drift is regular: Proposition \ref{nlprop:aver} implies that
\bee\label{nlRRR}
\mbox{for each $j\in\CC$ the function $\RR_j$ is  $C^1$-smooth with respect to $J=(J_k)_{k\in\CC}$.}
\eee   
\begin{theo}
\label{nltheo:av}
The measure $Q_0$ is a law of the process $J^{0}(\cdot)$ which is a unique weak solution of the averaged equation   (\ref{nlaver}) with the initial conditions $\DD(J(0))=\DD(I(u_0))$.
Moreover,
\bee\label{nlJconv}
\DD(J^{\eps}(\cdot)) \rightharpoonup \DD(J^0(\cdot))   \quad\mbox{as}\quad \eps\ra 0 \quad\mbox{on}\quad C([0,T],\mR^N).
\eee
This convergence is uniform in $N$. 
For all $j\in\CC$ we have
\begin{equation}
\label{nlest_averJ}
 \MO\sup\limits_{\tau\in [0,T]} e^{2\al J^0_j(\tau)} < C \quad\mbox{and}\quad
\int\limits_0^T \PR(J_j^0(\tau)<\delta)\, d\tau \ra 0 \mbox{ as } \delta\ra 0,
 \end{equation}
  where the latter convergence is  uniform in $N$.
\end{theo}
{\it Proof.} The proof of convergence (\ref{nlJconv}) follows a scheme suggested in \cite{KuPi}, \cite{Kuk12} while the latter  works  use the averaging method developed in \cite{Kh},\cite{FW}.
Main difficulties of our situation compared to \cite{Kh} are similar to those in \cite{KuPi}, \cite{Kuk12} and  manifest themselves in the proof of Lemma \ref{nllem:lem} below. Equation (\ref{nlang'}) has a singularity when $J_k=0$, and for $J$, such that the rotating frequencies $\chp_{J_j}H_0$ are rationally dependent,  system  (\ref{nlac'})-(\ref{nlang'}) enters into resonant regime.   To overcome these difficulties we note that singularities and resonances have Lebesgue measure zero and prove the following lemma, which affirms that the probability of the event that actions $J^{\eps}$ for a long lime belong to a set of small Lebesgue measure  is small. A similar idea was used in  \cite{FW03,FW06}, where  was established the stochastic averaging principle for different systems with weak resonances (\cite{FW03,FW06}) and singularities (\cite{FW06}). See also \cite{FW}, Chapters 9.2 and 9.3.
\begin{align}\nonumber
&\mbox{Let $\La\subset\mZ^d$ be independent from $N$ and satisfies $\La\subset\CC(N)$ for $N\geq N_\La$. Denote by $M$}  \\
\label{nlLalala}
&\mbox{ the number of nodes in $\La$. Further on we assume that $N\geq N_\La$.}
\end{align}
\begin{lem}
\label{nllem:smallact}
Let $\JJ^{\eps}:=(J^{\eps}_k)_{k\in\Lambda}$ and a set $E^\eps\in\mR^M_{+0}$ be such that its Lebesgue measure $|E^\eps|\rightarrow 0$ as $\eps\rightarrow 0$. Then
\begin{equation}
\int\limits_0^T \PR \big(  \JJ^{\eps}(\tau) \in E^\eps\big) \, d\tau \rightarrow 0 \quad\mbox{as}\quad \eps \rightarrow  0  \quad\mbox{uniformly in $N$.} 
\end{equation}
\end{lem}
The proof of Lemma \ref{nllem:smallact} is based on Krylov's estimates (see \cite{Kry1}) and the concept of local time. It follows a scheme suggested in \cite{Sh} (see also  \cite{KuSh}, Section 5.2.2). 

Another difficulty, which is the principal difference between our case and those of all works mentioned above, is that we need to establish the uniformity in $N$ of the convergence (\ref{nlJconv}).  For this purpose we  use the uniformity of estimates and convergences  of Corollary \ref{nlcor:est} and Lemmas \ref{nllem:smallact},\ref{nllem:lem}, and the fact that the averaged equation   for the infinite system of rotators has a unique weak solution (in a suitable class). 

Now let us formulate the following averaging lemma which is the main tools of the proof of the theorem.
\begin{lem}
\label{nllem:lem}
Take a function $P\in \LL_{loc}(\mC^N)$ which depends on $v=(v_j)_{j\in\CC}\in\mC^N$ only through $(v_j)_{j\in\Lambda}\in\mC^M$. Let it has at most a polynomial growth at infinity. Then, writing $P(v)$ in the action-angle coordinates $P(v)=P(J,\psi)$, we have 
\begin{equation*}
\MO \sup\limits_{\tau\in [0,T]} \left| \int\limits_0^\tau P(J^{\eps}(s), \psi^{\eps}(s)) - \lan P \ran (J^{\eps}(s)) \, ds \right| \rightarrow 0 \mbox{ as } \eps \rightarrow 0 \mbox{ uniformly in $N$.}
\end{equation*}
\end{lem}
Similarly one can prove that 
\begin{equation}
\label{nllem''}
\MO \sup\limits_{\tau\in [0,T]} \left| \int\limits_0^\tau P(J^{\eps}(s), \psi^{\eps}(s)) - \lan P \ran (J^{\eps}(s)) \, ds \right|^2 \rightarrow 0 \mbox{ as } \eps \rightarrow 0 \mbox{ uniformly in $N$.}
\end{equation} 
We establish Lemmas \ref{nllem:smallact} and \ref{nllem:lem} in Section \ref{nlsec:lemmas}.
\ssk 

Now we will prove that $Q_0$ is a law of a weak solution of (\ref{nlaver}).  It sufficies to show (see \cite{KarShr}, chapter 5.4) that for any $j,k,l\in\CC$ the  processes 
\begin{equation}
\label{nlmarting}
Z_{j}(\tau):= J_{j}(\tau) - \int\limits_0^{\tau} (\RR_{j}(J(s))+\TT_{j}) \, ds, \quad
Z_{k} Z_{l}(\tau) - 2\delta_{kl} \TT_{k}\int\limits_0^{\tau} J_{k}(s) \, ds
\end{equation}
are square-integrable martingales with respect to the measure $Q_0$ and the natural filtration of $\sigma$-algebras in $C([0,T], \mR^N)$.
We establish it for the first process, for the second the proof is similar, but one should use (\ref{nllem''}) (for the first one we do not need this).
Consider the process
\begin{equation}
\label{nlKKK}
K_{j}^{\eps_k}(\tau):=J_{j}^{\eps_k}(\tau)- \int\limits_0^\tau (\RR_{j}(J^{\eps_k}(s)) +\TT_{j})\, ds.
\end{equation} 
Then, according to (\ref{nlac'}), 
\begin{equation}
\nonumber
K_{j}^{\eps_k}(\tau)=M_{j}^{\eps_k}(\tau) + \Theta_{j}^{\eps_k}(\tau),
\end{equation}
where
$M_{j}^{\eps_k}$ is a martingal and by (\ref{nlavdrift}) we have
\bee
\label{nlsxodgener}
\Theta_{j}^{\eps_k}(\tau)=\int\limits_0^\tau\big( (i\nabla_{j} H_2 +g_{j})\cdot v_j^{\eps_k}  - \lan(i\nabla_{j} H_2 +g_j)\cdot v_{j}^{\eps_k} \ran  
+ \sqrt\eps r_{j}\cdot v_{j}^{\eps_k} + (d_{jj}^1-\TT_{j})  \big) \,ds.
\eee
Due to Corollary \ref{nlcor:est} and Theorem \ref{nllem:prest}.3, we have
\bee
\label{nlmalenkie}
\MO \sup\limits_{0\leq \tau \leq T }|r_{j} \cdot v^{\eps_k}_{j}| \leq C, \quad |d_{jj}^1-\TT_{j}|\leq C\sqrt\eps. 
\eee
Then, applying Lemma \ref{nllem:lem}, we get
\bee
\label{nlsxodgener'}
\MO\sup\limits_{0\leq \tau\leq T} |\Theta_{j}^{\eps_k}(\tau) |\rightarrow 0 \quad \mbox{as}\quad \eps_k\rightarrow 0.
\eee
Consequently, 
\begin{equation}
\label{nlK=M}
\lim\limits_{\eps_k\rightarrow 0}\DD\big(K_{j}^{\eps_k}(\cdot)\big)=\lim\limits_{\eps_k\rightarrow 0}\DD\big( M_{j}^{\eps_k}(\cdot)\big)
\end{equation}
in the sense that if one limit exists then the another exists as well and the two are equal.

Due to  (\ref{nllp}) and the Skorokhod Theorem,  we can find random processes $L^{\eps_k}(\tau)$ and $L(\tau)$, $0\leq \tau \leq T$, such that $\DD\big(L^{\eps_k}(\cdot)\big)=\DD\big(J^{\eps_k}(\cdot)\big)$, $\DD\big(L(\cdot)\big)=Q_0$ and $$L^{\eps_k}\rightarrow L \quad \mbox{in} \quad C([0,T], \mR^N) \quad \mbox{as} \quad  \eps_k\rightarrow 0 \quad\mbox{a.s.}$$
 Then by (\ref{nlKKK}) the left-hand side limit in (\ref{nlK=M}) exists and equals
\begin{equation}
\label{nlmartingals}
L_{j}(\tau) - \int\limits_0^{\tau} (\RR_j(L(s))+\TT_{j}) \, ds.
\end{equation}

Due to (\ref{nlest'}), the family of martingales $\{M_{j}^{\eps_k},\,k\in\mN\}$ is uniformly square integrable. Due to (\ref{nlK=M}), they converge in distribution to the process (\ref{nlmartingals}). Then the latter is a square integrable martingal as well. Thus, each limiting point $Q_0$ is a weak solution of the averaged equation   (\ref{nlaver}).

Since the initial conditions $u_0$ are independent from $\eps$, Theorem \ref{nllem:prest}.4 implies that $\DD(J(0))=\DD(I(u_0))$. In \cite{YW} Yamada and Watanabe established the uniqueness of a weak solution for an equation with a more general dispersion matrix then that for (\ref{nlaver}), but with a Lipschitz-continuous drift. Their proof can be easily generalized to our case by the stopping time arguments. We will not do this here since in Proposition \ref{nlN} we will consider more difficult infinite-dimensional situation.  
 
The uniqueness of a weak solution of (\ref{nlaver}) implies that all the limiting points (\ref{nllp}) coincide and we obtain the convergence (\ref{nlJconv}).
The first estimate in (\ref{nlest_averJ}) follows from Corollary \nolinebreak \ref{nlcor:est} and the second one follows from Lemma \ref{nllem:smallact}.
\ssk

Now we will prove the uniformity in $N$ of the convergence (\ref{nlJconv}). Recall that it is understood in the sense that for any  $\La\subset\mZ^d$ as in (\ref{nlLalala})  we have
\bee\label{nlJJc}
\DD\big((J^{\eps}_{j}(\cdot))_{j\in\La}\big) \raw \DD\big((J^{0}_{j}(\cdot))_{j\in\La}\big) \quad\mbox{as }\eps\ra 0 
 \mbox{ on } C([0,T],\mR^M)\mbox{ uniformly in }N.
\eee
It is well known that the weak convergence of probability measures on a separable metric space  is equivalent to convergence in  the dual-Lipschitz norm, see Theorem 11.3.3 in \cite{Dud}. Analysing the proof of this theorem, we see that in order to establish the uniformity in $N$ of the convergence (\ref{nlJJc}) with respect to the dual-Lipschitz norm, it suffices to show that for any bounded continuous functional $h: (J_j(\cdot))_{j\in\La}\in C([0,T],\mR^M) \mapsto \mR$, we have
\bee\label{nlJJJconv}
\MO h(J^{\eps})\ra\MO h(J^{0})\quad \mbox{as} \quad \eps\ra 0 \quad \mbox{uniformly in } N,
\eee
where we have denoted $h(J):=h\big((J_j(\cdot))_{j\in\La}\big)$. In order to prove (\ref{nlJJJconv}),
first we pass to the limit $N\ra\infty$. Recall that $\CC^\infty=\cup_{N\in\mN}\CC(N)$. 
Denote $J^{\eps,N}=(J^{\eps,N}_j)_{j\in\CC^\infty}$, where 
\bee
\label{nlprodolzhenie}
J_j^{\eps,N}:=
\left\{
\begin{array}{cl}
J_j^{\eps},\quad &\mbox{if}\quad j\in\CC=\CC(N), \\
0, \quad &\mbox{if}\quad j\in\CC^\infty\sm\CC.
\end{array}
\right.
\eee 
Using the  uniformity in $N$ of estimate (\ref{nlest'}), we get that the family of measures 
$$\{\DD(J^{\eps,N}(\cdot)),\; 0<\eps\leq 1,\; N\in\mN\}$$ 
 is tight on a space $C([0,T],\mR^\infty)$.
Take any limiting point $Q_0^\infty$ such that $\DD(J^{\eps_k,N_k}(\cdot))\raw Q_0^\infty$ as $\eps_k\ra 0, \, N_k\ra\infty$. Recall that the initial conditions $u_0$ satisfy {\it HI(ii)}. Denote the vector of actions corresponding to $u_0^\infty$ by $I_0^\infty=I(u_{0}^\infty)\in\mR_{0+}^\infty$.
\begin{prop}
\label{nlN}
The measure $Q_0^\infty$ is a law of the process $J^{0,\infty}(\tau)$ which is a unique weak solution of the averaged equation   for the infinite system of rotators
\bee
\label{nlaverN}
d J_{j} = (\RR_{j}(J)+\TT_{j}) \,d\tau + \sqrt{2J_{j}\TT_{j}}\,d\wid\beta_{j}, \quad j\in\CC^\infty, \quad \DD(J(0))=\DD(I^\infty_0),
\eee
in the class of processes satisfying the first estimate from (\ref{nlest_averJ}), for all $j\in\CC^\infty$. 
Moreover, $\DD(J^{\eps,N}(\cdot))\raw \DD(J^{0,\infty}(\cdot))$ as $\eps\ra 0,\, N\ra\infty$ on $C([0,T],\mR^\infty)$.  
\end{prop} 
Before proving this proposition we will establish (\ref{nlJJJconv}).
Proposition \ref{nlN} implies 
\bee
\label{nlepsN}
\MO h(J^{\eps})\ra \MO h(J^{0,\infty}) \quad \mbox{as} \quad\eps\ra 0,\,N\ra\infty.  
\eee
In view of convergence (\ref{nlJconv}) which is already proven for every $N$,  (\ref{nlepsN}) implies that  $\MO h(J^{0})\ra \MO h(J^{0,\infty})$ as $N\ra\infty$.
Consequently, for all $\delta>0$ there exist $N_1\in\mN$ and $\eps_1>0,$ such that for every $N\geq N_1,\, 0\leq \eps < \eps_1$, we have 
$$
|\MO h(J^{\eps})-\MO h(J^{0,\infty})|< \delta/2.
$$  
Then, for $N$ and $\eps$ as above, 
\bee
\label{nlras}
|\MO h(J^{\eps})-\MO h(J^{0})|\leq |\MO h(J^{\eps})-\MO h(J^{0,\infty})| +
|\MO h(J^{0,\infty})-\MO h(J^{0})|< \delta. 
\eee
Choose $\eps_2>0$ such that for every $0<\eps<\eps_2$ and $N<N_1$ we have
\bee
\label{nldvas}
|\MO h(J^{\eps})-\MO h(J^{0})|< \delta.
\eee
Then, due to  (\ref{nlras}), (\ref{nldvas}) holds for all $N$ and $\eps<\eps_1\wedge\eps_2$.
Thus, we obtain (\ref{nlJJJconv}). The proof of the theorem is completed. \qed 
\ssk

{\it Proof of Proposition \ref{nlN}.} To prove that $Q_0^\infty$ is a law of a weak solution of (\ref{nlaverN}), it suffices to show that the processes (\ref{nlmarting}) are square-integrable martingales with respect to the measure $Q_0^\infty$ and the natural filtration of $\sigma$-algebras in $C([0,T], \mR^\infty)$ (see \cite{Yor}). The proof of that literally coincides with the corresponding proof for the finite-dimensional case, one should just replace the limit $\eps_k\ra 0$ by $\eps_k\ra 0, N_k\ra\infty$ and the space $C([0,T], \mR^N)$ by $C([0,T], \mR^\infty)$. Estimate of Corollary \ref{nlcor:est} joined with Fatou's lemma implies that the obtained weak solution belongs to the desired class of processes. To prove that a weak solution of  (\ref{nlaverN}) is unique, it suffices to show that the pathwise uniqueness of solutions holds (see \cite{Yor,RSZ}). Let  $J(\tau)$ and $\hat J(\tau)$ be two solutions of (\ref{nlaverN}), defined on the same probability space, corresponding to the same Brownian motions and initial conditions, distributed as $I_0^\infty$, and satisfying the first estimate from (\ref{nlest_averJ}). Let $w(\tau):=J(\tau)-\hat J(\tau).$ 
Following literally the proof of Theorem 1 in \cite{YW}, for every $j\in\CC^\infty$ and any $\tau\geq 0$ we get the estimate 
\bee\label{nldif}
\MO |w_j(\tau)|\leq \MO\int\limits_{0}^\tau |\RR_j(J(s))-\RR_j(\hat J(s))|\, ds. 
\eee
Define for $R>0$ and $q>0$ a stopping time
$$\tau_R=\inf\{\tau\geq 0:\,\exists j\in\CC^\infty\mbox{ satisfying } J_j(\tau)\vee \hat J_j(\tau)\geq R(|j|^q+1)\}.$$ 
For any $\tau\geq 0$ we have
\begin{align}\nonumber
\PR(\tau_R\leq \tau) &\leq \sum\limits_{j\in\CC^\infty}\PR(\sup\limits_{0\leq s\leq\tau }J_j(s)\geq R(|j|^q+1))+ 
 \sum\limits_{j\in\CC^\infty}\PR(\sup\limits_{0\leq s\leq\tau } \hat J_j(s)\geq R(|j|^q+1))\\
\label{nlstaverinfty}
&\leq C\sum\limits_{j\in\CC^\infty}e^{-2\al R (|j|^q+1)} \ra 0 \quad\mbox{as}\quad R\ra\infty.
\end{align}
 For $L\in\mN$ denote $|w|_L:=\sum\limits_{|j|\leq L} e^{-|j|} |w_j|$. Using the Taylor expansion, it is possible to show that, in view of (\ref{nlRRR}) and assumption {\it Hg(i)}, the derivatives $\chp_{J_k}\RR_j(J)$ have at most a polynomial growth of some power $m>0$, which is uniform in $j,k\in\CC^\infty$.  Since for any $\tau<\tau_R$ and $k\in\CC^\infty$ satisfying $|k|\leq L+1$ we have $J_k(\tau),\hat J_k(\tau)\leq  R((L+1)^q+1)$, then  estimate (\ref{nldif}) implies 
\begin{align}\nonumber
\MO |w(\tau\wedge\tau_R)|_L &\leq C \sum\limits_{|j|\leq L}e^{-|j|}\MO\int\limits_0^{\tau\wedge\tau_R}\Big(1+\sum\limits_{k:|k-j|\leq 1}(J_k+\hat J_k)^m\Big) \sum\limits_{k:|k-j|\leq 1} |w_j|\, ds \\ \nonumber
&\leq C(R) (L+1)^{mq}\MO\int\limits_0^{\tau\wedge\tau_R} \Big( |w|_L  + 
 e^{-L}\sum\limits_{|k|= L+1} |w_k|\Big)  \, ds \\ \nonumber 
&\leq C_1(R) (L+1)^{mq} \int\limits_0^{\tau}\big( \MO |w(s\wedge\tau_R)|_L  + e^{-L}L^{d-1} \big) \, ds,
\end{align}
where we used $\MO \sum\limits_{|k|= L+1} |w_k|\leq CL^{d-1}$. Applying the Gronwall-Bellman inequality, we obtain 
$$\MO |w(\tau\wedge\tau_R)|_L\leq L^{d-1}e^{-L+C_1(R)(L+1)^{mq}\tau}.$$
Choosing $q< 1/m$, we  obtain that $\MO |w(\tau\wedge\tau_R)|_L\ra 0$ as $L\ra\infty$ and, consequently,  $\MO |w_j(\tau\wedge\tau_R)|=0$ for all $j\in\CC^\infty$. Sending $R\ra\infty$, in view of (\ref{nlstaverinfty}) we get that $\MO |w_j(\tau)|=0$ for any $\tau\geq 0$ and $j\in\CC^\infty$.
\qed
\msk

Let us now investigate the dynamics in the original  $(I,\ph)$-variables. Let $u^{\eps}(\tau)$ be a solution of (\ref{nlini_e})-(\ref{nlini_c}), written in the slow time and $I^\eps(\tau)=I(u^\eps(\tau))$ be the corresponding vector of actions. 
 By Theorems \ref{nllem:prest}.4 and \ref{nltheo:av} we have
\bee
\nonumber
\lim\limits_{\eps\rightarrow 0} \DD(I^{\eps}(\cdot))=\lim\limits_{\eps\rightarrow 0} \DD(J^{\eps}(\cdot)) = Q_0 \quad \mbox{on}\quad C([0,T],\mR^N). 
\eee
 Since the estimate of Theorem \ref{nllem:prest}.4 and the convergence (\ref{nlJconv}) are uniform in $N$, then the convergence $\DD(I^{\eps}(\cdot))\raw Q_0$ is also uniform in $N$.
Thus, we get
\begin{theo}
\label{nltheo:fin_dyn} The assertion of Theorem \ref{nltheo:avIintro} holds. Moreover, for any $j\in\CC$
\begin{equation}
\label{nlest_aver}
 \MO\sup\limits_{\tau\in [0,T]} e^{2\al I^0_j(\tau)}< C \mbox{ and }
\int\limits_0^T \PR(I_j^0(\tau)<\delta)\, d\tau \ra 0 \mbox{ as } \delta\ra 0,
 \end{equation}
 where the latter convergence is uniform in $N$.
\end{theo}  
Let us define a local energy of a $j$-th rotator as 
$$
H^\eps_j(u)=\frac12 F_j(|u_j|^2) + \frac{\sqrt\eps}{4}\sum\limits_{k:|j-k|=1} G(|u_j-u_k|^2).
$$
Consider the vectors 
$
\hat H^\eps(u):=(H^\eps_j(u))_{j\in\CC} 
$
and
$
\hat F(I):=\frac12 (F_j(2I_j))_{j\in\CC}.
$
\begin{cor}\label{nllocen}Let  $I^0(\tau)$ be a unique weak solution of system (\ref{nlav_I})-(\ref{nlav_Ic}). Then
$$\DD\big(\hat H^\eps(u^\eps(\cdot))\big) \raw  \DD\big(\hat F(I^0(\cdot))\big) 
\quad\mbox{as}\quad \eps\ra 0 \quad\mbox{on}\quad C([0,T],\mR^N),$$ 
uniformly in $N$.
\end{cor}
{\it Proof.}
The second estimate of Theorem \ref{nllem:prest}.4 implies that the process $u^\eps$ satisfies the first estimate of Corollary \nolinebreak \ref{nlcor:est}. Since the potential $G$ has at most a polynomial growth, we get  
\bee\label{nl4653}
\lim\limits_{\eps\rightarrow 0} \DD\big(\hat H^\eps(u^\eps(\cdot))\big) =\lim\limits_{\eps\rightarrow 0}\DD\big(\hat F(I^\eps(\cdot))\big) \quad \mbox{on}\quad C([0,T],\mR^N)
\eee
in the sense that if one limit exists then another one exists as well and the two are equal. Moreover, if one convergence holds uniformly in $N$ then another one also holds uniformly in $N$.
It remains to note that, due to Theorem \ref{nltheo:fin_dyn},  we have 
$\DD\big(\hat F(I^\eps(\cdot))\big)\raw\DD\big(\hat F(I^0(\cdot))\big)$ as $\eps\ra 0$ uniformly in $N$.
\qed

\subsection{Joint distribution of actions and angles}
\label{nlsec:measure}

Here we prove Theorem \ref{nltheo:measure}, which describes the limiting  joint dynamics of actions and angles. Let, as usual, $u^{\eps}(\tau)$ be a solution of (\ref{nlini_e})-(\ref{nlini_c}), written in the slow time, and let  $I^\eps(\tau)=I(u^\eps(\tau)),\ph^\eps(\tau)=\ph(u^\eps(\tau))$.   
Denote by $\mu^{\eps}_\tau=\DD(I^{\eps}(\tau), \ph^{\eps}(\tau))$ the law of $u^\eps(\tau)$ in action-angle coordinates.  For any function $h(\tau)\geq 0$ satisfying $\int\limits_0^T h(\tau) \, d\tau=1$, set $\mu^{\eps}(h):=\int\limits_0^T h(\tau) \mu^{\eps}_\tau \, d\tau$. Moreover, denote $m^{0}(h):=\int\limits_0^T h(\tau) \DD(I^{0}(\tau)) \, d\tau$, where $I^{0}(\tau)$ is a weak solution of (\ref{nlav_I})-(\ref{nlav_Ic}). 
\begin{theo}
\label{nltheo:measure}
For any continuous function $h$ as above, we have
$$
\mu^{\eps}(h)\rightharpoonup m^{0}(h)\times d\ph \quad \mbox{as} \quad \eps\rightarrow 0 \quad \mbox{uniformly in } N. 
$$
\end{theo} 
{\it Proof.}
Let us first consider the case $h=(\tau_2-\tau_1)^{-1} \mI_{[\tau_1,\tau_2]}$, where $\mI_{[\tau_1,\tau_2]}$ is an indicator function of the interval $[\tau_1,\tau_2]$. Take a set $\La$ as in (\ref{nlLalala}) and a function  $P\in \LL_b(\mR^N\times\mT^N)$ which depends on $(I,\ph)=(I_j,\ph_j)_{j\in\CC}\in\mR^N\times\mT^N$ only through $(I_j,\ph_j)_{j\in\La}$. Let us first treat the case when the function $P(u):=P(I,\ph)(u)$ belongs to $\LL_{loc}(\mC^N)$ (this can fail since the vector-function $\ph(u)$ has a discontinuity  when $u_j=0$ for some $j\in\CC$, so the function $ P(u)$ may be also discontinuous there). Let $v^{\eps}(\tau)$ be a solution of (\ref{nlvs}) and $J^\eps(\tau),\psi^\eps(\tau)$ be the corresponding vectors of actions and angles.
Due to Theorem \nolinebreak\ref{nllem:prest}.4, we have 
$$\int\limits_{\tau_1}^{\tau_2} \lan\mu^{\eps}_\tau,P\ran \, d\tau=
\MO \int\limits_{\tau_1}^{\tau_2} P(u^{\eps}(\tau)) \, d\tau \quad \mbox{is close to} \quad
\MO\int\limits_{\tau_1}^{\tau_2}P( v^{\eps}(\tau)) \, d\tau \quad \mbox{uniformly in $N$}.$$ 
Due to  Lemma \ref{nllem:lem}, the integral $\ds{\MO\int\limits_{\tau_1}^{\tau_2} P( v^{\eps}(\tau)) \, d\tau}$ is close to 
 $\ds{\MO\int\limits_{\tau_1}^{\tau_2}\lan P \ran(J^{\eps}(\tau)) \, d\tau}$ uniformly in $N$.
 Due to Theorem \ref{nltheo:av}, the last integral is uniformly in $N$ close to 
$$\MO\int\limits_{\tau_1}^{\tau_2}\lan P \ran(J^{0}(\tau)) \, d\tau=\MO\int\limits_{\mT^N}\int\limits_{\tau_1}^{\tau_2}P(J^{0}(\tau), \ph) \, d\tau d\ph=(\tau_2-\tau_1)\lan m^{0}(h)\times d\ph,P\ran.$$
If the function $P(u)\notin \LL_{loc}(\mC^N)$, we approximate it by functions $P_\del(u)\in\LL_{loc}(\mC^N)$,
$$
P_\del(u) = P(u) k_\del([I(u)]), \qquad [I(u)]:=\min\limits_{j\in\La} I_j(u), 
$$
where the function $k_\del$ is smooth, $0\leq k_\del\leq 1$, $k_\del(x)=0$ for $x\leq \del$ and $k_\del(x)=1$ for $x\geq 2\del$. 
Then we let $\del\ra 0$ as $\eps\ra 0$ and use the estimate of  Lemma \ref{nllem:smallact} and (\ref{nlest_averJ}).

In the case of a continuous  function $h$, we approximate it by piecewise constant functions. 
\qed

\subsection{Stationary measures}
\label{nlsec:st_measures}
In this section we prove  Theorem \ref{nltheo:stmintro} which describes the limiting  behaviour of a stationary regime of (\ref{nlini_e}).
\ssk

{\bf The effective equation and proof of Theorem  \ref{nltheo:stmintro}.i.}
The averaged equation   (\ref{nlaver}) is irregular: its dispersion matrix is not Lipschitz continuous, so we do not know if (\ref{nlaver}) is mixing or not. We are going to lift it to so-called effective equation which is regular and mixing.

Let us define an operator  $\Psi_\theta: v=(v_j)_{j\in\CC}\in\mC^N\mapsto\mC^N$ of rotation by an angle $\theta=(\theta_j)_{j\in\CC}\in\mT^N$, i.e. $(\Psi_\theta v)_j=v_je^{i\theta_j}$.
We rewrite the function $\RR_j$ from (\ref{nlusre}) as
\begin{equation}
\label{nlef_drift}
\RR_j(J)=\lan g_j(v)\cdot v_j \ran = \int\limits_{\mT^{N}} g_j(\Psi_\theta v)\cdot (e^{i\theta_j} v_j)\, d\theta= \KK_j(v)\cdot v_j,
\end{equation}
where 
$\ds{\KK_j(v):=\int\limits_{\mT^{N}} e^{-i\theta_j}g_j(\Psi_\theta v)\, d\theta}$
and $ d\theta$ is a normalized Lebesgue measure on the torus $ \mT^{N}$.
Consider the {\it effective equation} 
\begin{equation}
\label{nlef}
d v_j = \KK_j(v)\,d\tau + \sqrt{\TT_j}d\beta_j, \quad j\in\CC,
\end{equation}
where $\beta_j$, as usual, are standard complex independent Brownian motions. It is well known that a stochastic equation of the form (\ref{nlef}) has a unique solution which is defined globally (see \cite{Khb}), and that it is mixing (see \cite{Khb},\cite{Ver},\cite{VerPol}). The following proposition explains the role of the effective equation.
\begin{prop}
\label{nltheo:lifting}
(i)
Let $v(\tau)$, $\tau\geq 0$ be a weak solution of the effective equation (\ref{nlef}) and $J(\tau)=J(v(\tau))$ be the corresponding vector of actions. Then
$J(\tau)$, $\tau\geq 0$ is a weak solution of the averaged equation   (\ref{nlaver}).

(ii) Let $J^0(\tau)$, $\tau\geq 0$ be a weak solution of the averaged equation   (\ref{nlaver}). Then for any vector $\theta=(\tht_j)_{j\in\CC}\in\mT^{N}$ there exists a weak solution $v(\tau)$ of the effective equation (\ref{nlef})  such that 
\bee\label{nlliftt}
\DD(J(v(\cdot)))=\DD(J^0(\cdot)) \mbox{ on } C([0,\infty),\mR^N) \quad \mbox{and} \quad v_j(0)=\sqrt{2J^0_j(0)}e^{i\theta_j},\,j\in\CC.
\eee  
\end{prop}
{\it Proof. (i)}
Due to (\ref{nlef_drift}) and (\ref{nlef}), the actions $J(\tau)$ satisfy
\bee
\label{nleff_ac}
d J_j=(\RR_j(J) + \TT_j)\,d\tau + \sqrt{\TT_j}v_j\cdot\,d\beta_j, \quad j\in\CC.
\eee
 The drift and the diffusion matrix of equation (\ref{nleff_ac}) coincide with those of the averaged equation   (\ref{nlaver}). Consequently, $J(\tau)$ is a solution of the (local) martingale problem associated with the averaged equation    (see \cite{KarShr}, Proposition. 5.4.2). So, due to \cite{KarShr}, Proposition 5.4.6, we get that $J(\tau)$ is a weak solution of the averaged equation   (\ref{nlaver}).
 
 {\it (ii)} Let $v(\tau)$  be a solution the effective equation with the initial condition as in (\ref{nlliftt}).  Then, due to {\it (i)}, the process $J(\tau):=J(v(\tau))$ is a weak solution of the averaged equation   and $J(0)=J^0(0)$. Since the weak solution of the averaged equation    is unique, we obtain that $\DD(J(\cdot))=\DD(J^0(\cdot))$. Consequently, $v(\tau)$ is the desired process.
 \qed
 
Let $m$ be the unique stationary measure of the effective equation. Denote the projections to the spaces of actions and angles by $\Pi_{ac}:v\in\mC^N\mapsto\mR^N_{+0}\ni I$ and $\Pi_{ang}:v\in\mC^N\mapsto\mT^N\ni\psi$ correspondingly.
Denote 
\bee\label{nlprojection}
\pi:=\Pi_{ac*}m.
\eee
\begin{cor}
\label{nlcor:avmix}
The averaged equation   (\ref{nlaver}) is mixing, and $\pi$ is its unique stationary measure. More precisely,
for any its solution $J(\tau)$ we have
$\DD(J(\tau))\raw\pi$ as $\tau\ra\infty.$
\end{cor}
Corollary \ref{nlcor:avmix} implies Theorem \ref{nltheo:stmintro}.i.

{\it Proof}. 
First we claim that $\pi$ is a stationary measure of the averaged equation. Indeed, take a stationary distributed solution  $\wid v(\tau)$ of the effective equation, $\DD(\wid v(\tau))\equiv m$. By Proposition \ref{nltheo:lifting}.i, the process $J(\wid v(\tau))$ is a stationary weak solution of the averaged equation. It remains to note that (\ref{nlprojection}) implies $\DD\big(J(\wid v(\tau))\big)\equiv \pi.$ 

Now we claim that any solution $J^0(\tau)$ of the averaged equation converges in distribution to $\pi$ as $\tau\ra\infty$. 
For some $\tht\in\mT^N$  take  $v(\tau)$ from Proposition \ref{nltheo:lifting}.ii.
Due to the mixing property of the effective equation, $\DD(v(\tau))\raw m$ as $\tau\ra\infty$ and, consequently, $\DD(J^0(\tau))=\DD(J(v(\tau)))\raw \Pi_{ac*}m=\pi$ as $\tau\ra\infty$.
\qed 
\msk
  
{\bf Proof of Theorem \ref{nltheo:stmintro}.ii.}
First we will show that 
\bee\label{nlconvact}
\Pi_{ac*}\wid\mu^\eps\rightharpoonup \pi\quad \mbox{as}\quad  \eps\ra 0.
\eee
We will work in the $v$-variables. 
Note that equation (\ref{nlvs}) is mixing since it is obtained by a $C^2$-smooth time independent change of variables from equation (\ref{nlini_e}), which is mixing.  
Denote by $\wid\nu^\eps$ its unique stationary measure. 
Due to Theorem \ref{nllem:prest}.4, to establish (\ref{nlconvact}) it suffices to show that
\bee
\label{nlvsowlos}
\Pi_{ac*}\wid\nu^\eps\rightharpoonup \pi \quad \mbox{as}\quad  \eps\ra 0.
\eee 
Let $\wid v^\eps(\tau)$ be a stationary solution of equation (\ref{nlvs}), $\DD(\wid v^\eps(\tau))\equiv \wid\nu^\eps$, and $\wid J^\eps(\tau)=J(\wid v^\eps(\tau))$ be the corresponding vector of actions. Similarly to Proposition \ref{nlprop:tight} we get that the set of laws $\{\DD(\wid J^\eps(\cdot)),\; 0<\eps\leq 1\}$ is tight in $C([0,T],\mR^N)$. Let $\wid Q_0$ be its limiting point as $\eps_k\ra 0.$ Obviously, it is stationary in $\tau$. The same arguments that was used in the proof of Theorem \ref{nltheo:av} imply  
\bpp\label{nlprop:stact}
The measure $\wid Q_0$ is a law of the process $\wid J^0(\tau)$, $0\leq \tau \leq T$, which is a stationary weak solution of
the averaged equation   (\ref{nlaver}).   
\epp
Since $\pi$ is the unique stationary measure of  the averaged equation,  we have $\DD(\wid J^0(\tau))\equiv\pi$. Consequently, we get (\ref{nlvsowlos}) which implies (\ref{nlconvact}). 

Let $\wid u^\eps(\tau)$ be a stationary solution of equation (\ref{nlini_e}) and $\wid I^\eps(\tau),\wid \ph^\eps(\tau)$ be the corresponding  vectors of actions and angles. By the same reason as in Theorem \ref{nltheo:measure}, we have
\bee\label{nljdstm}
\wid \mu^\eps(h)\raw \wid m^0(h)\times d\ph \quad\mbox{as}\quad \eps\ra 0,
\eee
where $\wid \mu^\eps(h)$ and $ \wid m^0(h)$ are defined as $\mu^\eps(h)$ and $m^0(h)$, but with the processes 
$I^\eps(\tau),\ph^\eps(\tau)$ and $I^0(\tau)$ replaced by the processes $\wid I^\eps(\tau),\wid\ph^\eps(\tau)$ and $\wid J^0(\tau)$ correspondingly. Since a stationary regime does not depend on time, we get (\ref{nlintroconv}): 
\bee\label{nljdstm1}
\DD(\wid I^\eps(\tau), \wid \ph^\eps(\tau)) \raw \pi\times d\ph \quad\mbox{as}\quad \eps\ra 0.
\eee

Assume now that the averaged equation   for the infinite system of rotators (\ref{nlaverN}) has a unique stationary measure $\pi^\infty$ in the class of measures satisfying $\sup\limits_{j\in\CC^\infty}\lan \pi^\infty, J_j\ran <\infty$.
Let us define $\wid J^{\eps,N}$ as in (\ref{nlprodolzhenie}), but with $J^\eps$ replaced by $\wid J^\eps$. Then the set of laws $\{\DD(\wid J^{\eps,N}(\cdot)),\; 0<\eps\leq 1,N\in\mN\}$ is tight in $C([0,T],\mR^\infty)$. Let $\wid Q^\infty_0$ be its limiting point as $\eps_k\ra 0,N_k\ra\infty$. Similarly to Proposition \ref{nlN} we get
\bpp\label{nlprop:Nst}
The measure $\wid Q_0^\infty$ is a law of the process $\wid J^{0,\infty}(\tau)$, $0\leq \tau \leq T$, which is a stationary weak solution of equation (\ref{nlaverN}), satisfying the first estimate from (\ref{nlest_averJ}) for all $j\in\CC^\infty.$
\epp
Thus, we obtain that a marginal distribution of the measure $\wid Q_0^\infty$ as $\tau=const$ is a stationary measure of eq. (\ref{nlaverN}) from the class of measures above. So that it coincides with $\pi^\infty$ and we have $\DD(\wid J^{\eps,N}(\tau))\raw \pi^\infty $ as $\eps\ra 0, N\ra\infty$.
Then, arguing as in Theorem \ref{nltheo:av}, we get that the convergence (\ref{nlvsowlos}) is uniform in $N$. As in the proof of Theorem \ref{nltheo:measure}, this implies that the convergence (\ref{nljdstm}) and,  consequently, the convergence  (\ref{nljdstm1}) are also uniform in $N$. 
\ssk

{\bf Proof of Theorem \ref{nltheo:stmintro}.iii.} Due to the mixing property of (\ref{nlini_e}), we have $\DD(I^{\eps}(\tau))\rightharpoonup \Pi_{ac*}\wid\mu^\eps$ as $\tau\ra\infty$. Then item {\it (ii)} of the theorem implies that
$\Pi_{ac*}\wid\mu^\eps\raw \pi$ as $\eps\ra 0$. On the other hand, Theorem \ref{nltheo:fin_dyn} implies that $\DD(I^\eps(\tau))\rightharpoonup \DD(I^0(\tau))$ as $\eps\ra 0$ for any $\tau \geq 0$, where $I^0(\tau)$ is a weak solution of equation (\ref{nlav_I})-(\ref{nlav_Ic}). Then item {\it (i)} of the theorem implies that $\DD(I^0(\tau))\rightharpoonup \pi$ as $\tau\ra\infty$. 
The proof of the theorem is completed.
\qed

\begin{rem}
\label{nlrem:stmef1} It is possible to show  that the effective equation is rotation invariant: if $v(\tau)$ is its weak solution, then for any $\xi\in\mT^{N}$ we have that $\Psi_\xi v$ is also its weak solution. Since it has the unique stationary measure $m$, we get that $m$ is rotation invariant. Consequently,  $\Pi_{ang*}m=\,d\ph.$ That is why the convergence (\ref{nlstmef1}) is equivalent to (\ref{nlintroconv}).
\end{rem}

\subsection{Examples}
\label{nlsec:example}

{\bf 1.} Consider a system with linear dissipation, i.e. $p=2$ and $g_j(u)= - u_j + \sum\limits_{k:|k-j|=1} b_{jk} u_k$, where $b_{jk}\in\mC$. If $|b_{jk}|$ are sufficiently small uniformly in $j$ and $k$ then assumption {\it Hg} is satisfied (see Example \ref{nlexam:conditions}). Since $\left\lan u_k\cdot u_j\right\ran=0$ for $k\neq j$, we have  $\RR_j(I)=-2 I_j$. Then the averaged equation   (\ref{nlav_I}) turns out to be diagonal and takes the form
\bee\label{nltratata}
d I_j= (-2I_j + \TT_j) d\tau + \sqrt{2\TT_j I_j} \,d\wid\beta_j, \quad j\in\CC.
\eee
The unique stationary measure of (\ref{nltratata}) is  
$$
\pi(dI)=\prod_{j\in\CC} \frac{2}{\TT_j} \mI_{\mR_+}(I_j) e^{-2I_j/\TT_j} dI_j.
$$
The averaged equation   for the infinite system of rotators is diagonal and, consequently, has a unique stationary measure. Thus, the convergence (\ref{nlintroconv}) holds uniformly in \nolinebreak $N$.
\ssk

{\bf 2.} Let $d=1$ and $\CC=\{1,2,\ldots,N\}$.  Put for simplicity $p=4$ and choose
$$
g_j(u)=\frac{1}{4}\Big( |u_{j+1}|^{2} u_j-|u_{j-1}|^2 u_j -|u_j|^2 u_j \Big),
$$   
where $1\leq j\leq N$, $u_0=u_{N+1}:=0$.
By the direct computation one can verify that $g_j$ satisfies the condition {\it Hg}.
We have $R_j(I)=\left\lan  g_j(u)\cdot u_j \right\ran =  I_{j+1} I_j -   I_{j-1} I_j- I_j^2$, and the averaged equation   (\ref{nlav_I}) takes the form
$$
d I_j = \Big( \frac12 (2I_{j+1} I_j -   2I_{j-1} I_j) - I_j^2 + \TT_j\Big)\,d\tau + \sqrt{2I_j\TT_j}\,d\wid\beta_j. 
$$ 
Its r.h.s. consists of two parts: 
$$d I_j/d\tau =\wid\nabla \Theta(j) + \mbox{Ter}(j),$$ 
where $\Theta(j):=2I_{j+1} I_j$,  $\wid\nabla\Theta(j):= \frac12(\Theta(j) - \Theta(j-1))$ is the discrete gradient of $\Theta$,  and $\mbox{Ter}(j):=  - I_j^2 + \TT_j  + \sqrt{2I_j\TT_j}\,d\wid\beta_j/d\tau$.  
Analogically to the concept of the flow of energy (see \cite{Leb}, Section 5.2) we call the function $\Theta(j)$ the {\it flow of actions}. The term $\wid \nabla\Theta (j)$ describes the transport of actions through the $j$-th site while the term $\mbox{Ter}(j)$  can be considered as an input of a (new) stochastic thermostat interacting with the $j$-th node.
In the same way one can treat the case $p=2q$, where $q\in\mN,\,q>2$.
\ssk

\section{Auxiliary propositions}
\label{nlsec:lemmas}

In this section we  prove  Lemmas \ref{nllem:smallact} and \ref{nllem:lem}. 

\subsection{Proof of Lemma \ref{nllem:smallact}}
\label{nlsec:smallact}
For the brevity of notations we skip the index $\eps$ everywhere, except the set $E^\eps$. 
Let us rewrite 
 (\ref{nlac'}) for $k\in\Lambda$ as an equation with real noise
\begin{equation}
\label{nlrn}
d\JJ = A^\JJ \,d\tau + \sigma \,d\hat\beta, \quad \mbox{where}\quad\JJ:=(J_k)_{k\in\Lambda},\;A^\JJ:=(A^J_k)_{k\in\Lambda},
\end{equation}
$\sigma$ is $M\times 2N$ matrix with real entires and $\hat\beta=(\Ree \beta_k,\Imm\beta_k)_{k\in\CC}$.
Denote by $a=(a_{kl})_{k,l\in\La}$ the diffusion matrix for (\ref{nlrn}), divided by two,  $a:=\frac12\sigma\sigma^T.$ It is $M\times M$-matrix with real entires $a_{kl}=S^J_{kl}/2$, $k,l\in\La$, where  $S^J_{kl}$ is calculated in (\ref{nlSSS}).  
Then Theorem \nolinebreak\ref{nllem:prest}.3 implies that
\begin{equation}
\label{nlaaa}
| a_{kl}-\TT_k \delta_{kl}\frac{|v_k|^2}{2} |\leq C\sqrt{\eps}|v_k||v_l|. 
\end{equation}

{\bf Step 1.} For $R>0$ denote by $\tau_R$ the stopping time 
$$
\tau_R=\inf\{\tau\geq 0: \|\JJ(\tau)\|_{\mR^M}\vee  \|A^\JJ(\tau)\|_{\mR^M} \geq R\},
$$
 where $\|\cdot\|_{\mR^M}$ stands for the Euclidean norm in $\mR^M$, $\JJ(\tau)=\JJ(v(\tau))$, $A^\JJ(\tau)=A^\JJ(v(\tau))$, and $v(\tau)$ is a solution of (\ref{nlvs}). 
A particular case of  Theorem \nolinebreak 2.2.2 in \cite{Kry1} provides that
\bee
\label{nlkry}
\MO\int\limits_0^{\tau_R \wedge T} e^{-\int\limits_0^\tau \|A^\JJ(s)\|_{\mR^M}\, ds }\mI_{E^\eps}(\JJ(\tau)) (\det a(\tau))^{1/M} 
\, d\tau \leq C(R,M) |E^\eps|^{1/M}, 
\eee
where $a(\tau)=a(v(\tau))$.
Denote the event
$\Omega_\nu(\tau)=\{\det a(\tau)< \nu\}.
$
We have 
\begin{align}\nonumber
\int\limits_0^T \PR(\JJ(\tau)\in E^\eps)\, d\tau&=\MO \int\limits_0^T \mI_{E^\eps}(\JJ(\tau)) \, d\tau \leq \MO \int\limits_0^{\tau_R \wedge T} \mI_{E^\eps}(\JJ(\tau)) \mI_{\ov\Omega_\nu}(\JJ(\tau)) \Big(\frac{\det a(\tau)}{\nu}\Big)^{1/M} \, d\tau  \\ 
\label{nlkryy}
&+ \int\limits_0^T \PR(\Omega_\nu(\tau)) \, d\tau+ T\PR(\tau_R<T)=:\YY_1+\YY_2+\YY_3.
\end{align}
Due to (\ref{nlkry}),
\bee
\label{nlda4a3}
\YY_1\leq \frac{e^{TR}}{\nu^{1/M}}   \MO\int\limits_0^{\tau_R \wedge T} e^{-\int\limits_0^\tau \|A^\JJ(s)\|_{\mR^M}\, ds } \mI_{E^\eps} (\JJ(\tau))(\det a(\tau))^{1/M} \, d\tau \leq C(R,M)\Big(\frac{|E^\eps|}{\nu}\Big)^{1/M}.
\eee
Take $\nu=\sqrt{|E^\eps|}$. Choosing $R$ sufficiently large and $\eps$ sufficiently small, we can make the terms $\YY_1$ and $ \YY_3$ arbitrary small uniformly in $N$. Indeed, for $\YY_1$ this follows from (\ref{nlda4a3}), while for $\YY_3$ this follows from Corollary \ref{nlcor:est} and estimate (\ref{nlest'}). So, to finish the proof of the lemma it remains to show that if
 $\nu(\eps)\rightarrow 0$ with $\eps\ra 0$ then
\bee\label{nlommegasmall}
 \YY_2=\int\limits_0^T \PR(\Omega_\nu(\tau)) \, d\tau \rightarrow 0  \quad\mbox{when}\quad \eps \rightarrow 0 \quad\mbox{uniformly in }N. 
\eee
{\bf Step 2.} The rest of the proof is devoted to the last convergence.
Note that by (\ref{nlaaa})  
$$
\det a = \prod\limits_{k\in\Lambda} (\TT_kJ_k) + \sqrt\eps\Delta_1, 
$$
where $\MO \sup\limits_{0\leq\tau\leq T} |\Delta_1| \leq C $ by Corollary \ref{nlcor:est}. The constant $C$ does not depend on $N$ because the dimension $M$ does not depend on it. 
Then
$$
\PR(\Omega_\nu) \leq \PR \big(\prod\limits_{k\in\Lambda} (\TT_kJ_k) < \nu+\sqrt\eps|\Delta_1| \big)\leq 
\sum\limits_{k\in\Lambda}\PR\big(J_k< \TT_k^{-1}(\nu+\sqrt\eps|\Delta_1|)^{1/M}\big).
$$
Thus, to establish (\ref{nlommegasmall}), it is sufficient to show that 
\begin{equation}
\label{nlJ<delta}
\int\limits_0^T \PR \big( \sqrt{J_j(\tau)} < \delta\big) \, d\tau \rightarrow 0 \mbox{ when } \delta \rightarrow 0 \mbox{ uniformly in }N \mbox{ and $\eps$ sufficiently small.}
\end{equation} 

{\bf Step 3.} To prove the last convergence we use the concept of the local time.
Let $h\in C^2(\mR)$ and its  second derivative has at most polynomial growth at the infinity.
We consider the process $h_\tau:=h(J_j(\tau))$. Then, by the Ito formula, 
$$d h_\tau = A^h \,d\tau +  \sigma^h  d\hat\beta, $$
where 
$$A^h = h'(J_j) A^\JJ_j+h''(J_j) a_{jj} = h'(J_j)(A_j\cdot v_j +d_{jj}^1) + h''(J_j)a_{jj},$$
and the $1\times 2N$-matrix $\sigma^h(\tau)=(\sigma^h_k(\tau))$ is out of the interest.

Due to Theorem  \ref{nllem:prest}.3 and (\ref{nlaaa}), for sufficiently small $\eps$ we have
\begin{equation}
\label{nlnonzero}
d_{jj}^1 \geq \frac 78 \TT_j, \quad |a_{jj}|\leq \frac{3J_j}{2} \TT_j. 
\end{equation}
Let $\Theta_\tau(b,\omega)$ be the local time for the process $h_\tau$.
Then for any Borel set $\GG\subset\mR$ we have
$$
\int\limits_0^T \mI_{\GG}(h_\tau) \sum\limits_{k} |\sigma_k^h|^2 \, d\tau 
= 2\int\limits_{-\infty}^{\infty} \mI_{\GG}(b)\Theta_T(b,\omega) \, db. 
$$ 
On the other hand, denoting $(h_\tau-b)_+:=\max(h_\tau-b,0)$, we have 
$$
(h_T-b)_+ = (h_0-b)_+ + \int\limits_0^T \mI_{(b,\infty)} (h_\tau)\sigma^h d\hat\beta
+ 
\int\limits_0^T \mI_{(b,\infty)}(h_\tau) A^h \, d\tau + \Theta_T(b,\omega). 
$$
Consequently,
$$
\MO \int\limits_0^T \mI_{\GG }(h_\tau) \sum\limits_{k} |\sigma_k^h|^2 \, d\tau 
=2\MO \int\limits_{-\infty}^{\infty}  \mI_{\GG}(b)  \Big((h_T-b)_+ - (h_0-b)_+ - 
\int\limits_0^T \mI_{(b,\infty)}(h_\tau) A^h \,d\tau\Big) \,  db.
$$
The left-hand side is non negative, so 
\begin{equation}
\label{nlloctime}
\MO \int\limits_{-\infty}^{\infty}  \mI_{\GG}(b)\int\limits_0^T \mI_{(b,\infty)}(h_\tau) A^h \, d\tau db
\leq \MO \int\limits_{-\infty}^{\infty}  \mI_{\GG}(b)  \big((h_T-b)_+ - (h_0-b)_+ \big) \,  db.
\end{equation}
Let us apply relation (\ref{nlloctime}) with $\GG=(\xi_1,\xi_2)$, $\xi_2>\xi_1>0$ and  a function $h(x)\in C^2(\mR)$ that coincides with $\sqrt x$ for $x\geq \xi_1$ and vanishes for $x\leq 0$. Due to  Corollary \ref{nlcor:est}, the right-hand side of (\ref{nlloctime}) is bounded by $(\xi_2-\xi_1)C$. Then
\begin{equation}
\label{nlda4a2}
\MO \int\limits_{\xi_1}^{\xi_2}  \int\limits_0^T \mI_{(b,\infty)}(\sqrt{J_j}) \left( \frac{A_j\cdot v_j + d_{jj}^1}{2\sqrt{J_j}} - \frac{a_{jj}}{4\sqrt{J_j^3}}\right) \, d\tau db
\leq (\xi_2-\xi_1)C.
\end{equation}
In view of estimate (\ref{nlest'}) we have
$$\ds{\MO \int\limits_{\xi_1}^{\xi_2} \int\limits_0^T \frac{|A_j\cdot v_j| }{2\sqrt{J_j}} \, d\tau db \leq  (\xi_2-\xi_1)C }.$$
Moving this term to the right-hand side of (\ref{nlda4a2}),  applying (\ref{nlnonzero}) and sending $\xi_1$ to $0^+$, we get
$$
\MO \int\limits_{0}^{\xi_2}  \int\limits_0^T \mI_{(b,\infty)}(\sqrt{J_j}) 
J_j^{-1/2} \, d\tau db \leq C\xi_2.
$$
Note that
\begin{eqnarray}
\nonumber
\MO \int\limits_{0}^{\xi_2}  \int\limits_0^T \mI_{(b,\infty)}(\sqrt{J_j}) 
J_j^{-1/2} \, d\tau db &\geq& 
\frac{1}{\delta} \MO \int\limits_{0}^{\xi_2}  \int\limits_0^T \mI_{(b,\delta)}(\sqrt{J_j}) \, d\tau db 
 \\
 \nonumber
  &=& 
 \frac{1}{\delta} \int\limits_{0}^{\xi_2} \int\limits_0^T \PR(b<\sqrt{J_j}<\delta)\, d\tau db.
\end{eqnarray}
Consequently,
$$
\frac{1}{\xi_2} \int\limits_{0}^{\xi_2}  \int\limits_0^T \PR(b<\sqrt{J_j}<\delta)\, d\tau db \leq C\delta.
$$
Tending $\xi_2\rightarrow 0^+$ we obtain that
$$
\int\limits_0^T \PR \big( 0<\sqrt{J_j} < \delta\big) \, d\tau \rightarrow 0 \mbox{ when } \delta \rightarrow 0 \mbox{ uniformly in }N \mbox{ and $\eps$ sufficiently small}.
$$
{\bf Step 4.} To establish (\ref{nlJ<delta}) it remains to show that 
\begin{equation}
\label{nlzzzzero}
\int\limits_0^T \PR \big( |v_j(\tau)|=0) \, d\tau = 0 \mbox { for all $N$, $j\in\CC$  and $\eps$ sufficiently small. } 
\end{equation}
Writing a $j$-th component of equation (\ref{nlvs}) in the real coordinates $v^x_j:=\Ree v_j$ and $ v^y_j:=\Imm v_j$, we obtain the following two-dimensional system:
\bee\label{nlreal}
dv_j^x=\Ree \wid A_j\,d\tau+\Ree(Wd B)_j,\quad
dv_j^y=\Imm \wid A_j\,d\tau+\Imm(Wd B)_j,
\eee
where $\wid A_j:= \eps^{-1} i\nabla_j H_0+ i\nabla_j H_2+ g_j+ \sqrt\eps r_j.$ 
By the direct computation we get that the diffusion matrix for (\ref{nlreal}) with respect to the real Brownian motion $(\Ree \beta_k, \Imm\beta_k)_{k\in\CC}$ is
$$a^j:=\begin{pmatrix}
d_{jj}^1+\Ree d_{jj}^2 & \Imm d_{jj}^2 \\
 \Imm d_{jj}^2 & d_{jj}^1-\Ree d_{jj}^2
\end{pmatrix}.$$ 
Theorem \ref{nllem:prest}.3 implies that for $\eps$ sufficiently small, $\det a^j(\tau)$ is separated from zero uniformly in $\tau$. 
For $R>0$ define a stopping time
$$
\wid\tau_R=\inf\{\tau\geq 0: |v_j(\tau)|\vee |\wid A_j(\tau)|\geq\eps^{-1}R\}.
$$
Then, similarly to (\ref{nlkryy}) and  (\ref{nlda4a3}), we have
\begin{align}
\nonumber
\MO \int\limits_0^T \mI_{[0,\delta)}(|v_j(\tau)|) \, d\tau &\leq  C e^{\eps^{-1}TR}\MO \int\limits_0^{\wid\tau_R \wedge T} e^{-\int\limits_0^\tau |\wid A_j(s)|\, ds} \mI_{[0,\delta)}(|v_j(\tau)|)   (\det a^j(\tau))^{1/2} \, d\tau \\
\label{nlreallll}
&+ T\PR(\wid\tau_R<T) \leq C(R,\eps^{-1})\sqrt\delta
+  T\PR(\wid\tau_R<T).
\end{align}
Letting first $\delta\ra 0$ and then $R\ra\infty$ while $\eps$ is fixed, we arrive at (\ref{nlzzzzero}).
\qed
\subsection{Proof of Lemma \ref{nllem:lem}}
\label{nlsec:lem}

For the purposes of the proof we first introduce some notations.
For events $\Gamma_1, \Gamma_2$ and a random variable $\xi$ we denote 
$$
\mbox{\bf E}_{\Gamma_1}\,\xi:=\MO(\xi\mI_{\ov\Gamma_1}) \mbox{ and } \PR_{\Gamma_1}(\Gamma_2):= \PR(\Gamma_2\cap\ov\Gamma_1).
$$
Let us emphasize that in these definitions we consider an expectation and a probability on the complement of $\Gamma_1$. 
By $\kappa(r),\kappa_1(r),\ldots$ we denote various functions of $r$ such that $\kappa(r)\rightarrow 0$ as $r\rightarrow\infty$. By $\kappa_\infty(r)$ we denote functions $\kappa(r)$ such that $\kappa(r)=o(r^{-m})$ for each $m>0$. We write $\kappa(r)=\kappa(r;b)$ to indicate that $\kappa(r)$ depends on a parameter $b$. Functions $\kappa_\infty(r),\kappa(r),\kappa(r;b),\ldots$ never depend on $N$ and may depend on $\eps$ only through $r$, and we do not indicate their  dependence on the dimension $d$, power $p$ and time $T$. Moreover, they can change from formula to formula.
\ssk

{\bf Step 1.} For the brevity of notation we skip the index $\eps$.
Denote by $\wid \La$ the neighbourhood of radius $1$ of $\La$: 
\begin{equation}
\label{nlwidla}
\wid\La:=\{n\in\CC\big| \, \mbox{there exists } k\in\La \mbox{ satisfying } |n-k|\leq 1\}. 
\end{equation}  
Fix $R>0$. Set
\bee
\label{nl4etirep}
\Omega_R = \{ \max\limits_{k\in\wid\La}\sup\limits_{0\leq \tau \leq	T} |J_k(\tau)|\vee |A^\psi_k(\tau)| \geq R\}.
\eee
Due to Corollary \nolinebreak\ref{nlcor:est} and estimate (\ref{nlest'}),
 \begin{equation}
\label{nlodin}
\PR(\Omega_R)\leq \kappa_\infty(R).
\end{equation}
The polynomial growth of the function $P$ implies 

$$
\MO_{\ov\Om_R} \sup\limits_{\tau\in [0,T]} \left| \int\limits_0^\tau P(J(s),\psi(s))\,ds \right| \leq \kappa_\infty(R),
$$
and the function $\lan P \ran (J(s))$ satisfies a similar relation. 
Thus it is sufficient to show that for any $R\geq 0$
$$
\UU:= \MOOM \sup\limits_{\tau\in [0,T]} \left| \int\limits_0^\tau P(J(s), \psi(s)) - \lan P \ran(J(s)) \, ds \right|\ra 0 \quad\mbox{as}\quad\eps\ra 0\quad\mbox{uniformly in }N.
$$
For this purpose we consider a partition of the interval $[0,T]$ to subintervals of length $ \nu$ by the points 
$$
\tau_l=\tau_0 +  l \nu, \quad 0\leq l \leq L,\quad L=[T/\nu]-1,
$$ 
where the (deterministic) initial point $\tau_0 \in [0,\nu)$ will be chosen later.
Choose the diameter of the partition as
$$\nu=\eps^{7/8}.$$
Denote 
$$
\ds{\eta_l=\int\limits_{\tau_l}^{\tau_{l+1}} P(J(s), \psi(s)) - \lan P \ran(J(s))  \, ds.}
$$
Then
$$
\ds{\UU\leq  \MOOM \sum\limits_{l=0}^{L-1} |\eta_l| + \nu C(R).}
$$
\bee\label{nlYYY}
\mbox{Denote $Y(J)=(Y_{k}(J))_{k\in\CC}:=\big(\chp_{J_k} H_0(J)\big)_{k\in\CC}\in\mR^N$ and $Y(\tau):=Y(J(\tau))$.}
\eee
We have
\begin{eqnarray}
\nonumber
|\eta_l| &\leq& \left|\int\limits_{\tau_l}^{\tau_{l+1}} P\big(  J(s), \psi(s) \big) - P \big(J(\tau_l), \psi(\tau_l)+ \eps^{-1}Y(\tau_l)(s-\tau_l) \big) \, ds \right|  \\
\nonumber
 &+& \left|\int\limits_{\tau_l}^{\tau_{l+1}} P \big(J(\tau_l), \psi(\tau_l)+ \eps^{-1}Y(\tau_l)(s-\tau_l) \big)  - \lan P \ran\big(J(\tau_l)\big) \, ds \right|  \\
\label{nlygreki}
 &+& \left|\int\limits_{\tau_l}^{\tau_{l+1}} \lan P \ran\big(J(\tau_l)\big) - \lan P \ran\big(J(s)\big) \, ds \right| =: \YY_l^1 + \YY_l^2 + \YY_l^3. 
\end{eqnarray}

{\bf Step 2.} In the next proposition we will introduce "bad" events, outside of which actions are separated from zero, change slowly, and the rotation frequencies $Y(J(\tau_l))$ are not resonant. We will choose the initial point $\tau_0$ in such a way that probabilities  of these events will be small, and it will be sufficient to estimate $\YY^1_l,\YY^2_l,\YY^3_l$ only outside these events. Recall that $\wid\La$ is defined in (\ref{nlwidla}).
\bpp \label{nlbe}
There exist events $\FF_l,$ $0\leq l\leq L-1$, such that outside $\FF_l\cup \Om_R$ 
\begin{align}\nonumber
&i)\;\forall k\in \La\;  \sup\limits_{\tau_l \leq \tau \leq \tau_{l+1}} J_k(\tau)\geq \frac12 \eps^{1/24}, \quad 
ii)\;\forall k\in\wid\La\; \sup\limits_{\tau_l \leq \tau \leq \tau_{l+1}} |J_k(\tau) - J_k(\tau_l)| \leq \nu^{1/3}, \\ \nonumber
&iii)\; \Big|\frac{1}{\eps^{-1}\nu}\int_0^{\eps^{-1}\nu} P\big(J(\tau_l),\psi(\tau_l) +Y(\tau_l)s\big)\, ds - \lan P\ran\big(J(\tau_l)\big)\Big| \leq \kappa(\eps^{-1};R),
\end{align}
where the function $\kappa$ is independent from $0\leq l\leq L-1$.
There exists $\tau_0$ such that
\bee
\label{nlsdal4}
L^{-1} \sum\limits_{l=0}^{L-1} \PR_{\Omega_R}(\FF_l) \leq \kappa(\eps^{-1};R).
\eee
\epp
Before proving this proposition we will finish the proof of the lemma. 
Outside $\Omega_R$ we have $\YY_l^i\leq \nu C(R)\leq C_1(R)/L$. Fix $\tau_0$ as in Proposition \ref{nlbe}. Then from (\ref{nlsdal4})  we obtain 
$$
\sum\limits_{l=0}^{L-1}(\MOOM-\MOF) \YY_l^i \leq \frac{C(R)}{L} \sum\limits_{l=0}^{L-1} \PR_{\Omega_R} (\FF_l)\leq C(R)\kappa(\eps^{-1};R) 
= \kappa_1(\eps^{-1}; R), \quad i=1,2,3.
$$  
Thus, it is sufficient to show that for any $R\geq 0$ we have
$$\ds{\sum\limits_{l=0}^{L-1}\MOF (\YY_l^1+\YY_l^2+\YY_l^3)}\ra 0 \quad\mbox{as}\quad \eps\ra 0\quad\mbox{uniformly in } N.$$
\ssk

{\bf Step 3.} Now we will estimate each term $\YY_l^i$ outside the "bad" event  $\FF_l\cup\Omega_R$.

{\it Terms $\YY_l^1.$}
We will need the following 
\begin{prop} 
\label{nlprop:event}
For every $k\in\Lambda$ and each $0 \leq l\leq L-1$, we have
\begin{equation}
\label{nlevent}
\PR_{\FF_l\cup\Omega_R} \Big( \sup\limits_{\tau_l \leq \tau \leq \tau_{l+1}} 
|\psi_k(\tau)- \big( \psi_k(\tau_l) + \eps^{-1} Y_k(\tau_l)(\tau-\tau_l)\big)| \geq \eps^{1/24} \Big) \leq \kappa_\infty(\eps^{-1}),
\end{equation}
where the function $\kappa_\infty$ is independent from $k,l$. 
\end{prop}
{\it Proof.}
Let us denote the event in the left-hand side of (\ref{nlevent}) by $\Gamma$.
According to (\ref{nlang'}),
\begin{eqnarray}
\nonumber
\PR_{\FF_l\cup\Omega_R}(\Gamma) &\leq&  
\PR_{\FF_l\cup\Omega_R} \left(  \eps^{-1}\sup\limits_{\tau_l \leq \tau \leq \tau_{l+1}} \left| \int\limits_{\tau_l}^{\tau} Y_k(s)- Y_k(\tau_l)  \, ds \right| \geq \frac13\eps^{1/24} \right) \\
\nonumber
 &+&  \PR_{\FF_l\cup\Omega_R}\left(\sup\limits_{\tau_l \leq \tau \leq \tau_{l+1}} \left| \int\limits_{\tau_l}^\tau  \frac{A^\psi_k}{|v_k|^2} \, ds \right| \geq \frac13\eps^{1/24} \right) \\
 \nonumber
 &+& \PR_{\FF_l\cup\Omega_R}\left( \sup\limits_{\tau_l \leq \tau \leq \tau_{l+1}} \left| \int\limits_{\tau_l}^\tau  \frac{ iv_k} {|v_k|^2} \cdot (WdB)_k  \right| \geq \frac13\eps^{1/24} \right) \\
 \nonumber
  &=:& \PR_{\FF_l\cup\Omega_R}(\Gamma_1)+\PR_{\FF_l\cup\Omega_R}(\Gamma_2)+\PR_{\FF_l\cup\Omega_R}(\Gamma_3).
\end{eqnarray}
$\Gamma_1:$ Due to (\ref{nlH000}),  $Y_k(J) \in\LL_{loc}(\mR^N)$. Since it depends on $J$ only through $J_n$ with $n$ satisfying $|n-k|\leq 1$, we get  
\begin{eqnarray}
\nonumber
\PR_{\FF_l\cup\Omega_R}(\Gamma_1)\leq \PR_{\FF_l\cup\Omega_R}\big( \max\limits_{n:|n-k| \leq 1} \sup\limits_{\tau_l \leq \tau \leq \tau_{l+1}} |J_n(\tau)- J_n(\tau_l)| \geq C(R) \eps^{1+1/24}\nu^{-1} \big).
\end{eqnarray}
 If $\eps$ is sufficiently small, we have $C(R)\eps^{1+1/24}\nu^{-1}>\nu^{1/3}$ (recall that $\nu=\eps^{7/8}$). Then, due to Proposition \ref{nlbe}.ii, we get 
$$
\PR_{\FF_l\cup\Omega_R}(\Gamma_1) = 0 \quad \mbox{for} \quad \eps\ll 1.
$$
$\Gamma_2:$ Proposition \ref{nlbe}.i implies
\begin{eqnarray}
\nonumber
\PR_{\FF_l\cup\Omega_R}(\Gamma_2) \leq \PR_{\FF_l\cup\Omega_R} \big( \sup\limits_{\tau_l \leq \tau \leq \tau_{l+1}} |A^\psi_k| \geq 
\frac13\eps^{1/24+ 1/24}\nu^{-1} \big)=0 \quad\mbox{for}\quad \eps\ll 1, 
\end{eqnarray}
 since outside  $\Om_R$ we have $|A^\psi_k|\leq R$, in view of (\ref{nl4etirep}).
\ssk

$\Gamma_3:$ In view of (\ref{nlvariation}), the Burkholder-Davis-Gandy inequality jointly with Theorem \nolinebreak\ref{nllem:prest}.3, and Proposition \ref{nlbe}.i imply that 
\begin{eqnarray}
\nonumber
\MOF  \sup\limits_{ \tau_l \leq \tau \leq \tau_{l+1}} 
\left| \int\limits_{\tau_l}^{\tau} \frac{ iv_k} {|v_k|^2} \cdot ( W_k dB) \right|^{2m}\leq C(m) \MOF \left(\int\limits_{\tau_l}^{\tau_{l+1}} \frac{1}{|v_k|^2}\, ds  \right)^m \leq C(m)\nu^m\eps^{-m /24},
\end{eqnarray}
for any $m>0$.
From Chebyshev's inequality it  follows that
$$
\PR_{\FF_l\cup\Omega_R}(\Gamma_3) \leq C(m) \nu^m\eps^{-m(1/24+ 2/24)} \quad\mbox{for any } m>0.$$
Thus, $\PR_{\FF_l\cup\Omega_R}(\Gamma_3) = \kappa_\infty(\eps^{-1}).$
\qed
\ssk

Estimates {\it (i)} and {\it (ii)} of Proposition \ref{nlbe} imply that outside $\FF_l\cup \Om_R$, for any  $k\in\La$
\begin{equation}
\label{nlsmallact}
\sup\limits_{\tau_l\leq \tau\leq\tau_{l+1}}\big||v_k(\tau)|-|v_k(\tau_l)|\big|\leq\frac{\sqrt 2 |J_k(\tau)-J_k(\tau_l)|}{\sqrt{J_k(\tau)}+\sqrt{J_k(\tau_l)}}\leq \nu^{1/3}\eps^{-1/48}=\eps^{13/48}.
\end{equation}
Since $P\in\LL_{loc}(\mC^N)$, then Proposition \ref{nlprop:event} and  (\ref{nlsmallact}) imply that 
$$\PR_{\FF_l\cup\Omega_R}\big(\YY_l^1\geq \nu C(R)(\eps^{1/24}+\eps^{13/48})\big)\leq \kappa_\infty(\eps^{-1}).$$
Then we get
$$
\MOF\YY_l^1\leq \nu C(R)(\eps^{1/24}+\eps^{13/48}) +\nu C(R)\kappa_\infty(\eps^{-1})= \nu\kappa(\eps^{-1};R).
$$

{\it Terms $\YY_l^2$.}
Put $\hat s:=\eps^{-1}(s-\tau_l)$. Then
Proposition \ref{nlbe}.iii implies that  outside $\FF_l\cup\Omega_R $ 
\begin{equation*}
\YY_l^2=\nu\left|\frac{1}{\eps^{-1}\nu}\int\limits_{0}^{\eps^{-1}\nu} P\big(J(\tau_l), \psi(\tau_l)+Y(\tau_l)\hat s \big) \, d\hat s  - \lan P \ran\big( J(\tau_l)\big)\right|\leq \nu \kappa(\eps^{-1};R).
\end{equation*}

{\it Terms $\YY_l^3$.} Proposition \ref{nlprop:aver}.i jointly with (\ref{nlsmallact}) implies that outside $\FF_l\cup\Omega_R$ we have
$$
\YY_l^3 \leq \nu C(R) \eps^{13/48}.
$$

{\bf Step 4.} Summing by $l$, taking the expectation and noting that $L\nu\leq T$, we get 
\begin{equation}
\nonumber
\sum\limits_{l=0}^{L-1}\MOF (\YY_l^1+\YY_l^2+\YY_l^3) \leq L(\nu\kappa(\eps^{-1}; R)+   \nu C(R)\eps^{13/48}) \rightarrow 0 \mbox{ as } \eps\rightarrow 0,
\end{equation}
 uniformly in $N$. The proof of the lemma is complete.
\qed

{\it Proof of Proposition \ref{nlbe}.} 
We will construct the set $\FF_l$ as a union of three parts. The first two are $\EE_l:=\cup_{k\in\La}\EE_l^k$ and $Q_l:=\cup_{k\in\wid\La} Q_l^k$, where
\begin{equation}
\label{nlsets}
\EE^k_l:=\{ J_k (\tau_l) \leq \eps^{1/24} \}, 
\quad Q^k_l:=\{ \sup\limits_{\tau_l \leq \tau \leq \tau_{l+1}} |J_k(\tau) - J_k(\tau_l)| \geq \nu^{1/3} \}.
\end{equation}
Outside $Q_l$ we have {\it (ii)} and, if $\eps$ is small, outside $\EE_l\cup Q_l$ we get  {\it (i)}: for aevery $k\in\La$
\bee\nonumber
\sup\limits_{\tau_l \leq \tau \leq \tau_{l+1}} J_k(\tau)\geq \eps^{1/24} - \nu^{1/3}\geq \frac12\eps^{1/24}, \quad\mbox{if }\eps\ll 1.
\eee

Now we will construct the event $\Omega_l^{\eps,R}$, which will form the third part of $\FF_l$. Let us accept the following notation: 
$$\mbox{for a vector $Z=(Z_j)_{j\in\CC}\in\mR^N$ we denote $Z^\La:=(Z_j)_{j\in\La}\in\mR^M$}.$$

For any fixed $J\in \mR^N_{+0}$ the function  $P(J,\psi)$ is Lipschitz-continuous in angles $\psi\in\mT^N$. From \cite{Tm} it follows that the Fourier series of a Lipschitz-continuous function of $\psi\in\mT^N$ converges uniformly in $\psi$. Then, using standard method (e.g., see in \cite{MS}), we obtain that for every $\delta>0$ and $R'>0$   there exists a Borel set  $E^{\delta,R'}\subset\{ x=(x_k)_{k\in\La}\in\mR^M: \|x\|_{\mR^M}\leq R'\}$ with the Lebesgue measure $|E^{\delta,R'}|\leq\delta$, such that  
for any $Z=(Z_k)_{k\in\CC}\in\mR^N$ satisfying $Z^\La\notin E^{\delta,R'}$ and $ \|Z^\La\|_{\mR^M}\leq R',$ we have
\begin{equation}
\label{nlav}
\left|\frac{1}{t}\int_0^t P(J,  \psi+  Z s) \, ds - \lan P\ran(J) \right| \leq \kappa(t;J,\del, R'),
\end{equation} 
for all $J\in\mR^N_{+0}$ and $\psi\in\mT^N$. Moreover, since $P\in\LL_{loc}(\mC^N)$, then we can choose the function $\kappa$ to be independent from $J$ for $J\in B_R^\La$, where 
$$B_R^\La:=\{J=(J_k)_{k\in\CC}\in\mR^N_{0+}:\,\max\limits_{k\in\La} J_k\leq R\},$$
i.e. $\kappa=\kappa(t;R,\del,R').$ 
The rate of convergence in (\ref{nlav}) depends on $\del$. Choose a function
$\del=\del(\eps)$, such that $ \del(\eps)\ra 0$  as $\eps\ra 0 $ so slow that
\begin{equation}
\label{nlav'}
\left|\frac{1}{\eps^{-1}\nu}\int_0^{\eps^{-1}\nu} P(J, \psi+  Z s) \, ds - \lan P\ran(J) \right| \leq \kappa(\eps^{-1};R, R')
\end{equation} 
for all  $ J\in B^\La_R,\,\psi\in\mT^N$ and $Z$ as above.
\ssk

Let us choose $R'=R'(R)=\sup\limits_{\ov \Om_R} \sup\limits_{0 \leq \tau \leq T}\|Y^\La(\tau)\|_{\mR^M}$.  Let 
$$
\Omega_l^{\eps, R}:=\{ Y^\La(\tau_l)\in E^{\delta(\eps),R'(R)} \}.
$$
Then outside $\Omega_l^{\eps, R}\cup\Omega_R$ we get $Y^\La(\tau_l)\notin  E^{\delta(\eps),R'(R)}$ and $\|Y^\La(\tau_l)\|_{\mR^M}\leq R'(R).$ Since outside $\Om_R$ we have $J(\tau_l)\in B^\La_R$, then, due to (\ref{nlav'}), outside $\Omega_l^{\eps,R}\cup\Omega_R$ we get {\it (iii)}.

Let $\FF_l:=\EE_l\cup Q_l \cup \Omega_l^{\eps,R}$. 
Then outside $\FF_l\cup\Om_R$ items {\it (i),(ii)} and {\it (iii)} hold true.  
\ssk

Now we will estimate the probabilities of $\EE_l,Q_l$ and $\Omega_l^{\eps,R}$. 
\begin{prop}
\label{nlQ_l} {\it (i)} We have
$\PR(Q_l)\leq \kappa_\infty(\nu^{-1})$, where $\kappa_\infty$ is independent from $l$.

{\it (ii)} There exists an initial point $\tau_0\in [0,\nu)$ such that 
$$
L^{-1} \sum\limits_{l=0}^{L-1} \PR_{\Omega_R}(\EE_l\cup\Omega_l^{\eps,R}) = \kappa(\eps^{-1}; R).
$$
\end{prop}
Propositions \ref{nlQ_l} implies  (\ref{nlsdal4}):
\bee
\nonumber
L^{-1} \sum\limits_{l=0}^{L-1} \PR_{\Omega_R}(\FF_l) \leq \kappa(\eps^{-1};R) + \kappa_\infty(\nu^{-1})
=\kappa_1(\eps^{-1};R).
\eee
\qed

{\it Proof of Proposition \ref{nlQ_l}}.
{\it (i)} Let us take $\rho>\sqrt{\nu}$. Then, due to (\ref{nlac'}), for any $k\in\wid\La$ 
\begin{eqnarray}
\nonumber
\PR \big(\sup\limits_{\tau_l \leq \tau \leq \tau_{l+1}} |J_k(\tau) - J_k(\tau_l)| \geq \rho \big) &\leq&
 \PR \left( \sup\limits_{\tau_l \leq \tau \leq \tau_{l+1}} 
\left| \int\limits_{\tau_l}^{\tau} A^J_k\, ds \right| \geq \rho/2\right) \\
\nonumber
&+&
\PR \left( \sup\limits_{\tau_l \leq \tau \leq \tau_{l+1}} 
\left| \int\limits_{\tau_l}^{\tau} v_k \cdot (WdB)_k \right| \geq \rho/2\right)\\ 
\nonumber
&=:& \PR(\Gamma_1) + \PR(\Gamma_2).
\end{eqnarray}
Due to estimate (\ref{nlest'}), we have
\begin{eqnarray}
\nonumber
\PR (\Gamma_1) \leq \PR (\nu \sup\limits_{\tau_l \leq \tau \leq \tau_{l+1}}  | A_k^J| \geq \rho/2) 
\leq \kappa_\infty(\nu^{-1}).
\end{eqnarray}
In view of (\ref{nlvariation}),  the Burkholder-Davis-Gundy inequality jointly with (\ref{nlest'}) implies 
\begin{eqnarray}
\label{nlBDG}
\MO \sup\limits_{\tau_l \leq \tau \leq \tau_{l+1}} \left| \int\limits_{\tau_l}^{\tau}  v_k  \cdot (WdB)_k \right|^{2m} 
 \leq C(m) \MO \left(\int\limits_{\tau_l}^{\tau_{l+1}} S_{kk}^J \, ds \right)^m 
\leq C_1(m)\nu^m,    
\end{eqnarray}
for every $m >0$.
Consequently, $\PR(\Gamma_2) \leq C(m)\nu^m \rho^{-2m}$. Choosing $\rho=\nu^{1/3}$ we get $\PR(\Gamma_2)\leq \kappa_\infty(\nu^{-1})$. It remains to sum up the probabilities by $k\in\wid\La$.
\ssk

{\it (ii)} Denote $\AAA(\tau):=(\EE\cup\Omega^{\eps,R})(\tau)$, where the last set is defined similarly to 
$\EE_l\cup\Omega_l^{\eps,R}$ but at the moment of time $\tau$ instead of $\tau_l$.   
Recall that $Y^\La(J)$ depends on $J$ only through $J^{\wid\La}:=(J_k)_{k\in\wid\La}$.  Denote by $\wid M$ the number of nodes in $\wid \La$ and let
$$ E^{\eps,R}_J:=\big\{J^{\wid\La} \in \mR^{\wid M}_{+0} : Y^\La (J) \in E^{\delta(\eps),R'(R)} \mbox{ and }J_k\leq R\quad \forall k\in\wid\La \big\}.$$
In view of assumption {\it HF} which states that the functions $f'_j$ have only isolated zeros, it is not difficult to show that  the convergence $|E^{\delta(\eps),R'(R)}|\ra 0$ as $\eps\ra 0$ implies that  $|E^{\eps,R}_J|\rightarrow 0$ as $\eps\rightarrow 0$. 
Note that $\ov\Omega_R\cap\Omega^{\eps, R}(\tau)\subset \{J^{\wid\La}(\tau)\in E^{\eps,R}_J\}$.  Then Lemma \ref{nllem:smallact} implies
$$
\int\limits_0^T \PR_{\Omega_R}\big(\AAA(\tau)\big) \, d\tau\leq 
\int\limits_0^T \PR_{\Omega_R}\big(\EE(\tau)\big) \, d\tau + \int\limits_0^T \PR\big(J^{\wid\La}(\tau)\in E^{\eps,R}_J\big) \, d\tau
  \rightarrow 0 \quad \mbox{as }  \eps \rightarrow 0,
$$
uniformly in $N$. It remains to note that there exists a deterministic point $\tau_0\in[0,\infty)$ such that
\begin{align}
\nonumber
\int\limits_0^T \PR_{\Omega_R}\big(\AAA(\tau)\big) \, d\tau &\geq \int\limits_0^\nu \sum\limits_{l=0}^{L-1}\PR_{\Omega_R}\big(\AAA(l\nu+s)\big)\,ds \\
\nonumber 
&\geq \nu\sum\limits_{l=0}^{L-1}\PR_{\Omega_R}\big(\AAA(l\nu+\tau_0)\big)
\geq T(L+1)^{-1} \sum\limits_{l=0}^{L-1} \PR_{\Omega_R}\big(\AAA(\tau_l)\big).
\end{align}
\qed

\section{The change of variables: proof of Theorem \ref{nllem:prest}}
\label{nlsec:change}
In this section we accept the following notations. Let $a=(a_s)\in\mC^M$ be a vector-function $a=a(b)$, where $b=(b_r)\in\mC^K$, and $s,r$ be multi-indices. By $\ds{\frac{\chp a}{\chp b}}$ we denote the $M\times K$-matrix with entires  
$
\ds{\Big(\frac{\chp a}{\chp b}\Big)_{jn}:=\frac{\chp a_j}{\chp b_n}}$. By $\ds{\frac{\chp a}{\chp b}^T}$ we denote the transposed matrix,  $\ds{\Big(\frac{\chp a}{\chp b}^T\Big)_{jn}=\frac{\chp a_n}{\chp b_j}}$.  
By $\Id_M$ we denote the identity matrix of the size $M\times M$. 
\subsection{The Hamiltonian and the equation}
\label{nlsec:change_H}
Here we find the canonical transformation $v\mapsto u$, $(J,\psi)\mapsto (I,\ph)$, and calculate Hamiltonian (\ref{nlham0}) and equation (\ref{nlini_e}) in the new variables $v$.
We find the transformation as the time-1-map $\Gamma$  of the Hamiltonian flow $X^s_{\sqrt{\eps}{\Phi}}$ given by the Hamiltonian $\sqrt{\eps}{\Phi}$.  
Denote 
$$\wid F(J):=\frac12\sum\limits_{j\in\CC} F_j (|v_j|^2) \quad \mbox{and}\quad \wid G(J,\psi):= \frac14\sum\limits_{|j-n|=1} G(|v_j-v_{n}|^2),$$ 
where $v=v(J,\psi)$. By the Taylor expansion we have
\begin{eqnarray}
\label{nlham}
\HH(J,\psi) &=& H\circ \Gamma (J,\psi) = \\ 
\nonumber
&=& \wid F(J) + \sqrt{\eps} \Big( \wid{G}(J,\psi) + \big\{ \wid F, \Phi \big\}(J,\psi) \Big) + \\
\nonumber
&+& \eps \Big( \big\{\wid{G}, \Phi \big\}(J,\psi) + \int_0^1(1-s)\big\{ \big\{ H,\Phi \big\}, \Phi \big\}\circ X_{\sqrt{\eps}\Phi}^s \,ds \Big),
\end{eqnarray}
where 
$$
\{h_1,h_2 \}=2i\sum\limits_{j\in\CC}(\chp_{v_j}h_1 \chp_{\ov v_j}h_2 - \chp_{\ov v_j} h_1 \chp_{v_j}h_2)
=\sum\limits_{j\in\CC}(\chp_{\psi_j}h_1 \chp_{J_j}h_2 - \chp_{J_j} h_1 \chp_{\psi_j}h_2)
$$ 
denotes the Poisson bracket.
We wish to choose $\Phi$ in such a way that the homological equation 
\begin{equation}
\label{nlgomol}
\wid{G}(J,\psi) + \big\{ \wid F, \Phi \big\}(J,\psi) = h(J) 
\end{equation}
holds for some function $h$.
Let us denote 
\begin{equation}
\nonumber
\theta_{jn}:=\psi_j-\psi_{n} \quad \mbox{for} \quad j,n\in\CC.
\end{equation}
The potential $G$ admits the representation 
\begin{equation}
\label{nlG}
G(|v_{j}-v_{n}|^2)=G(|\sqrt{2J_{j}} e^{i{\theta_{jn}}} - \sqrt{2J_{n}}|^2)=:G(J_{j},J_n,\theta_{jn}).
\end{equation}
We seek the function $\Phi$ in the form
$\Phi=\sum\limits_{|j-n|=1}\Phi_{jn}(v_{j},v_n)=\sum\limits_{|j-n|=1}\Phi_{jn}(J_{j},J_n,\theta_{jn})$. Expanding $G$ and $\Phi$ in the Fourier series, we obtain
\begin{eqnarray}
\nonumber
G(J_{j},J_n,\theta_{jn})=\sum\limits_{k\in\mZ}G_k(J_{j},J_n)e^{ik\theta_{jn}}, \quad
 \Phi_{jn}(J_{j},J_n,\theta_{jn})=\sum\limits_{k\in\mZ} \Phi_{jnk}(J_{j},J_n)e^{ik\theta_{jn}}.
\end{eqnarray}
Now let us solve equation (\ref{nlgomol}).
Calculating the Poisson bracket we have
\begin{equation}
\nonumber
\big\{ \wid F, \Phi \big\}(J,\psi)=-\sum\limits_{j\in\CC} f_j (2J_j) \frac{\partial \Phi}{\partial \psi_j}=-
\sum\limits_{|j-n|=1}\chp_{\tht_{jn}}\Phi_{jn}(f_j-f_n).
\end{equation}
 Then the left-hand side of (\ref{nlgomol}) is equal to
\begin{equation}
\label{nlgomol1}
\sum\limits_{k\in\mZ}\sum\limits_{|j-n|=1} \Big( -ik\Phi_{jnk}(f_j-f_n) + \frac14 G_{k}(J_{j},J_n) \Big)e^{ik\theta_{jn}}.
\end{equation}
We choose $\Phi$ in such a way that each bracket in (\ref{nlgomol1}) except that with $k=0$ vanishes. That is
\begin{equation}
\label{nlPhik}
\Phi_{jnk}=\frac14 \frac{G_{k}(J_{j},J_n)}{ik(f_{j}-f_{n})}, \quad k\in\mZ/\{0\},\quad \Phi_{jn0}=0.
\end{equation}
From representation (\ref{nlG}) it follows that $G(J_j,J_n,\theta_{jn})=G(J_j,J_n,-\theta_{jn})$, so $G_{k}(J_j,J_n)=G_{-k}(J_j,J_n)$. Then (\ref{nlPhik}) implies
\begin{equation}
\label{nlPhij}
\Phi_{jn}=\frac14\sum\limits_{k\in\mZ/\{0\}} \frac{G_{k}(J_{j},J_n)e^{ik\theta_{jn}}}{ik(f_{j}-f_{n})}=\frac14\frac{\int\limits_0^{\theta_{jn}}G^0(J_j,J_n,\theta)\,d\theta}{f_{j}-f_{n}}, \quad \Phi=\sum\limits_{|j-n|=1} \Phi_{jn},
\end{equation}
where 
\bee\label{nlG000}
G^0:=G-\lan G\ran.
\eee
Due to assumption {\it HF}, the denominator of (\ref{nlPhij}) is separated from zero.
Thus, choosing $\Phi$ as above and putting
\begin{equation}
\label{nlhh}
h=\lan\wid G \ran,
\end{equation} 
we get that the homological equation (\ref{nlgomol}) holds. 
Taking the next order term of Taylor's expansion of Hamiltonian (\ref{nlham}) and applying (\ref{nlgomol}) joined with (\ref{nlhh}) to the terms of order $\sqrt\eps$ and $\eps$, we have
\begin{eqnarray}
\nonumber
\HH(J,\psi)&=&\wid F(J) + \sqrt\eps \lan \wid G \ran(J) + \frac{\eps}{2} \big\{  \lan \wid{G} \ran + \wid G,\Phi \big\} + \\
\nonumber
&+&\frac{\eps\sqrt \eps}{2} \Big( \big\{\wid{G}, \Phi \big\}_2 + \int_0^1(1-s)^2 \big\{ H,\Phi \big\}_3 \circ X_{\sqrt{\eps}\Phi}^s \,ds \Big) \\
\label{nlham1}
&=:& H_0(J) + \eps H_2(J,\psi) + \eps\sqrt \eps H_> (J,\psi),  
\end{eqnarray}  
where $\displaystyle{ \{h,\Phi\}_k:=\{\ldots{\{h,\Phi\},\Phi},\ldots,\Phi\}}$ denotes the Poisson bracket with $\Phi$  taken $k$ times. Thus, we arrive at (\ref{nlham1'}).

\begin{prop}
\label{nlprop:est}
The function $\Phi(v)$ is $C^3$-smooth. Let $a,b,c\in\{v,\ov v\}$. Then for every $k,l,m\in\CC$ satisfying the relation  
$| k-l | \vee| m-k |\vee |m-l| \leq 1$  we have 
\begin{equation}
\label{nlpropest}
\left| \frac{\chp\Phi}{\chp \psi_{k} }\right|, 
\left| \frac{\chp\Phi}{\chp a_{k} }\right|,  \left| \frac{\chp^2\Phi}{\chp a_{k}\chp b_{l}}\right|,
 \left| \frac{\chp^3\Phi}{\chp a_{k}\chp b_{l}\chp c_{m}}\right| \leq C. 
\end{equation} 
For other $k,l,m \in\CC$ the second and the third derivatives are equal to zero. 
\end{prop}
We will prove this proposition after the end of the proof of the theorem. 
Denote
$$v^s:=X_{\sqrt\eps \Phi}^s(v) \quad\mbox{and}\quad u^s:=X_{-\sqrt\eps \Phi}^s(u).$$
We have
\begin{equation}
\label{nlchange}
v_j^s=v_j +  \sqrt\eps \int\limits_0^s i\nabla_{j}\Phi(v^\tau) \, d\tau, \quad
u_j^s=u_j - \sqrt\eps \int\limits_0^s i\nabla_{j}\Phi(u^\tau) \, d\tau.
\end{equation}
In particular, since the transformation $v\mapsto u$ is a time-1-map of the Hamiltonian flow $X_{\sqrt\eps \Phi}^s$, we have $u=X_{\sqrt\eps \Phi}^1(v)=v^1,
\, v=X_{-\sqrt\eps \Phi}^1(u)=u^1$, so that
\begin{equation}
\label{nlchange1}
u_j=v_j +  \sqrt\eps \int\limits_0^1 i\nabla_{j}\Phi(v^\tau) \, d\tau, \quad
v_j=u_j - \sqrt\eps \int\limits_0^1 i\nabla_{j}\Phi(u^\tau) \, d\tau.
\end{equation}
In view of (\ref{nlchange1}) and Proposition \ref{nlprop:est}, we see that  the change of variables is $C^2$-smooth. This implies that the function $\HH(v)$ is also $C^2$-smooth. Jointly with Proposition \ref{nlprop:aver}, from Proposition \ref{nlprop:est} it follows that that the function $H_2(v)$ is $C^2$-smooth.  Moreover, Proposition \ref{nlprop:aver} implies that the function $H_0(v)$ is $C^4$-smooth. Then from (\ref{nlham1}) we conclude that the function $H_>(v)$ is $C^2$-smooth.
\ssk

Let us now rewrite equation (\ref{nlini_e}) in the $v$-variables. 
 Applying Ito's formula to the function $v(u(t))$, where $u(t)$ satisfies equation (\ref{nlini_e}), we get
\begin{equation}
\label{nldV}
\dot v = \frac{\chp v}{\chp u} \dot u + \frac{\chp v}{\chp \ov u}\dot{\ov u} +\eps\sum\limits_{k\in\CC} \TT_k\frac{\chp^2 v}{\chp u_k \chp \ov u_k},
\end{equation}
since the coefficients, corresponding to the second other derivatives vanish.
Denote $B= (\beta,\ov\beta)^T$, where $\beta=(\beta_j)_{j\in\CC}$ are, as usual, standard complex independent Brownian motions. Using the Hamiltonian representation (\ref{nlini}) of equation (\ref{nlini_e}), we rewrite (\ref{nldV}) in more details
\begin{equation}
\label{nlDV}
\dot v = i\nabla \HH(v) + \eps \frac{\chp v}{\chp u} g(u)+\eps \frac{\chp v}{\chp \ov u} \ov g(u)  + \eps\sum\limits_{k\in\CC} \TT_k \frac{\chp^2 v}{\chp u_k \chp \ov u_k} +  \sqrt{\eps} W \dot B,
\end{equation}
where the new dispersion matrix $W(v)$ is of the size $N \times 2 N$ and consists of two $N\times N$ blocks $W=(W^1,W^2)$, so that 
$W\dot B=W^1\dot\beta +W^2\dot{\ov\beta}$. The blocks have the form
\bee\label{nlWWWWW}\ds{ W^1:= \frac{\chp v}{\chp u}\mbox{diag}\,(\sqrt{\TT_j})}\quad\mbox{and}\quad
\ds{W^2:=\frac{\chp v}{\chp \ov u}\mbox{diag}\,(\sqrt{\TT_j})}.
\eee
Let us represent the dissipative  and Ito's terms from (\ref{nlDV}) as  a leading part and a remainder of higher order in $\eps$. In view of  (\ref{nlchange1}), we have $\ds{\frac{\chp v}{\chp u}-\Id_N,\frac{\chp v}{\chp \ov u}, \frac{\chp^2 v}{\chp u_k \chp \ov u_k}\volna \sqrt\eps.}$
Denote
\bee
\label{nlsmord}
r^D(v):=\eps^{-1/2}\Big(\big( g(u)- g(v)\big) + \Big(\frac{\chp v}{\chp u} - \Id_N\Big)g(u)+\frac{\chp v}{\chp \ov u} \ov g(u)\Big)\mbox{ and }
r^I(v):=\eps^{-1/2}\sum\limits_{k\in\CC}\TT_k\frac{\chp^2 v}{\chp u_k \chp \ov u_k},
\eee
where $u=u(v)$.
Since the transformation $u\mapsto v$ is $C^2$-smooth, the function $r^D$ is $C^1$-smooth while the function $r^I$ is continuous.
Let $$r=(r_j)_{j\in\CC}:=r^D + r^I + i\nabla H_>.$$ 
Substituting this relation in (\ref{nlDV}), we arrive at (\ref{nlv}).

\subsection{Some estimates}
\label{nlsec:change_est}

Here we prove auxiliary propositions essential to establish items {\it 1-4} of the theorem. Recall that the constant $\ga$ satisfies the estimate $1/2<\gamma<1$. Till the end of Section \ref{nlsec:change} we always indicate the dependence of constants on $\ga$ and do not indicate their dependence on the power $q$ (see below).

First let us establish the following two corollaries of Proposition \ref{nlprop:est}.
\begin{cor}
\label{nlprop:ekviv}
For any $q\geq 1$ there exists a constant $C(\gamma)$, such that for every $0 \leq s \leq 1$  we have
$$
\|v\|_{j,q} - \sqrt\eps C(\gamma)\leq \|v^s\|_{j,q} \leq \|v\|_{j,q} + \sqrt\eps C(\gamma).
$$
\end{cor}
{\it Proof.} 
This follows from  Proposition \ref{nlprop:est} by summing in $k\in\CC$ the increments $|v_k^s-v_k|,$ given by (\ref{nlchange}), raised to the power $q$, and multiplied by  coefficients $\gamma^{|j-k|}$.
\qed

Let us introduce the following notations.
By $\wid\CC:=\{1,2\}\times\CC$ we denote the set of multi-indices $\wid j=(j_0,j)$, where  $j_0\in \{1,2\}$ and $j=(j_1,\ldots,j_d)\in\CC$. By $U$ and $V$  we denote the vectors from $\mC^{2N}$
$$
 U=(U_{\wid j})_{\wid j\in \wid\CC}=(u, \ov u)^T\qquad\mbox{and}\quad V=(V_{\wid j})_{\wid j\in \wid\CC}=(v,\ov v)^T, 
$$
where $U_{(1,j)}=u_j$, $U_{(2,j)}=\ov u_j$ and  $V_{(1,j)}=v_j$, $V_{(2,j)}=\ov v_j$ . 

Introduce a $2N\times 2 N$-matrix $\ds{\frac{\chp U^s}{\chp U}}$ with entires $\ds{\Big(\frac{\chp U^s}{\chp U}\Big)_{\wid k \wid l}}$, where $\wid k,\wid l\in\wid\CC$. It will be convenient to write it in the form
\begin{equation}
\label{nl04}
\frac{\chp U^s}{\chp U}=
\begin{pmatrix}
\frac{\chp u^s}{\chp u} & \frac{\chp u^s}{\chp \ov u} \\
\frac{\chp \ov u^s}{\chp u} & \frac{\chp \ov u^s}{\chp \ov u}
\end{pmatrix}, 
\end{equation}
where
$\ds{\Big(\frac{\chp U^s}{\chp U}\Big)_{(1,k)(1,l)}=\Big(\frac{\chp u^s}{\chp u}\Big)_{kl}}$, 
$\ds{\Big(\frac{\chp U^s}{\chp U}\Big)_{(1,k)(2,l)}=\Big(\frac{\chp u^s}{\chp \ov u}\Big)_{kl}}$, etc. So 
the first indices $k_0,l_0\in\{1,2\}$ numerate the blocks of the matrix (\ref{nl04}).
Using (\ref{nlchange}) it is easy to check that $\ds{\frac{\chp U^s}{\chp U}}$ satisfies the equation
\begin{equation}
\label{nlDF}
\frac{\chp U^s}{\chp U}=-\sqrt\eps \int\limits_0^s D^2\Phi(u^\tau)\frac{\chp U^\tau}{\chp U} \,d\tau + \Id_{2 N}, \quad
D^2\Phi:=2i\begin{pmatrix}
\frac{\partial^2\Phi}{\partial \ov u \partial u} & \frac{\partial^2\Phi}{\partial \ov u^2} \\
-\frac{\partial^2\Phi}{\partial u^2} & -\frac{\partial^2\Phi}{\partial  u \partial \ov u}
\end{pmatrix},
\end{equation}
where
$
D^2\Phi=(D^2\Phi_{\wid k \wid l})_{\wid k, \wid l\in\wid\CC}$ and $\ds{(D^2\Phi)_{(1,k)(1,l)}=\frac{\partial^2\Phi}{\partial \ov u_k \partial u_l}, \quad (D^2\Phi)_{(1,k)(2,l)}=\frac{\partial^2\Phi}{\partial \ov u_k \partial \ov u_l}},
$ etc.

Introduce the family of norms in $\mC^{2N}\ni V=(V_{\wid k})_{\wid k\in \wid\CC}$: for $j\in\CC$ and $q\geq 1$  we define
\begin{equation}
\label{nlnorm2}
\big\bracevert V \big\bracevert_{j,q}^q = \sum\limits_{k\in\CC} \gamma^{|k-j|} (|V_{(1,k)}|^q + |V_{(2,k)}|^q ).
\end{equation}
For a $2N\times 2N$-matrix $A$ with complex entires by $\big\bracevert A \big\bracevert_{j,q}$ 
 we will denote its operator norm, corresponding to the norm in $\mC^{2N}$ above. Likewise for a $N\times N$-matrix $A$ we define  its norm $\| A \|_{j,q}$.
\begin{cor}
\label{nllem:matrix}
Let $a,b \in \{u,v\}$ or $a,b \in \{\ov u,\ov v\}$. Then for any $q\geq 1$ there exists a constant $C$ such that for all $0\leq s\leq 1$ and $j\in\CC$ we have
$$
\big\bracevert D^2\Phi \big\bracevert_{j,q}\leq C, \; \Big\|\frac{\chp a^s}{\chp b}-\Id_{N}\Big\|_{j,q},\; \Big\|\frac{\chp a^s}{\chp \ov b}\Big\|_{j,q} \leq C\sqrt\eps.
$$ 
\end{cor}
{\it Proof.}  
The matrix $D^2\Phi$ consists of four blocks of the size $N\times N$. Consider  the upper left-hand block. By Proposition \ref{nlprop:est}  the only nonzero entries  are  $\ds{{\frac{\chp^2\Phi}{\chp \ov u_k\chp u_m}}}$ with $|k-m|\leq 1$, and they are bounded by the same constant. Since the other blocks have a similar structure, we obtain 
$\big\bracevert D^2\Phi \big\bracevert_{j,q} \leq C$.
Due to (\ref{nlDF}), we get
$$
\Big\bracevert \frac{\chp U^s}{\chp U}  \Big\bracevert_{j,q} \leq \sqrt\eps C \int\limits_0^s\Big\bracevert \frac{\chp U^\tau}{\chp U}  \Big\bracevert_{j,q} \,d\tau +1 \Rightarrow \Big\bracevert \frac{\chp U^s}{\chp U}  \Big\bracevert_{j,q} \leq C_1.
$$
Using (\ref{nlDF}) once more, we obtain
\bee
\label{nlgrbl}
\Big\bracevert \frac{\chp U^s}{\chp U} - \Id_{2 N} \Big\bracevert_{j,q} \leq \sqrt\eps C \int\limits_0^s\Big\bracevert \frac{\chp U^\tau}{\chp U}  \Big\bracevert_{j,q} \,d\tau \leq C_1\sqrt\eps.
\eee
The matrix $\ds{ \frac{\chp u^s}{\chp u} - \Id_{ N}}$ is the upper left-hand $N\times N$ block of the matrix $\ds{\frac{\chp U^s}{\chp U} - \Id_{2N}}$. Then the wanted estimate 
$\ds{\Big\| \frac{\chp u^s}{\chp u} - \Id_{ N}\Big\|_{j,q} \leq C\sqrt{\eps}}$
follows from (\ref{nlgrbl}). The estimates for the  other matrices  can be obtained similarly. 
\qed
\ssk

We will also need the following elementary propositions.
\begin{prop}
\label{nlprop:bounded}
Let $A=(a_{kl})_{k,l\in\CC}$ be a matrix with complex entires of the size $N\times N$. Assume that there exists a constant $C_0$ such that for some $q\geq 1$ and any $j\in\CC$ we have $\|A\|_{j,q}\leq C_0$. Then $|a_{kl}|\leq C_0$ for all $k,l\in\CC$.
\end{prop}
{\it Proof}. Let us fix any $k,l\in\CC$ and take the vector $v=(v_m)_{m\in\CC}$ such that $v_m=\del_{ml}$. Then we have $\|v\|^q_{k,q}= \ga^{|l-k|}< 1$. Consequently, $\|Av\|^q_{k,q} = \sum\limits_{m\in\CC} |a_{ml}|^q\ga^{|k-m|}\leq C_0^q$, and it follows that $|a_{kl}|\leq C_0$.
\qed
\begin{prop} 
\label{nlrem:conditions}
Estimates of assumptions {\it HF} and {\it HG} imply that for any $x\geq 0$ and any $w,z\in\mC$ we have 

{\bf 1.} $
|\chp^{n+m}_{w^n \ov w^m} f(|w|^2)|\leq C f(|w|^2), \quad \mbox{where}\quad  1\leq m+n\leq 3; 
$

{\bf 2.} $
\big|\chp^{m_1 + l_1 + m_2 + l_2}_{w^{m_1} \ov w^{l_1} z^{m_2} \ov z^{l_2}} [G(|w-z|^2)]\big| \leq C\big(1+|w|^{p-1}+|z|^{p-1}\big),\mbox { where }  1 \leq m_1 + l_1 + m_2 + l_2 \leq 3;
$

{\bf 3.} $|G(x)|\leq C(1+x^{p/2})$.
\end{prop}
{ \it Proof}. By the direct computation. \qed

\subsection{Properties of the transformation}
\label{nlsec:change_Prop}

Here we prove items {\it 1-4} of the theorem. 
We give the proofs  in the following order: {\it 4,3 2a, 1, 2b}. 


{\bf Item 4.} 
The first estimate follows from (\ref{nlchange1}) joined with Proposition \ref{nlprop:est}. To get the second, we write the change of variables in the action-angle coordinates
\bee
\nonumber
I_j=J_j - \sqrt\eps\int\limits_0^1 \chp_{\psi_j}\Phi (J^\tau,\psi^\tau) \, d\tau, 
\eee
and apply Proposition \ref{nlprop:est} once more.
\ssk

{\bf  Item 3.} 
Recall that the matrices $W^1$ and $W^2$ have the form (\ref{nlWWWWW}).  
Due to Corollary \nolinebreak \ref{nllem:matrix}, for all $j\in\CC$ and $q \geq 1$ we have
$$
W^1=\mbox{diag}(\sqrt{\TT_k})+\wid W^1, \mbox{ where } \|\wid W^1\|_{j,q}\leq C\sqrt\eps, \mbox{ and } \|W^2\|_{j,q}\leq C\sqrt\eps.
$$
Then
$$
\|\ov{W^1}W^{1T}-\diag(\TT_k)\|_{j,q}, \|\ov{W^2}W^{2T}\|_{j,q},\|{W^2}W^{1T}\|_{j,q},\|W^1W^{2T}\|_{j,q} \leq C\sqrt\eps,
$$
so the desired estimate follows from Proposition \ref{nlprop:bounded}.
\ssk

{\bf Item 2a.} Recall that, due to (\ref{nlham1}), 
$$H_2=\frac12 \{\wid G+\lan \wid G \ran , \Phi\}.$$ 
Then, obviously, the first assertion of {\it 2a} holds. The second follows from Propositions \nolinebreak\ref{nlrem:conditions}.2 and  \ref{nlprop:est}: for all $k\in\CC$ we have  
\begin{equation*}
\left| \chp_{\ov v_{k}}\{\wid G, \Phi\} \right|,\left| \chp_{\ov v_{k}}\{\lan \wid G \ran, \Phi\} \right| \leq C\sum\limits_{n:|n-k|\leq 2}|v_n|^{p-1}+C.
\end{equation*}

{\bf  Item 1.} We will need the following
\begin{prop}
\label{nlprop:raznost}
Let the function $h(\psi)=h((\psi_k)_{k\in\CC})$ be $C^1$-smooth and depends on the argument $\psi$ only through the differences of its neighbouring components $(\tht_{kn})_{|k-n|=1},$ where $\tht_{kn}=\psi_k-\psi_n$.
Then 
\footnote{The variables $(\tht_{kn})_{|k-n|=1}$ are not independent: indeed, $\frac{\chp\tht_{kn}}{\chp\tht_{nk}}=-1.$} 
$$
\left| \sum\limits_{k\in\CC} \gamma^{|j-k|}\chp_{\psi_k} h \right| \leq  
(1-\gamma)\sum\limits_{|k-n|=1} \gamma^{|j-k|} |\chp_{\theta_{kn}} h|.
$$
\end{prop}  
{\it Proof}.
Using the equality 
$\chp_{\psi_{k}}=\frac12\sum\limits_{n:|k-n|=1}(\chp_{\tht_{kn}}-\chp_{\tht_{nk}})$, 
we get
\begin{eqnarray}
\label{nlrazn1}
\sum\limits_{k\in\CC} \gamma^{|j-k|}\chp_{\psi_k} h =
 \frac12\sum\limits_{|k-n|=1} \gamma^{|j-k|}(\chp_{\theta_{kn}}-\chp_{\tht_{nk}})h= \frac12\sum\limits_{|k-n|=1}(\gamma^{|j-k|}-\gamma^{|j-n|})\chp_{\tht_{kn}}h.  
\end{eqnarray}
Since $1/2<\ga<1$, for $k$ and $n$ satisfying $|k-n|=1$ we have
\begin{eqnarray}
\label{nlrazn2}
\big|1-\ga^{|j-n|-|j-k|}\big| =
\left\{ 
\begin{array}{cl}
1-\ga, \quad &\mbox{if} \quad |j-n|-|j-k|>0, \\
\ga^{-1} - 1\leq 2(1-\ga), \quad &\mbox{otherwise.}
\end{array}  
\right.
\end{eqnarray}
Since $\ga^{|j-k|}-\ga^{|j-n|}=\ga^{|j-k|}(1-\ga^{|j-n|-|j-k|})$,  (\ref{nlrazn1}) jointly with (\ref{nlrazn2}) implies the desired estimate.
\qed
\ssk

Now we will deduce item {\it 1} from Proposition \ref{nlprop:raznost} and  item {\it 2a}. 
Using the notation $G_{ml}(v_m,v_l):=G(|v_m-v_l|^2)$,  let us represent the Hamiltonian $H_2$ in the form 
$$
H_2=\frac{1}{8}\sum\limits_{|m-l|=1} \{\lan G_{ml} \ran+ G_{ml}, \sum\limits_{|k-n|=1}\Phi_{kn} \} =: \sum\limits_{|m-l|=1} H_2^{ml}.
$$ 
Since the functions $G_{ml}$ and $\Phi_{kn}$ depend on the angles $\psi=(\psi_k)_{k\in\CC}$ only through the differences of their neighbouring components $(\theta_{kn})_{|k-n|=1}$, then the function $H_2^{ml}$ depends on $\psi$ in the same way, so that it  satisfies the conditions of Proposition \ref{nlprop:raznost}. Using the identity $i\nabla_k h\cdot v_k=-\chp_{\psi_k} h$, which holds for any $C^1$-smooth function $h$, we get 
\bee\label{nlinablah}
|(i\nabla H_2^{ml} \cdot v )_j | = \Big|\sum\limits_{k\in\CC} \gamma^{|j-k|}\chp_{\psi_k}H_2^{ml}\Big| \leq
(1-\gamma)\sum\limits_{|k-n|=1}\gamma^{|j-k|} |\chp_{\theta_{kn}}H_2^{ml}|. 
\eee 
Note that the function $H_2^{ml}(v)$ depends on $v=(v_s)_{s\in\CC}$ only through $v_s$ with $s$ satisfying 
\bee\label{nlkaka}
|m-s|\wedge |l-s|\leq 1.
\eee
Then the derivative  $\chp_{\psi_{s}}H_2^{ml}$ may not vanish only if $s$ satisfies (\ref{nlkaka}). This implies 
\bee\label{nl9865}
|\chp_{\theta_{kn}}H_2^{ml}|\leq  \sum\limits_{s:|m-s|\wedge |l-s|\leq 1} |\chp_{\psi_s}H_2^{ml}|.
\eee
Since,  in view of the inequality $1/2<\ga<1$, for $k$ and $s$ satisfying (\ref{nlkaka}) we have $\gamma^{|j-k|}/\gamma^{|j-s|}\leq C$, from (\ref{nl9865}) we obtain
\bee\label{nl98651}
\gamma^{|j-k|}  |\chp_{\theta_{kn}}H_2^{ml}| 
\leq C\sum\limits_{s:|m-s|\wedge |l-s|\leq 1}  \gamma^{|j-s|}|\chp_{\psi_s}H_2^{ml}|. 
\eee
Since the derivative
$\chp_{\tht_{kn}}H_2^{ml}$ may not vanish only for $k$ and $n$ satisfying (\ref{nlkaka}), then (\ref{nlinablah}) joined with (\ref{nl98651}) implies
\bee\label{nlhmlll}
|(i\nabla H_2^{ml} \cdot v )_j | \leq 
(1-\gamma) C \sum\limits_{k:|m-k|\wedge |l-k|\leq 1} \gamma^{|j-k|} |\chp_{\psi_k}H_2^{ml}|.
\eee
Obviously, the estimate of item {\it 2a} holds if replace the Hamiltonian $H_2$ by $H_2^{ml}$, so that 
\begin{equation}
\label{nlactangdop}
|\chp_{\psi_k} H_2^{ml}|=|\nabla_k H_2^{ml}\cdot v_k|\leq C\sum\limits_{n:|n-k|\leq 2}|v_n|^p+C.
\end{equation}
Then (\ref{nlhmlll}) implies
\begin{eqnarray}
\nonumber
|( i\nabla H_2 \cdot v )_j | &\leq& (1-\gamma)C\sum\limits_{|m-l|=1}\sum\limits_{k: |m-k|\wedge |l-k|\leq 1} \gamma^{|j-k|}\Big(\sum\limits_{n:|n-k|\leq 2}|v_n|^p+1 \Big)  \\
\nonumber
&\leq& (1-\gamma)C_1\sum\limits_{|m-l|=1}\sum\limits_{k: |m-k|\wedge |l-k|\leq 3}\gamma^{|j-k|} |v_k|^p + C_2(\ga)  \\
\nonumber
&\leq&
(1-\gamma)C_3\sum\limits_{k\in\CC}\gamma^{|j-k|}|v_k|^p +C_2(\ga)=
(1-\gamma)C_3\|v\|^p_{j,p} + C_2(\gamma).
\end{eqnarray}

{\bf Item 2b.} Remind that $r=i\nabla H_> + r^I + r^D$, so
\begin{equation}
\nonumber
\|r\|_{j,q}\leq\|i\nabla H_>\|_{j,q}+\|r^I\|_{j,q}+\|r^D\|_{j,q}.
\end{equation}

{\bf Step 1.} Firstly we will show that for any $j\in\CC$ and $q\geq 1$
\begin{equation}
\label{nlinablaH}
\|i\nabla H_>\|_{j,q} \leq C\|v\|_{j,q(p-1)}^{p-1}+ C_1(\gamma).
\end{equation} 
Note that by Taylor's expansion 
$$
\frac12 \{\wid G, \Phi\}_2 + \frac{\sqrt{\eps}}{2} \int\limits_0^1 (1-s)^2 \{\wid G,\Phi\}_3 \circ X_{\sqrt\eps\Phi}^s \, ds=
\int\limits_0^1 (1-s) \{\wid G,\Phi\}_2 \circ X_{\sqrt\eps\Phi}^s \, ds.
$$
Then we have
$$
H_> =  
\frac12 \int\limits_0^1 (1-s)^2 \{F , \Phi\}_3 \circ X_{\sqrt\eps\Phi}^s \, ds+  
\int\limits_0^1 (1-s) \{\wid G, \Phi\}_2 \circ X_{\sqrt\eps\Phi}^s \, ds.
$$
Since, due to (\ref{nlgomol}) and (\ref{nlhh}), we have $ \{\wid F,\Phi\}_3  =  \{\lan \wid G \ran - \wid G,\Phi\}_2$, we obtain
\begin{eqnarray}
\nonumber
H_> &=&  
\frac12 \int\limits_0^1 (1-s)^2 \{\lan \wid G \ran -\wid G, \Phi\}_2 \circ X_{\sqrt\eps\Phi}^s \, ds+  
\int\limits_0^1 (1-s) \{\wid G, \Phi\}_2 \circ X_{\sqrt\eps\Phi}^s \, ds
= \\
\nonumber
&=& \frac12 \int\limits_0^1 (1-s^2) \{\wid G, \Phi\}_2 \circ X_{\sqrt\eps\Phi}^s \, ds + 
\frac12 \int\limits_0^1 (1-s)^2 \{\lan \wid G \ran, \Phi\}_2 \circ X_{\sqrt\eps\Phi}^s \, ds \\
\nonumber
&=:& 
\int\limits_0^1 Y^s\circ X_{\sqrt\eps\Phi}^s \, ds.
\end{eqnarray}
Due to  Propositions \ref{nlrem:conditions}.2 and  \ref{nlprop:est}, for $w_k\in\{v_k,\ov v_k\}$ and all $k\in\CC$ we have
\begin{equation}
\nonumber
\left| \chp_{w_{k}}\{\lan\wid G\ran, \Phi\}_2 \right| , \left| \chp_{w_{k}}\{\wid G, \Phi\}_2 \right|\leq C\sum\limits_{l:|k-l|\leq 3}|v_l|^{p-1} +C_1. 
\end{equation}
Then
\begin{equation}
\label{nldY}
\left|\frac{\chp Y^s}{\chp v_{k}}\right|,\left|\frac{\chp Y^s}{\chp \ov v_{k}}\right| \leq  C\sum\limits_{l:|k-l|\leq 3}|v_l|^{p-1} +C_1  \quad\mbox{for any}\quad  k \in \CC.
\end{equation}
Estimate (\ref{nldY}) implies
$$
\left\|\frac{\chp Y^s}{\chp v}\right\|_{j,q}^q, \left\|\frac{\chp Y^s}{\chp \ov v}\right\|^q_{j,q}\leq 
 C\sum\limits_{k\in\CC}\ga^{|j-k|}|v_k|^{q(p-1)} +C_1(\gamma)=C\|v\|_{j,q(p-1)}^{q(p-1)}+C_1(\gamma).
$$
Thus, we get
\begin{equation}
\label{nlDY}
\left\|\frac{\chp Y^s}{\chp v}\right\|_{j,q},\, \left\|\frac{\chp Y^s}{\chp \ov v}\right\|_{j,q} \leq  C\|v\|_{j,q(p-1)}^{p-1}+C_1(\gamma).
\end{equation}
Exchanging the derivative and the integral we find 
\begin{align}
\nonumber
\nabla H_>(v)= 2 \chp_{\ov v}\int\limits_0^1 Y^s(v^s) \, ds 
 = 2 \int\limits_0^1\left( \frac{\chp Y^s}{\chp v^s} 
\frac{\chp v^s}{\chp \ov v} 
+  
 \frac{\chp Y^s}{\chp \ov v^s} 
\frac{\chp \ov v^s}{\chp \ov v} \right) \, ds .
\end{align}
By (\ref{nlDY}), Corollaries \ref{nllem:matrix} and  \ref{nlprop:ekviv}  we obtain (\ref{nlinablaH}):
\begin{align}
\nonumber
\|i\nabla H_> \|_{j,q} &\leq C \int\limits_0^1 \left( \left\|\frac{\chp Y^s}{\chp v^s}\right\|_{j,q} 
  +  
 \left\|\frac{\chp Y^s}{\chp \ov v^s}\right\|_{j,q} 
 \right) \, ds \\
\nonumber
&\leq C_1\int\limits_0^1\|v^s\|_{j,q(p-1)}^{p-1} \,ds + C_2(\gamma) \leq C_3\|v\|_{j,q(p-1)}^{p-1} +C_4(\gamma).
\end{align}

{\bf Step 2.} Let us show that 
\begin{equation}
\label{nltri}
\|r^D\|_{j,q}\leq  C\|v\|_{j,q(p-1)}^{p-1}+ C_1(\gamma).
\end{equation}
Due to (\ref{nlsmord}), we have $r^D=r^D_1+r^D_2$, where $$
r^D_1=\eps^{-1/2}\big( g(u)-g(v) \big) , \; r^D_2=\eps^{-1/2}\left(\left(\frac{\chp v}{\chp u}-\Id_{ N}\right)g(u) + \frac{\chp v}{\chp \ov u}\ov g(u)\right). 
$$
Due to assumption {\it Hg(i)}  and item {\it 4} of the theorem, we have 
$$|r^D_{1k}|\leq C\Big( 1+ \sum\limits_{l:|k-l|\leq 1}(|u_l|^{p-1}+|v_l|^{p-1})\Big) \quad\mbox{for all $k \in\CC$}.$$ 
Then, Corollary \ref{nlprop:ekviv} implies 
\begin{equation}
\label{nlrd1}
\|r^D_1\|_{j,q}^q\leq C\sum\limits_{k\in\CC}\ga^{|j-k|}(|u_k|^{q(p-1)}+|v_k|^{q(p-1)}) +C_1(\ga) \leq  C_2\|v\|_{j,q(p-1)}^{q(p-1)} +C_3(\ga).
\end{equation}
Similarly, applying  Corollaries \ref{nllem:matrix} and \ref{nlprop:ekviv},  we get 
\bee\label{nlrD2}
\|r^D_2\|^q_{j,q}\leq C\|g(u)\|^q_{j,q}\leq  C_1\|v\|_{j,q(p-1)}^{q(p-1)} +C_2(\ga).
\eee
Combining (\ref{nlrd1}) with (\ref{nlrD2}), we obtain (\ref{nltri}).

{\bf Step 3.} Let us show that  
\begin{equation}
\label{nl4etire}
\|r^I\|_{j,q}\leq C(\gamma). 
\end{equation}
Denote $D^2\Phi^s:=D^2\Phi\circ X^s_{\sqrt\eps\Phi}$ and $\ds{A^s:=\sum\limits_{k\in\CC}\TT_k\frac{\chp^2 U^s }{\chp u_k \chp \ov u_k}.}$
Due to (\ref{nlDF}), we have 
\begin{align}
\nonumber
A^s&=
 -\sqrt\eps\sum\limits_{k\in\CC}\TT_k\frac{\chp}{\chp \ov u_{ k} }\int\limits_{0}^s(D^2\Phi)^\tau\frac{\chp U^\tau}{\chp  u_k}\, d\tau \\
\nonumber
&= -\sqrt\eps\sum\limits_{k\in\CC}\TT_k\sum\limits_{\wid m \in\wid\CC}
\int\limits_{0}^s \chp_{U_{\wid m}^\tau}(D^2\Phi)^\tau\frac{\chp U^\tau}{\chp u_k}
\frac{\chp U^\tau_{\wid m}}{\chp \ov  u_k}\, d\tau - \sqrt\eps\sum\limits_{k\in\CC}\TT_k
\int\limits_{0}^s (D^2\Phi)^\tau\frac{\chp^2 U^\tau}{\chp u_k \chp \ov u_k}
\, d\tau  \\
\label{nltrois}
&=:\sqrt\eps\Big(\Gamma- \int\limits_{0}^s (D^2\Phi)^\tau A^\tau\, d\tau\Big), \quad \mbox{where $\Gamma = (\Gamma_{\wid l})_{\wid l\in\wid \CC}.$}
\end{align}
Let us show that 
\begin{equation}
\label{nlgama1}
\big\bracevert \Gamma \big\bracevert_{j,q} \leq C(\ga).
\end{equation}
Fix $\wid l=(l_0,l)\in\wid\CC$ and define the set of indices   $\SSS(\wid l)\subset\wid \CC\times\wid\CC$, where
$$\SSS(\wid l):=\{\wid r=(r_0,r),\wid m=(m_0,m)\in\wid\CC: |r-l|\vee|m-l|\vee|m-r| \leq 1\}.$$
Due to Proposition \ref{nlprop:est}, the function 
 $\chp_{U_{\wid m}}(D^2\Phi)_{\wid l\wid r}$ may not vanish only if $\wid r, \wid m\in \SSS(\wid l)$. 
Then 
\begin{align}
\nonumber
\Gamma_{\wid l}&=
 -\sum\limits_{(\wid m, \wid r)\in\SSS (\wid l)}
\int\limits_{0}^s \chp_{U_{\wid m}^\tau}(D^2\Phi)^\tau_{\wid l \wid r}\sum\limits_{k\in\CC}\TT_k\frac{\chp U_{\wid r}^\tau}{\chp u_k}
\frac{\chp U^\tau_{\wid m}}{\chp \ov u_k}\, d\tau \\
\label{nl2946}
&=
-\sum\limits_{(\wid m, \wid r)\in\SSS (\wid l)}\int\limits_{0}^s \chp_{U_{\wid m}^\tau}(D^2\Phi)^\tau_{\wid l \wid r}\left[\frac{\chp U^\tau}{\chp u}\diag(\TT_k)
\left(\frac{\chp U^\tau}{\chp \ov u} \right)^T\right]_{\wid r \wid m}\, d\tau.
\end{align}
The $2 N\times 2 N$ matrix  $\ds{\frac{\chp U^\tau}{\chp \ov u}\diag(\TT_k)
\left(\frac{\chp U^\tau}{\chp  u}\right)^T}$ consists of four blocks of the type
$$
\ds{\frac{\chp a^\tau}{\chp  b}\diag(\TT_k)
\left(\frac{\chp  c^\tau}{\chp  d}\right)^T},
\quad \mbox{where} \quad  a,b,c,d \in \{ u, \ov u\}.
$$ 
By Corollary \ref{nllem:matrix}, for every $j\in\CC$ and $q\geq 1$ we have 
$$
\left\|\frac{\chp a^\tau}{\chp  b}\diag(\TT_k)
\left(\frac{\chp  c^\tau}{\chp  d}\right)^T\right\|_{j,q} \leq C.
$$
Proposition \ref{nlprop:bounded} implies that the elements of this matrix are bounded by the same constant.
Thus, for every $\wid r,\wid m\in\wid\CC$ we have
$
\ds{\left|\left[\frac{\chp U^\tau}{\chp \ov u}\diag(\TT_k)
\left(\frac{\chp U^\tau}{\chp  u}\right)^T\right]_{\wid r \wid m}\right|\leq C}.
$

Proposition \ref{nlprop:est} implies that
$
|\chp_{U_{\wid m}^\tau}(D^2\Phi)^\tau_{ \wid l \wid r}|\leq C
$ 
for all pairs $(\wid m,\wid r)\in\SSS(\wid l)$. Since the cardinality of the set $\SSS(\wid l)$ is bounded uniformly in $\wid l\in\wid\CC$, in view of (\ref{nl2946}) we have $|\Gamma_{\wid l}| \leq C$, so that we get (\ref{nlgama1}).

Due to Corollary \ref{nllem:matrix}, we have $\big\bracevert D^2\Phi \big\bracevert_{j,q}\leq C.$ Then (\ref{nltrois}) jointly with (\ref{nlgama1}) implies
$$
\big\bracevert A^s \big\bracevert_{j,q} \leq \sqrt\eps \big\bracevert \Gamma \big\bracevert_{j,q} + \sqrt\eps C \int\limits_{0}^s \big\bracevert A^\tau \big\bracevert_{j,q} \, d\tau  \Rightarrow \big\bracevert A^s \big\bracevert_{j,q} \leq \sqrt\eps C(\gamma).
$$
In paticular, this implies (\ref{nl4etire}). The proof of  the theorem is complete.  
\qed 
\ssk

{\it Proof of Proposition  \ref{nlprop:est}. }
According to (\ref{nlPhij}), we have 
\begin{equation}
\label{nlY/Z}
\Phi=\frac14 \sum\limits_{|j-n|=1} \frac{Y_{jn}}{f_j-f_{n}}, \quad \mbox{where} \quad 
Y_{jn} = \int\limits_0^{\theta_{jn}} G^0(J_{j}, J_n, \theta) \, d\theta.
\end{equation}
Denote 
\bee\label{nlDDD}
D^q:= \chp^{q}_{v_j^{m_j}\ov v_j^{l_j}v_n^{m_n}\ov v_n^{l_n}},\quad \mbox{where}\quad q:=m_j+l_j+m_n+l_n,\quad m_j,l_j,m_n,l_n\in\mN\cup\{0\}.
\eee
Let  
$$G_{jn}(v_j,v_n):=G(|v_j-v_n|^2)\quad\mbox{and}\quad  G_{jn}^0:=G_{jn}-\lan G_{jn}\ran.$$ 
We will need the following proposition.
\bpp
\label{nlprop:igrek}
Assume that
$G_{jn}\in C^{q_0+1}(\mC^2)$, where $q_0\in\mN$.
Then 

{\it (i)} $Y_{jn}\in C^{q_0}(\mC^2)$.

{\it (ii)} Assume that for an operator $D^q$ as above, with $q\leq q_0$,  for all $v_j,v_n\in\mC$ and some function $K:\mR^2_{+0}\mapsto\mR$ we have  $|D^qG_{jn}(v_j,v_n)|\leq K(|v_j|,|v_n|)$. Then $|D^qY_{jn}(v_j,v_n)|\leq C K(|v_j|,|v_n|)$ for all $v_j,v_n\in\mC$.
\epp
Before proving Proposition \ref{nlprop:igrek} we will finish the proof of Proposition \ref{nlprop:est}.
The $C^3$-smoothness of the function $\Phi$ follows from Proposition \ref{nlprop:igrek}.{\it i} and assumptions {\it HF} and {\it HG}.
The assertion stating that the most components of the second and the third derivatives of the function $\Phi$ vanish is obvious. 
By (\ref{nlY/Z}), assumption {\it HF} and Proposition \ref{nlrem:conditions}.3 we have
$$
|\chp_{\psi_j}\Phi| \leq\frac12 \sum\limits_{n:|j-n|=1}\left|\frac{G^0(|v_{j}-v_{n}|^2)}{f_j-f_{n}}\right|\leq C,
$$
so that the first estimate of (\ref{nlpropest}) is proven. To establish the other, 
 in view of Proposition \nolinebreak\ref{nlrem:conditions}.1,  it suffices to show that for every $j,n\in\CC$ satisfying $|j-n|=1$ and for all $k, l,m\in \{j,n\}$ we have  
\begin{equation}
\label{nligrek}
\left| \frac{\chp Y_{jn}}{\chp a_{ k}} \right|,
\left| \frac{\chp^2 Y_{jn}}{\chp a_{k} \chp b_{l}} \right|,
\left| \frac{\chp^3 Y_{jn}}{\chp a_{k}\chp b_{l}\chp c_{m}} \right|   \leq C(1 + |v_j|^{p-1}+|v_{n}|^{p-1}),
\end{equation} 
where $a,b,c\in\{v,\ov v\}$. In view of Proposition \ref{nlrem:conditions}.2, estimate (\ref{nligrek}) is an immediate corollary of Proposition \ref{nlprop:igrek}.{\it ii} with $K(v_j,v_n)=C(1 + |v_j|^{p-1}+|v_{n}|^{p-1})$ and $q_0=3$. 
\qed 
\ssk

{\it Proof of  Proposition \ref{nlprop:igrek}.}
Since  we  will only use the functions $G_{jn},\,G^0_{jn},\,Y_{jn}$ and not  $G,\,G^0,\,Y$, then, abusing notations, instead of writing $G_{jn},\,G^0_{jn},\,Y_{jn}$, we will write just $G,\,G^0,\,Y$.

{\it (i)}
It is clear that $G^0\in C^{q_0+1}(\mC^2)$ while $Y\in C^{q_0+1}(\mC^2\cap\{v_j,v_k\neq 0\})$. 
Take an operator $D^q$ as in (\ref{nlDDD}) with $0\leq q\leq q_0$, where $D^0$ denotes the identity. 

First we assume $v_j,v_n\neq 0$,
 so that the function $D^{q}Y(v_j,v_n)$ is continuously differentiable. We start with writing down the Fourier series for the functions $D^{q}G^0$ and $D^{q}Y$. 
For this purpose we note that the Fourier's expansions of the functions $G^0$ and $Y$ have the form
\bee\label{nlG0Y}
G^0(v_j,v_n)=\sum\limits_{k\in\mZ\sm\{0\}} G_k(|v_j|,|v_n|)e^{ik (\psi_j-\psi_n)}\quad\mbox{and}\quad Y(v_j,v_n)=\sum\limits_{k\in\mZ\sm\{0\}} \frac{G_k(|v_j|,|v_n|)}{ik}e^{ik(\psi_j-\psi_n)},
\eee
where we have used the equality $G_k=G_{-k}$, so that $\ds{\sum\limits_{k\in\mZ\sm\{0\}} \frac{G_k}{ik}= 0}$.  
We will obtain the series for $D^{q}G^0$ and $D^{q}Y$  by formal differentiation of (\ref{nlG0Y}), with help of the formulas
\bee
\label{nlchpv}
\chp_{v_j}=e^{-i\psi_j}\Big(\frac{\chp_{|v_j|}}{2} + \frac{1}{2i|v_j|}\chp_{\psi_j} \Big) \quad \mbox{and} \quad
\chp_{\ov v_j}=e^{i\psi_j}\Big(\frac{\chp_{|v_j|}}{2} - \frac{1}{2i|v_j|}\chp_{\psi_j} \Big),
\eee
valid for $v_j\neq 0$. In order to justify the formal differentiation, one can calculate the Fourier coefficients for the functions $D^qG^0$ and $D^qY$ directly, using the integration by parts and formula (\ref{nlchpv}), and verify that the they coincide with the coefficients obtained below in (\ref{nlDh}).
Using (\ref{nlchpv}), for any $k\in\mZ\sm\{0\}$ we get 
$$
\chp_{v_j}(G_k e^{ik(\psi_j-\psi_n)}) = \wid G^1_k e^{i[(k-1)\psi_j - k\psi_n]}\quad\mbox{and}\quad\chp_{\ov v_j}(G_k e^{ik(\psi_j-\psi_n)}) =\wid G^2_k e^{i[(k+1)\psi_j - k\psi_n]},
$$
for some functions  $\wid G^1_k= \wid G^1_k (|v_j|,|v_n|)$ and $\wid G^2_k= \wid G^2_k (|v_j|,|v_n|)$, $\wid G_k^1,\wid G_k^2\in C^{q_0}(\mR^2_+)$. Similar relations hold for the derivatives with respect to $v_n,\ov v_n$. 
Arguing by induction, we obtain 
\bee
\label{nlDG}
D^q(G_ke^{ik(\psi_j-\psi_n)})=\hat G_k e^{i(s_j(k)\psi_j+s_n(k)\psi_n)}, 
\eee
where $s_j(k)=k+l_j-m_j,\; s_n(k)=l_n-m_n-k$ and $\hat G_k=\wid G_k(|v_j|,|v_k|),$ $\hat G_k\in C^{q_0+1-q}(\mR^2_+)$.
Thus, we get
\bee
\label{nlDh}
D^qG^0(v_j,v_n)= \sum\limits_{k\in\mZ\sm\{0\}} \hat G_k e^{i(s_j(k)\psi_j+s_n(k)\psi_n)} 
\mbox{ and } 
D^q Y(v_j,v_n)=\sum\limits_{k\in\mZ\sm\{0\}}  \frac{\hat G_k}{ik} e^{i(s_j(k)\psi_j+s_n(k)\psi_n)},
\eee
if $v_j,v_n\neq 0$. 
Now, using (\ref{nlDh}), we will write an integral representation of the function $D^qY$, which will permit us to accomplish the proof of the proposition.
For $r\geq 0$ and $\psi\in\mT$ let us define the lifting operator 
$$\VV_\psi r:=r e^{i\psi}\in\mC.$$
By the direct computation, from (\ref{nlDh}) we obtain 
\bee\label{nlDYAA}
D^qY(v_j,v_n)= A_j^q(v_j,v_n)=A_n^q(v_j,v_n) \quad\mbox{for all}\quad v_j,v_n\neq 0,
\eee
where 
\begin{align}\label{nlDY_j}
A_j^q(v_j,v_n)&=\int\limits_0^{\psi_j} D^qG^0(\VV_\psi|v_j|,v_n) e^{i(m_j-l_j)(\psi-\psi_j)}\,d\psi \\
\nonumber
&-\frac{1}{2\pi}\int_0^{2\pi} \int\limits_0^{\xi} D^qG^0(\VV_\psi|v_j|,v_n) e^{i(m_j-l_j)(\psi-\psi_j)}\,d\psi \,d\xi
\end{align}
and
\begin{align}\label{nlDY_n}
A_n^q(v_j,v_n)&=-\int\limits_0^{\psi_n} D^qG^0(v_j,\VV_\psi|v_n|) e^{i(m_n-l_n)(\psi-\psi_n)}\,d\psi \\
\nonumber
&+\frac{1}{2\pi}\int_0^{2\pi} \int\limits_0^{\xi} D^qG^0(v_j,\VV_\psi|v_n|) e^{i(m_n-l_n)(\psi-\psi_n)}\,d\psi \,d\xi.
\end{align} 
Since the function $D^qG^0$ is continuous, the function $A^q_j$ is continuous when $v_j\neq 0$ while the function $A^q_n$ is continuous if $v_n\neq 0$.
\bpp\label{nlprop:Acont}
If $0\leq q\leq q_0$ then the functions $A^q_j,A^q_n$ are continuous for all $v_j,v_n\in\nolinebreak\mC$. 
\epp  
Before proving Proposition \ref{nlprop:Acont} we will finish the proof of Proposition \ref{nlprop:igrek}. 
First we claim that the function $D^0Y=Y$ is continuous. Indeed, outside the planes $v_j=0$ and $v_n=0$ it follows from  continuity of the function $G^0$. Inside the planes we additionally use the relation $G^0(0,v_n)\equiv G^0(v_j,0)\equiv 0$ which is easy to check, in view of the definition (\ref{nlG000}) of the function $G^0$.  

Now we argue by induction.   Fix $0 \leq q < q_0$ and assume that the function $D^qY$ is continuous. First we claim that the function $D^{q+1}Y$ is also continuous, where $D^{q+1}:=\chp_{v_j}D^q$. 

It is clear that the derivative $\chp_{v_j}A_n^q$ is well-defined for all $v_j,v_n\in\mC$ and coincides with $A_n^{q+1}$. In view of (\ref{nlDYAA}) and Proposition \ref{nlprop:Acont}, the continuity of $D^qY$ implies that (\ref{nlDYAA}) holds for all $v_j,v_n\in\mC$, so that $D^{q+1}Y=\chp_{v_j}A_n^q=A_n^{q+1}$ for all $v_j,v_n\in\mC$. Then  Proposition \nolinebreak\ref{nlprop:Acont} implies that the function $D^{q+1}Y$ is continuous. 

The cases $D^{q+1}=\chp_{\ov v_j}D^q,\chp_{v_n}D^q$ and $\chp_{\ov v_n}D^q $ can be considered similarly.  In the last two situations one should differentiate the function $A_j^q$ instead of $A_n^q$. By induction axiom we obtain that the function $D^qY$ is continuous  for any $0\leq q\leq q_0$ and any operator $D^q$ of the form (\ref{nlDDD}). Consequently, $Y\in C^{q_0}(\mC^2).$
\ssk

{\it (ii)} In view of item {\it (i)}, the function $D^qY$ is continuous. Then Proposition \ref{nlprop:Acont} implies that (\ref{nlDYAA}) holds for all $v_j,v_n\in\CC$. Since, obviously, $|D^qG^0(\VV_\psi(|v_j|),v_n)|\leq CK(|v_j|,|v_n|),$ we have $|A_j^q(v_j,v_n)|\leq CK(|v_j|,|v_n|),$ so that the desired estimate holds. 
\qed
\ssk

{\it Proof of Proposition \ref{nlprop:Acont}}. We will prove only that the function $A_j^q$ is continuous, the continuity of $A_n^q$ can be obtained similarly. It suffices to show that for any $v_n\in\mC$ the function $A_j^q$ is continuous at the point $(0,v_n)$.  Since the function $D^qG^0$ is continuous, for small  $|v_j|$ we have that $A^q(v_j,v_n)$ is close to 
\bee\label{nlAAAAA}
D^qG^0(0,v_n)\Big(\int\limits_0^{\psi_j} e^{i(m_j-l_j)(\psi-\psi_j)}\,d\psi 
-\frac{1}{2\pi}\int_0^{2\pi} \int\limits_0^{\xi}  e^{i(m_j-l_j)(\psi-\psi_j)}\,d\psi \,d\xi\Big).
\eee 
Assume first that $m_j-l_j\neq 0$. By the direct computation we see that the integral from (\ref{nlAAAAA}) is independent from $\psi_j$ and equals to $i(l_j-m_j)^{-1}$. Consequently, $A^q(v_j,v_n)\ra i(l_j-m_j)^{-1} D^qG^0(0,v_n)$ as $v_j\ra 0$, so that   $A^q$ is continuous at the point $(0,v_n)$.

Let now $m_j=l_j$. We will show that in this case $D^qG^0(0,v_n)\equiv 0$. In view of (\ref{nlAAAAA}), this will imply the desired continuity of $A^q$. Since the function $D^qG^0$ is $C^1$-smooth, its Fourier coefficients are continuous with respect to $|v_j|,|v_n|$. It means that  the Fourier series for $D^qG^0$ has the form (\ref{nlDh}) for all $v_j,v_n\in\mC$, and not only for $v_j,v_n\neq 0$. Since at the point $(0,v_n)$ it can not depend on the angle $\psi_j$, then the Fourier coefficient $\hat G_k(0,|v_n|)$ may not vanish only if $k$ satisfies $s_j(k)=0$.  In view of $m_j=l_j$, the relation $s_j(k)=0$ is equivalent to $k=0$. Since in the Fourier series for $D^qG^0$ there is no term with $k=0$, we obtain that all the Fourier coefficients $\hat G_k(0,|v_n|)$ vahish, so that $D^qG^0(0,v_n)\equiv 0$.\qed

\section{Generalizations}
\label{nlsec:generalizations}
\subsection{Non dissipative case}
\label{nlsec:nondissipative}

Let us briefly discuss what happens if functions $g_j(u)$ do not have dissipative properties, i.e. if assumption {\it Hg(ii)} is not satisfied. In this case for $p=2$ the estimate, similar to that of Lemma \ref{nllem:est} holds, but is not uniform in time. Then for $p=2$ the main results are essentially the same except those concerning stationary measures: Theorems \ref{nltheo:fin_dyn} and \ref{nltheo:measure} hold true but Theorem \ref{nltheo:stmintro} fails. 
Their  proofs do not change.

\subsection{Defects}
\label{nlsec:defects}

In this section we briefly discuss the situation when some rotators are "defective": there exists a bounded region $\CC_D\subset\mZ^d$ which does not depend on $N$, such that the rotators situated there rotate in arbitrary directions, so their spins are not alternated. In this case the system of rotators has resonances of the first order, and we can not completely remove the leading order of the interaction potential by the canonical change of variables, as we did before. However, we are able to remove its part, responsible to the interaction between non defective rotators. 

Let us denote by $\ov\CC_{D}:=\{k:\,|k-\CC_{D}|\leq 1\}$ the "closure" of $\CC_{D}$, where $|k-\CC_{D}|=\inf\limits_{l\in \CC_{D}}|k-l|$, and let $M$ be the number of sites in $\CC\sm\ov\CC_D$. 
Denote also
$$\CC_{I}:=\{k:\,|k-\ov\CC_{D}|=1 \mbox{ or } 2\}\quad\mbox{and}\quad\CC_{G}:=\{k:\,|k-\ov\CC_{D}|\geq 3\},$$
where "I" means "Intermediate" and "G" means "Good".  For a vector $a=(a_j)_{j\in\CC}$ we will write 
$$a_{ND}:=(a_j)_{j\in\CC\setminus\ov\CC_D},$$
where "ND" means "Non Defective".

We make a global canonical change of variables, determined by a time-one map of the Hamiltonian flow 
 $X^s_{\sqrt\eps \Phi}$ with $\Phi=\sum\limits_{j,k\in\CC\sm\CC_D:|j-k|=1} \Phi_{jk}$, where $\Phi_{jk}$ is defined in (\ref{nlPhij}). Such transformation is well-defined since for $j,k\in\CC\sm\CC_D$ the denominator of  (\ref{nlPhij}) is separated from zero. We obtain a new Hamiltonian
\bee
\label{nlham_br}
\HH(J,\psi)=H_0(J)+\sqrt\eps H_1(J,\psi) + \eps H_2(J,\psi) + \eps\sqrt\eps  H_>(J,\psi),
\eee
where
$ 
\ds{H_1(J,\psi)=\frac12 \sum\limits_{j\mbox{ {\footnotesize or} }k\in\CC_D, |j-k|=1} G(|v_j-v_k|^2)}.
$ 


Let $u^\eps(t)$ be a unique solution of (\ref{nlini_e})-(\ref{nlini_c}) with the defective rotators and $I^\eps(t),\,\ph^\eps(t)$ be the corresponding vectors of actions and angles. By the change of variables above we obtain the processes $v^\eps(t), J^\eps(t)$ and $\psi^\eps(t)$. Arguing similarly to Lemma \ref{nllem:est} and using that the number of defective rotators does not depend on $N$ , we are still able to obtain uniform in $\eps, N, j$ and $t$ estimates on solution $v^\eps(t)$. 

In view of (\ref{nlham_br}), equation for non defective actions $J_{ND}(\tau)$, written in the slow time, turns out to be slow, and consequently the family of measures $\{\DD(J_{ND}^{\eps}(\cdot)), \; 0<\eps\leq 1\}$ is tight on $C([0,T],\, \mR^M)$. 
Take a subsequence $\eps_k$ such that $\DD(J_{ND}^{\eps_k}(\cdot)) \rightharpoonup  Q^0_{ND}$ as $\eps_k\rightarrow 0$. Since the transformation is $\sqrt{\eps}$ -close to identity, we get
\begin{equation}
\label{nllpd}
\DD(I_{ND}^{\eps_k}(\cdot)) \rightharpoonup  Q^0_{ND}\quad  \mbox{as} \quad \eps_k\rightarrow 0.
\end{equation}
We fix the subsequence $(\eps_k)$ for the next two theorems. 

\begin{theo}
\label{nltheo:fin_dynd}
The measure $ Q^0_{ND}$ is a distribution $\DD(I^0_{ND}(\cdot))$ of a weak solution $I^0_{ND}(\tau)=(I^0_j(\tau))_{j\in\CC\sm\ov\CC_D}$ of the system 
\begin{equation}
\label{nlaverd}
d I_{j} = (\RR_{j}(I)+\TT_{j}) \,d\tau + \sqrt{2I_{j}\TT_{j}}\,d\wid\beta_{j}, \quad {j}\in\CC_G,   
\end{equation}
 with the initial conditions $\DD(I_{ND}(0)) =\DD(I_{ND}(u_0))$, where $\wid\beta_{j}$ are standard real independent Brownian motions.
  Moreover, for any $j\in\CC\setminus\ov\CC_D$, we have
\begin{equation}
\nonumber
 \MO\Big( \sup\limits_{\tau\in [0,T]} e^{2\al I^0_j(\tau)} \Big)< C \mbox{ and }
\int\limits_0^T \PR(I_j^0(s)<\delta)\, ds \ra 0 \mbox{ as } \delta\ra 0, 
 \end{equation}
where the latter convergence is uniform in $N$.
\end{theo}  
Note that system of equations (\ref{nlaverd}) is not closed: it depends on $I_j,\,j\in\CC_I$ for which we can not obtain a limiting relation. 
For a general case we can say nothing about  uniqueness of limiting point (\ref{nllpd}) and about the uniformity in $N$ of convergence in Theorem \ref{nltheo:fin_dynd}. However, if  the functions $\RR_j$ are diagonal, so that $\RR_j(I)=\RR_j(I_j)$ for all $j\in\CC_G$, then the properties above hold in some sense, see Example \ref{nlexd} below. In the next two theorems we observe a similar situation.
  
Denote $\nu_\tau^\eps:=\DD(I^\eps_{ND}(\tau),\ph^\eps_{ND}(\tau))$ and for any function $h(\tau)\geq 0$, satisfying  $\int\limits_0^T h(\tau) \, d\tau=1$, set $\nu^{\eps}(h):=\int\limits_0^T h(\tau) \nu^{\eps}_\tau \, d\tau$. Moreover, denote $n^{0}(h):=\int\limits_0^T h(\tau) \DD(I_{ND}^{0}(\tau)) \, d\tau$, where $I^{0}_{ND}$ is a solution of equation (\ref{nlaverd}), obtained in Theorem \ref{nltheo:fin_dynd}. 
\begin{theo}
\label{nltheo:measured}
For any continuous function $h$ as above, we have
$$
\nu^{\eps_k}(h)\rightharpoonup n^{0}(h)\times d\ph \quad \mbox{as} \quad \eps_k \rightarrow 0
 .
$$
\end{theo} 
Let $\wid\mu^\eps$ be a unique stationary measure of equation (\ref{nlini_e}) with the defective rotators. Let us denote by $\Pi_{I_{ND}}$ and $\Pi_{\ph_{ND}}$ the projections to the spaces of actions $I_{ND}$ and angles $\ph_{ND}$ correspondingly. The family of measures $\{\Pi_{I_{ND}*}\wid\mu^\eps,\,0<\eps\leq 1\})$ is tight. Take any limiting point $\Pi_{I_{ND}*}\wid\mu^{\eps_k}\raw \pi_{ND}$ as $\eps_k\ra 0$.
 \begin{theo}
\label{nltheo:stmd}
The measure $\pi_{ND}$ is a stationary measure of  equation (\ref{nlaverd}). Moreover, 
$$
(\Pi_{I_{ND}}\times\Pi_{\ph_{ND}})_* \wid\mu^{\eps_k} \rightharpoonup \pi_{ND}\times d\ph  \quad \mbox{as} \quad \eps_k \rightarrow 0. 
$$
\end{theo}
The proofs of Theorems \ref{nltheo:fin_dynd}, \ref{nltheo:measured} and \ref{nltheo:stmd} repeat the proofs of the corresponding theorems in the non-defective case.
\begin{example}
\label{nlexd} 
Let us consider the situation when the function $\RR_j(I)$ depends only on $I_j$ for all $j\in\CC_G$. For instance, this happens when the dissipation is diagonal, for more examples see Section \nolinebreak \ref{nlsec:example}. Then the system of equations (\ref{nlaverd}) is closed and depends only on $(I_j)_{j\in\CC_G}$. It has a unique weak solution and, consequently, the limiting measure $\Pi_{G*} Q^0_{ND}$ is unique.\footnote{Here $\Pi_G$ denotes the projection on  $\CC_G$} In this case the convergence $\DD\big((I_j(\cdot))_{j\in\CC_G})\big)\rightharpoonup \Pi_{G*} Q^0_{ND}$ as $\eps\ra 0$ holds and is uniform in $N$. Similarly, for the restrictions on $G$ of  measures from  Theorems \ref{nltheo:measured} and \ref{nltheo:stmd}, the corresponding convergences as $\eps\ra 0$ hold and are uniform in $N$. 
\end{example}

\appendix
\numberwithin{equation}{section}
\section{Appendix: The Ito formula in complex coordinates}
\label{nlcomplexIto}

Let $\{\Om,\,\FF,\,\PR;\,\FF_t\}$ be a filtered probability space and $v(t)=(v_k(t))\in\mC^N$ be a complex Ito process on this space of the form
\bee\nonumber
dv=b\,dt + W\,dB.
\eee
Here  $b(t)=(b_k(t))\in\mC^N$; $B=(\beta,\ov \beta)^T,$ $T$ denotes the transposition and $\beta=(\beta_k)\in\mC^N$, $\beta_k$ are standard independent complex Brownian motions; the $N\times 2N$-matrix $W$ consists of two blocks $(W_1,W_2)$, so that $W\,dB= W^1\, d\beta + W^2\, d\ov\beta$, where $W^{1,2}(t)=(W^{1,2}_{kl}(t))$ are $N\times N$ matrices with complex entires. The processes $b_k(t),\,W^{1,2}_{kl}(t)$ are $\FF_t$-adapted and  
assumed to satisfy usual growth conditions, needed to apply the Ito formula. Let  
\bee
\label{nlddd}
d^1_{kl}:=(\ov {W^1}W^{1T}+ \ov {W^2}W^{2T})_{kl} \quad\mbox{and}\quad d_{kl}^2:= ({W^2} {W^{1T}} +{W^1} {W^{2T}})_{kl}.
\eee
Denote by $(WdB)_k$ the $k$-th element of the vector $WdB$.
\begin{prop}
\label{nlprop:Ito}
Let $h:\,\mC^N\ra\mR$ be a $C^2$-smooth function. Then
\bee
\nonumber
\frac {dh(v(t))}{2}=
\sum\limits_{k}\frac{\chp h}{\chp \ov v_k}\cdot b_k\,  dt 
+ \sum\limits_{k,l}\left( \frac{\chp^2 h}{\chp \ov v_k \chp v_l}d^1_{kl} + \Ree \Big( \frac{\chp^2 h}{\chp v_k \chp v_l}d^2_{kl} \Big) \right)\, dt
+ \sum\limits_k \frac{\chp h}{\chp \ov v_k}\cdot  (W d B)_k .
\eee
\end{prop}
{\it Proof.} The result follows from the usual (real) Ito formula.
\qed 
\ssk

Consider the vectors of actions and angles $J=J(v)\in\mR^N_{0+}$  and $\psi=\psi(v)\in\mT^N$. Using formulas $\chp_{v_k}\psi_k=(2iv_k)^{-1}$ and $\chp_{\ov v_k}\psi_k=-(2i\ov v_k)^{-1}$, by Proposition \ref{nlprop:Ito} we get 
\bee\label{nlitoforactions}
dJ_k=(b_k\cdot v_k + d_{kk}^1)\,d t + d M^J_k, \quad 
d\psi_k=\frac{b_k\cdot(iv_k)- \Imm (\ov v_kv_k^{-1}d_{kk}^2)}{|v_k|^2}\,d t + d M^\psi_k,
\eee
where the martingales $M^J_k(t):=\int\limits_{t_0}^t v_k\cdot  ( W d B)_k $ and $M^\psi_k=\int\limits_{t_0}^t\frac{iv_k}{|v_k|^2}\cdot( W d B)_k$ for some $t_0<t$.
By the direct computation we obtain
\bpp\label{nlprop:diffusion}
The diffusion matrices for the $J$- and $\psi$-equations in (\ref{nlitoforactions}) with respect to the real Brownian motion $(\Ree \beta_k, \Imm \beta_k)$ have the form $S^J=(S^J_{kl})$ and $S^\psi=(S^\psi_{kl})$, where
\bee\label{nlSSS}
 S^J_{kl}=\Ree(v_k \ov v_l  d_{kl}^1+ \ov v_k \ov v_l d_{kl}^2) \quad\mbox{and}\quad S^\psi_{kl}=\Ree(v_k \ov v_l  d_{kl}^1 -\ov v_k \ov v_l d_{kl}^2 )(|v_k||v_l|)^{-2}.
\eee
The quadratic variations of $M^J_k$ and $M^\psi_k$ take the form
\bee\label{nlvariation}
[M^J_k]_t=\int\limits_{t_0}^t S^J_{kk}\, ds \quad\mbox{and} \quad [M^\psi_k]_t=\int\limits_{t_0}^t S^\psi_{kk} \, ds. 
\eee
\epp

\section{Appendix: Averaging}
\label{nlapp:aver}

Consider a complex coordinates $v=(v_j)\in\mC^N$ and the corresponding vectors of actions $J=J(v)$ and 
angles $\psi=\psi(v)$. Consider a function $P:\mC^N\mapsto\mR$ and write it in action-angle coordinates, $P(v)=P(J,\psi)$. Its averaging 
$$\lan P \ran:=\int\limits_{\mT^N} P(J,\psi)\, d\psi$$ 
is independent of angles and can be considered as a function $\lan P \ran(v)$ of $v$, or as a function $\lan P \ran\big((|v_j|)_j\big)$ of $(|v_j|)_j$, or as a function $\lan P \ran(J)$ of $J$. 
\bpp\label{nlprop:aver}
Let  $P\in\LL_{loc}(\mC^N).$ Then
 
{\it (i)} Its averaging $\lan P \ran\in\LL_{loc}(\mR^N_{+0})$ with respect to $(|v_j|)$.
 
{\it (ii)} If $P$ is $C^{2s}$-smooth then $\lan P\ran$ is $C^{2s}$-smooth with respect to $v$ and $C^{s}$-smooth with respect to $J$.
\epp
{\it Proof.} {\it (i)} Is obvious.

{\it (ii)} The first assertion is obvious. To prove the second consider the function $\hat P:x\in\mR^N\mapsto \mR$, $\hat P(x):=\lan P\ran|_{v=x}$. Then $\hat P(x)=\lan P\ran (J)$, where $J_j=x_j^2/2.$ The function $\hat P$ is $C^{2s}$-smooth and even in each $x_j$. Any function of finitely many arguments with this property is known to be a $C^s$- smooth function of the square arguments $x_j^2$ (see \cite{Whi}).  
\qed

\end{document}